\documentclass[12pt]{article}
\usepackage{hyperref}
\hypersetup{
    colorlinks=true,
    linkcolor=blue,
    citecolor=blue,
    filecolor=magenta,
    urlcolor=green,
}
\usepackage{bm}
\usepackage{natbib}
\usepackage{graphicx}
\usepackage{multirow}
\usepackage{enumitem}
\usepackage{amsmath,amsthm,amssymb,amsfonts}
\usepackage{mathrsfs}
\usepackage{authblk}
\usepackage{makecell}
\usepackage{subcaption}
\usepackage{float}
\usepackage{algorithm}
\usepackage{algorithmic}

\newtheorem{proposition}{Proposition}[section]
\newtheorem{theorem}{Theorem}[section]
\newtheorem{corollary}{Corollary}[section]
\newtheorem{assumption}{Assumption}[section]
\newtheorem{definition}{Definition}[section]
\newtheorem{remark}{Remark}[section]

\newtheorem{example}{Example}

\DeclareMathOperator{\Var}{Var}

\DeclareMathOperator{\E}{E}

\DeclareMathOperator*{\argmin}{arg\,min}

\def\f{Fr\'echet }
\def\bco{\iffalse}

\newcommand{\single}{\spacingset{1}}
\allowdisplaybreaks

\begin{document}

\def\spacingset#1{\renewcommand{\baselinestretch}%
{#1}\small\normalsize} \spacingset{1}

\title{\bf Regression Discontinuity Designs for Functional Data and Random Objects in Geodesic Spaces}
\author[1,*]{Daisuke Kurisu}
\author[2,*]{Yidong Zhou}
\author[3]{Taisuke Otsu}
\author[2,$^{\dagger}$]{Hans-Georg M\"uller}
\affil[1]{Faculty of Economics, The University of Tokyo, Japan}
\affil[2]{Department of Statistics, University of California, Davis, USA}
\affil[3]{Department of Economics, London School of Economics, UK}
\maketitle

\renewcommand*{\thefootnote}{\fnsymbol{footnote}}
\footnotetext[1]{The first two authors contributed equally to this work and are listed alphabetically.}
\footnotetext[2]{Corresponding author: hgmueller@ucdavis.edu.}

\bigskip
\begin{abstract}
Regression discontinuity designs (RDDs) are widely used for causal inference in observational studies with cutoff-based treatment assignment, primarily for Euclidean outcomes. We propose the geodesic regression discontinuity design (GRDD), which extends RDDs to complex non-Euclidean outcomes, including networks, compositional data, functional data, and other random objects in geodesic metric spaces. Since algebraic operations are unavailable in such spaces, we define the causal effect at the cutoff as the geodesic connecting the local Fr\'echet means of untreated and treated outcomes, recovering the classical local average treatment effect in the scalar case. Estimation is conducted intrinsically via local Fr\'echet regression to preserve geometric validity and interpretability. For inference, we adopt an extrinsic approach by embedding the metric space into a Hilbert space, enabling tractable asymptotic analysis. We establish asymptotic normality and develop bootstrap-based procedures for hypothesis testing and confidence intervals for the treatment effect magnitude. We also propose a data-adaptive bandwidth selection method tailored to RDDs in metric spaces and study its empirical performance. Applications include compositional voting outcomes in UK elections and daily CO concentration curves after the Taipei metro introduction, and we extend the framework to fuzzy designs with imperfect compliance.
\end{abstract}

\noindent%

\medskip
\noindent{\it MSC2020 subject classifications:} 62D20, 62R20
\vfill

\newpage
\newcommand{\double}{\spacingset{1.75}}
\double

\section{Introduction}
In causal inference, discontinuities in regression functions induced by an assignment variable can provide useful information for identifying causal effects. As a quasi-experimental design, the regression discontinuity design (RDD) has been widely applied in observational studies across fields such as economics, education, environmental studies, epidemiology, and political science to estimate causal effects at a cutoff point. In an RDD, treatment assignment is determined by whether a running variable crosses a known threshold, so that units on either side of the cutoff receive different treatments. Although the data are observational, this rule creates treatment and control groups that are locally comparable near the cutoff, making the design analogous to a randomized experiment in a neighborhood of the threshold. Causal effects are therefore identified by comparing outcomes for units just above and below the cutoff, with those on one side serving as counterfactuals for those on the other. For the conventional RDD with a scalar outcome, the causal parameters of interest are identified by differences in the left and right limits of the conditional mean functions, and various estimation and inference methods are available in the literature; see, e.g., \citet{imb:08}, \citet{catt:20}, \citet{catt:22}, an edited volume by \citet{catt:17}, and references therein.

This paper extends RDD analysis beyond the conventional setting of scalar outcomes to a broad class of complex non-Euclidean outcomes. Specifically, we develop a framework for causal inference that accommodates not only scalar or vector-valued responses but also networks, distributions, compositional data, functional data, and other types of random objects that reside in geodesic metric spaces. To achieve this goal, a major challenge is how to quantify a causal effect of interest in this generalized RDD setting, since algebraic operations are not available and taking differences is thus not feasible. To overcome this challenge, we express the causal effect at the cutoff as the geodesic connecting the local Fr\'echet means of the untreated and treated potential outcomes. This object reduces to the ordinary local average treatment effect in the scalar or vector case and thus provides a natural geometric extension of the classical RDD estimand.

A key feature of our framework is the distinction between intrinsic estimation and extrinsic inference. For estimation, we adopt an intrinsic approach based on local Fr\'echet regression, ensuring that all estimated quantities remain in the original metric space $\mathcal{M}$ and thus retain full interpretability as valid elements of $\mathcal{M}$. For statistical inference, however, the situation is fundamentally different: in general metric spaces, the lack of linear structure precludes the formulation of suitable central limit theorems and tractable limiting distributions. We therefore employ an extrinsic approach by embedding $\mathcal{M}$ into a Hilbert space \citep{scho:38,sejd:13,mull:20:10}, where the linear structure enables asymptotic analysis and hypothesis testing. We note that a key limitation of extrinsic approaches for estimation is their lack of reversibility: such embeddings are typically injective but not surjective, so that an inverse map from the Hilbert space to the original space generally does not exist. As a result, representing the treatment effect in the linear embedding space may involve elements that cannot be mapped back to $\mathcal{M}$ and therefore do not correspond to valid objects. This can lead to distortions even in relatively simple settings such as Wasserstein spaces \citep{bigo:17,caze:18,pego:22,chen:23:1} with an ensuing loss of interpretability. Our framework resolves this tension by using intrinsic methods for estimation to preserve validity and interpretability, while conducting inference in the extrinsic framework, since inference does not in any way rely on the inverse map back to $\mathcal{M}$.

Building on this formulation, we estimate the causal effect using local Fr\'echet regression, a regression method for metric space-valued responses that generalizes local linear regression \citep{mull:19:6}, and establish convergence rates for the proposed estimator. For practical implementation, we introduce a data-adaptive bandwidth selection procedure tailored to the boundary nature of regression discontinuity problems, based on one-sided cross-validation near the cutoff. Furthermore, we develop an inferential framework by deriving asymptotic normality for embedded estimators and proposing multiplier-bootstrap procedures for hypothesis testing and confidence intervals for the magnitude of the treatment effect.

After developing the above results for sharp RDDs under full compliance with the assignment rule, we extend our methodology to the fuzzy RDD setting with imperfect compliance. In the classical case of scalar outcomes, the average treatment effect for compliers at the cutoff can be identified by scaling the discontinuity in the outcome by the discontinuity in the treatment probability \citep{hahn:01}. Motivated by this framework, we propose a generalized notion of compliers' average treatment effect at the cutoff as a contrast of mappings of the local Fr\'echet means of the potential outcomes, establish its identification, and develop an estimation method for compliers' average treatment effect. Furthermore, we discuss identification and estimation of the geodesic from the local Fr\'echet mean of the untreated outcome to the treated one for compliers.

Metric statistics for complex data situated in metric spaces is an emerging field of statistics \citep{mull:24}, and the literature on causal inference for outcomes in metric spaces is rapidly growing. \citet{lin:23} and \citet{eck:24} studied estimation of the average treatment effect under the unconfoundedness assumption for distributional and functional outcomes, respectively, and their results were further extended by \citet{kuri:24} for a general class of geodesic metric spaces. For distributional outcomes, \citet{AtIm06}, \citet{guns:23}, \citet{GuHsLe24}, and \citet{toro:24} proposed extensions of the synthetic control and difference-in-differences methods. Using the notions of geodesic metric spaces, \citet{zhou:25} and \citet{kuri:25:1} developed general methodologies for the difference-in-differences and synthetic control methods, respectively, to accommodate not only functional or distributional outcomes but also other non-Euclidean random objects.

Very recently and independently of our work, \citet{va:25} investigated regression discontinuity designs for univariate distributional outcomes; see also \citet{fra12}, who proposed a quantile RDD estimator for scalar outcomes. In particular, \citet{va:25} studies the local average quantile treatment effect and develops corresponding nonparametric estimation and inference procedures, including asymptotic theory and bootstrap-based methods. Their approach exploits the representation of univariate distributions via quantile functions, which form a subset of a Hilbert space, and this structure is specific to univariate distributions. As a result, it does not readily extend to multivariate distributions or more general random objects such as networks, symmetric positive-definite matrices, or compositional data. In contrast, our framework formulates RDD for outcomes in general geodesic metric spaces and establishes identification of treatment effects via Fr\'echet means. As a special case, when the outcome is a univariate distribution equipped with the Wasserstein metric, our identification results recover the local average quantile treatment effect considered in \citet{va:25}.

We illustrate the proposed methodology primarily through an application to multiparty election outcomes in the United Kingdom, where the response is a vote-share composition across political parties and treatment is determined by prior Conservative victories. This application highlights both the intrinsic handling of compositional outcomes and the new inferential tools developed in this paper, including comparison with componentwise scalar RDD analyses. As a secondary illustration, we also consider functional air-pollution outcomes associated with the Taipei Metro expansion; this application is presented in the Supplement and is intended mainly to demonstrate how GRDD can be implemented for irregularly observed functional data.

This paper is organized as follows. Section~\ref{sec:pre} presents the basic concepts for geodesic metric spaces and motivating examples. Section~\ref{sec:GSRD} introduces geodesic RDDs for the sharp design case, establishes identification of the geodesic local average treatment effect, and studies estimation via local Fr\'echet regression. Section~\ref{sec:inf} develops inference for the magnitude of the treatment effect using Hilbert space embedding and a multiplier bootstrap. Section~\ref{sec:sim} presents the proposed bandwidth selector and evaluates finite-sample performance in simulations for network-valued outcomes. Section~\ref{sec:app} applies the methodology to UK parliamentary election data. Section~\ref{sec:FRD} extends the framework to fuzzy designs with imperfect compliance. Concluding remarks are provided in Section~\ref{sec:concl}.

\section{Preliminaries}\label{sec:pre}
In this section, we lay the geometric foundations for modeling and analyzing non-Euclidean outcomes, introduce uniquely geodesic spaces, and discuss motivating examples.  

\subsection{Metric and geodesic spaces}\label{sub:metric}
Let $(\mathcal{M}, d)$ be a metric space. A continuous mapping $\gamma: [0,1] \to \mathcal{M}$ is called a \textit{curve}, and its length is defined by $\ell(\gamma) = \sup \sum_{i=1}^k d\big(\gamma(t_{i-1}), \gamma(t_i)\big)$, where the supremum is taken over all finite partitions $0 = t_0 < t_1 < \cdots < t_k = 1$ of $[0,1]$. A curve is said to be \textit{rectifiable} if $\ell(\gamma) < \infty$.

The space $(\mathcal{M}, d)$ is a \textit{length space} if the distance between any two points $\alpha, \beta \in \mathcal{M}$ equals the infimum of the lengths of all rectifiable curves joining them: $d(\alpha, \beta) = \inf_\gamma \ell(\gamma)$, where the infimum ranges over all curves with $\gamma(0) = \alpha$ and $\gamma(1) = \beta$. If this infimum is attained for all such pairs, the space is called a \textit{geodesic space} \citep{brids:99}.

A \textit{geodesic} between two points is a curve $\gamma: [0,1] \to \mathcal{M}$ satisfying
\[d\big(\gamma(s), \gamma(t)\big) = |t - s|d(\alpha, \beta) \quad \text{for all } s,t \in [0,1],\]
with $\gamma(0) = \alpha$ and $\gamma(1) = \beta$. When such a geodesic exists and is unique for every pair $\alpha, \beta \in \mathcal{M}$, the space is said to be \textit{uniquely geodesic}. Hereafter, we denote the geodesic with the starting point $\gamma(0)=\alpha$ and the ending point $\gamma(1)=\beta$ as $\gamma_{\alpha,\beta}$.

\subsection{Examples of uniquely geodesic spaces}
We now introduce five representative examples of uniquely geodesic spaces commonly encountered in applications. We will revisit some of these example spaces to demonstrate the practical performance of the proposed approach in both simulation studies and real data analyses.

\begin{example}[Functional Data]\label{exm:fun}
Functional data arise when observations are naturally represented as curves or functions, typically indexed by time or space. Such data are common in longitudinal studies, environmental monitoring, and biomedical imaging \citep{rams:05,hsin:15,mull:16:3}. There is also a substantial literature on regression with functional responses; see, for example, \citet{fara:97}, \citet{chio:04} and \citet{morr:15}. Let $\mathcal{T}$ be a compact interval and consider the space $L^2(\mathcal{T})$ of square-integrable functions on $\mathcal{T}$. This Hilbert space is equipped with the inner product $\langle f, g \rangle = \int_\mathcal{T} f(t)g(t)\,dt$ and the associated $L^2$ metric $d_{L^2}(f, g) = \left( \int_\mathcal{T} \{f(t) - g(t)\}^2 \, dt \right)^{1/2}$. The space $L^2(\mathcal{T})$ is uniquely geodesic, with geodesics given by the path $\gamma_{f,g}(t) = (1 - t)f + t g$ for $t \in [0,1]$.
\end{example}

\begin{example}[Compositional Data]\label{exm:com}
Compositional data describe quantities that represent parts of a whole, such as proportions or percentages. Common examples include vote shares across political parties in election data, relative gene expression levels in transcriptomics, and dietary intake compositions in nutrition studies \citep{li:15}. Let $\Delta^{d-1} = \{ \mathbf{y} \in \mathbb{R}^d : y_j \geq 0, \sum_{j=1}^d y_j = 1 \}$ denote the unit simplex in $\mathbb{R}^d$. Through the square-root transformation $\mathbf{z} = \sqrt{\mathbf{y}}$, compositional data may be mapped to the positive orthant of the unit sphere $\mathcal{S}_+^{d-1}$ \citep{scea:11,scea:14}. The induced arc-length distance is given by $d_g(\mathbf{z}_1, \mathbf{z}_2) = \arccos(\mathbf{z}_1^\top \mathbf{z}_2)$ for $\mathbf{z}_1, \mathbf{z}_2 \in \mathcal{S}_+^{d-1}$. The geodesic joining $\mathbf{z}_1$ and $\mathbf{z}_2$ follows the shortest great-circle path on the sphere and is uniquely determined when the points are not antipodal.
\end{example}

\begin{example}[Symmetric Positive-Definite Matrices]\label{exm:spd}
Symmetric positive-definite (SPD) matrices are widely used to represent covariance structures, diffusion tensors, and kernel matrices in applications such as neuroimaging \citep{dryd:09}, signal processing \citep{arna:13}, and machine learning \citep{cher:16}. Let $\mathrm{Sym}_m^+$ denote the set of $m \times m$ SPD matrices. This space can be endowed with multiple geodesic metrics. A natural starting point is the Frobenius metric, $d_F(A, B) = \|A - B\|_F$, under which the geodesic between $A$ and $B$ is simply the linear interpolation $\gamma_{A, B}(t) = (1 - t)A + tB$ for $t \in [0,1]$. Alternatively, under the Log-Euclidean metric \citep{arsi:07} $d_{\mathrm{LE}}(A, B) = \|\log A - \log B\|_F$, the geodesic is given by $\gamma_{A,B}(t) = \exp\left((1 - t)\log A + t \log B\right)$. Both metrics induce a uniquely geodesic structure on $\mathrm{Sym}_m^+$. Other metrics such as the power metric \citep{dryd:09} and Log-Cholesky metric \citep{lin:19:1} also endow $\mathrm{Sym}_m^+$ with a uniquely geodesic structure.
\end{example}

\begin{example}[Networks]\label{exm:net}
Network data capture relationships or interactions between entities, such as social connections, transportation links, or gene regulatory networks. A common representation is via graph Laplacians \citep{kola:14,mull:22:11,seve:22}. Consider the space of undirected, weighted graphs with $m$ nodes and bounded edge weights, where each graph is encoded by its Laplacian matrix. Let $\mathcal{L}$ denote the collection of such Laplacians, and equip it with the Frobenius metric $d_F(L_1, L_2) = \|L_1 - L_2\|_F$. Since $\mathcal{L}$ forms a convex and closed subset of $\mathbb{R}^{m^2}$, the Frobenius metric yields a Euclidean geometry. Geodesics are given by straight-line paths: $\gamma_{L_1, L_2}(t) = (1 - t)L_1 + tL_2$.
\end{example}

\begin{example}[Univariate Probability Distributions]\label{exm:mea}
Probability distributions arise as natural data objects in diverse applications such as econometrics \citep{pete:22}, population pyramids \citep{hron:16}, and multi-cohort studies \citep{zhou:23}. Let $\mathcal{W}$ denote the space of Borel probability measures on a closed interval $\mathcal{I} \subset \mathbb{R}$ with finite second moments. This space becomes a geodesic metric space when equipped with the 2-Wasserstein metric \citep{ambr:08}, $d_{\mathcal{W}}(\mu, \nu) = \left(\int_0^1 \left\{ F_\mu^{-1}(s) - F_\nu^{-1}(s) \right\}^2 ds\right)^{1/2}$, where $F_\mu^{-1}$ and $F_\nu^{-1}$ are the quantile functions of $\mu$ and $\nu$, respectively. The unique geodesic connecting $\mu$ and $\nu$ under this metric is given by McCann's interpolant \citep{mcca:97}, $\gamma_{\mu, \nu}(t) = \{\mathrm{id}+t(F_{\nu}^{-1}\circ F_{\mu}-\mathrm{id})\}_\#\mu$ for $t \in [0,1]$, where $\mathrm{id}$ denotes the identity map, $F_{\mu}$ is the cumulative distribution function of $\mu$, and $\tau_\#\mu$ denotes the pushforward of $\mu$ under the map $\tau$. An alternative geometry is induced by the Fisher-Rao metric \citep{rao:45,sriv:07,dai:22}, which is particularly useful when probability densities exist and are smooth. For distributions with the Lebesgue densities $f_\mu$ and $f_\nu$, the Fisher-Rao distance is defined as $d_{\mathrm{FR}}(\mu, \nu) = \arccos\left( \int_{\mathcal{I}} \sqrt{f_\mu(x) f_\nu(x)}\, dx \right)$.
\end{example}

\section{Geodesic regression discontinuity designs}\label{sec:GSRD}
In this section, we develop the proposed methodology for geodesic regression discontinuity designs (GRDD), define the causal estimand of interest and establish its identification in uniquely geodesic metric spaces. We then present an estimation strategy based on local \f regression. For clarity of exposition, we focus on the sharp design, in which treatment is deterministically assigned based on whether the running variable crosses a known cutoff. An extension to fuzzy designs that allows for imperfect compliance is discussed in Section~\ref{sec:FRD}.

\subsection{Identification}\label{subsec:GSRD-id}
We begin with the canonical sharp regression discontinuity design. Let $(\Omega,\mathcal F,\mathbb P)$ be a probability space, let $R:\Omega\to\mathbb R$ denote the running variable, and let $c\in\mathbb R$ be a fixed cutoff. The cutoff assignment is $Z=1\{R\ge c\}$. In the sharp design, treatment receipt coincides with the cutoff assignment, so that $T=Z=1\{R\ge c\}$ almost surely.

For each treatment state $t\in\{0,1\}$, let $Y(t):\Omega\to\mathcal M$ denote the corresponding potential outcome, where $(\mathcal M,d)$ is a separable uniquely geodesic metric space such that $(\mathcal M,\mathcal B(\mathcal M))$ is a standard Borel space, with $\mathcal B(\mathcal M)$ denoting the Borel $\sigma$-field generated by the metric topology. The observed outcome is $Y=Y(T)$. Thus, units with running-variable values below the cutoff receive control, whereas those with values at or above the cutoff receive treatment. We assume that the essential range of $Y(t)$ is bounded, meaning that $\mathbb{P}\{d(\nu_0, Y(t))\le D\}=1$ for some $\nu_0\in\mathcal{M}$ and $D<\infty$. The fuzzy design considered in Section~7 relaxes the perfect-compliance condition $T=Z$, while retaining the deterministic cutoff assignment
$Z=1\{R\ge c\}$.

For a random object $Y\in\mathcal{M}$, the Fr\'echet mean of $Y$, which generalizes the notion of expectation to metric spaces, is defined as
\[\E_\oplus[Y]=\argmin_{\nu \in\mathcal{M}}\E[d^2(\nu, Y)].\]
The existence and uniqueness of the minimizer is guaranteed for Hadamard spaces \citep{stur:03} and the example spaces described in Section \ref{sec:pre}; see \citet{mull:19:6} for further background. The conditional Fr\'echet mean of $Y$ given a covariate $X\in\mathbb{R}^p$ \citep{mull:19:6} is analogously defined as
\[\E_\oplus[Y|X]=\argmin_{\nu \in\mathcal{M}}\E[d^2(\nu, Y)|X],\]
which reduces to the usual conditional expectation when $\mathcal{M} = \mathbb{R}$ and $d$ is the Euclidean distance.

We consider the following assumptions.

\begin{assumption}\label{asm:basic} \quad 
\begin{itemize}
\item[(i)] Under the stable unit treatment value assumption (SUTVA), the potential outcomes $Y(0), Y(1) \in \mathcal{M}$ are well-defined, and the observed outcome satisfies $Y = Y(0)$ if $T=0$ and $Y = Y(1)$ if $T=1$.
\item[(ii)] For each $t\in\{0,1\}$, there exists $\nu_0\in\mathcal{M}$ such that $\E[d^2(\nu_0, Y(t))|R=r]<\infty$ for every $r$ in a neighborhood of $c$. Moreover, the conditional Fr\'echet mean $\E_\oplus[Y(t)|R=r]$ exists uniquely for every such $r$ and is continuous at $r=c$.
\item[(iii)] We observe an independent and identically distributed sample $\{Y_i, R_i\}_{i=1}^n$ drawn from $(Y,R) \in \mathcal{M} \times \mathbb{R}$.
\end{itemize}
\end{assumption}

These assumptions parallel those in the classical RDD framework (e.g., \citealp{imb:08}), with the key difference that outcomes now lie in a general metric space $\mathcal{M}$. Because treatment is a deterministic function of the running variable, unconfoundedness $(Y(0),Y(1))\perp T|R$ holds automatically, while the overlap condition is violated by design. Assumption~\ref{asm:basic}(i) imposes SUTVA and therefore rules out interference and hidden variations of treatment, as is standard in RDD settings \citep[see, e.g.,][]{imb:05}. Assumption~\ref{asm:basic}(ii) is the metric-space analogue of the conventional continuity condition for conditional mean functions in RDD. It permits the counterfactual conditional Fr\'echet means at the cutoff to be recovered from their respective one-sided limits.

To define the causal estimand of interest, recall that the geodesic defined in Section \ref{sub:metric} with the starting point $\gamma(0)=\alpha$ and the ending point $\gamma(1)=\beta$ is denoted by $\gamma_{\alpha,\beta}$.

\begin{definition}[Geodesic local average treatment effect at the cutoff]\label{def:GLATE} The geodesic local average treatment effect (GLATE) at the cutoff $c$ is defined as the geodesic connecting the conditional \f means of the untreated and treated potential outcomes,
\begin{align*}
\tau_{\mathrm{GRDD}}:=\gamma_{ \E_\oplus[Y(0)|R=c],\E_\oplus[Y(1)|R=c]}.
\end{align*}
\end{definition}

In the classical Euclidean setting where $Y \in \mathbb{R}$, the sharp RDD estimand is given by the jump in the conditional mean function
\[\tau_{\mathrm{RDD}}:= \E[Y(1)|R=c] - \E[Y(0)|R=c].\]
The proposed GLATE can be interpreted as a natural extension of $\tau_{\mathrm{RDD}}$ to non-Euclidean outcomes, obtained by replacing Euclidean expectations with conditional Fr\'echet means and linear differences with geodesics. The resulting geodesic $\tau_{\mathrm{GRDD}}$ captures both the direction and magnitude of the treatment effect: its direction corresponds to the transition from $\E_\oplus[Y(0)|R=c]$ to $\E_\oplus[Y(1)|R=c]$, analogous to the sign of $\tau_{\mathrm{RDD}}$, while its length, $d(\E_\oplus[Y(0)|R=c], \E_\oplus[Y(1)|R=c])$, corresponds to the magnitude of the effect, analogous to the absolute value of $\tau_{\mathrm{RDD}}$.

\begin{remark}[Interpretation of GLATE]\label{rem:GLATE_interp}
In general metric spaces, the geodesic connecting the counterfactual Fr\'echet means is generally not equal to the Fr\'echet mean of the unit-level geodesics joining $Y_i(0)$ and $Y_i(1)$. Consequently, unlike the Euclidean setting, several distinct notions of treatment effects are possible. One possibility is the proposed GLATE, which is defined through the counterfactual Fr\'echet means and the geodesic connecting them. Another possibility would be to define treatment effects at the unit level through the geodesics joining $Y_i(0)$ and $Y_i(1)$ and then aggregate these geodesics across units. Such an approach would require additional structure to define an average of geodesics, for example by specifying a metric on the space of geodesics, and different choices may lead to different population-level estimands. The proposed GLATE therefore represents one particular notion of a population-level causal effect induced by the metric structure of the outcome space. It should be interpreted as a causal summary defined through counterfactual Fr\'echet means rather than as an aggregation of individual treatment-effect geodesics. While these two notions coincide in Euclidean spaces, they generally differ for metric space-valued outcomes, reflecting the richer geometric structure of non-Euclidean spaces.
\end{remark}

When the outcomes are univariate probability distributions equipped with the 2-Wasserstein metric (see Example~\ref{exm:mea}), the GLATE admits a natural interpretation via quantile functions. In this setting, geodesics correspond to linear interpolations between quantile functions, and the causal estimand introduced by \citet{va:25},
\[
\tau^{\mathrm{R3D}} := F^{-1}_{\E_\oplus[Y(1)|R = c]} - F^{-1}_{\E_\oplus[Y(0)|R = c]},
\]
can be viewed as the secant of the Wasserstein geodesic connecting the two distributions. This difference is well-defined due to the pseudo-linear structure of the univariate Wasserstein space and offers a convenient representation in quantile coordinates.

The geometric structure of our approach allows for broader generality. While conventional RDD methods may be adapted to distributional outcomes by summarizing them through Euclidean functionals (e.g., means or quantiles), such simplifications are often infeasible for more complex outcomes such as networks, compositional data, or manifold-valued objects. In contrast, the GLATE is intrinsically defined within the metric space, obviating the need for Euclidean (scalar or vector) representations. The following result establishes the identification of the GLATE.

\begin{theorem}\label{thm:GSRD-id}
Under Assumption~\ref{asm:basic}(i)--(ii), the causal estimand $\tau_{\mathrm{GRDD}}$ admits the representation
\begin{align}
\tau_{\mathrm{GRDD}} = \gamma_{\beta_0,\beta_1}, \label{eq:GSRD-id}
\end{align}
where $\beta_0 = \lim_{r \uparrow c}\E_\oplus[Y|R=r]$ and $\beta_1 = \lim_{r \downarrow c}\E_\oplus[Y|R=r]$.
\end{theorem}
This theorem says that the causal estimand, GLATE $\tau_{\mathrm{GRDD}}$ in Definition \ref{def:GLATE}, is identified by the statistical estimand, i.e. the geodesic $\gamma_{\beta_{0},\beta_{1}}$ from $\beta_{0}$ to $\beta_{1}$. We note that this identification result also applies to multivariate distributions endowed with the Fisher–Rao metric, whereas the extension of quantile-based RDD methods such as \citet{va:25} to the case of multivariate distributions remains a challenging open problem.

\subsection{Estimation}\label{subsec:GSRD-est}
To estimate the GLATE $\tau_{\mathrm{GRDD}}$, we employ the local \f regression (LFR) framework \citep{mull:19:6, mull:22:8}, which extends local linear regression \citep{fan:96} to accommodate responses that lie in general metric spaces. Here we impose a mild condition on the kernel function used to define the estimator.

\begin{assumption}\label{ass:ker}
The kernel function $k(\cdot)$ is supported on $[-1,1]$, zero outside its support, bounded, nonnegative, and positive and continuous on $(-1,1)$. 
\end{assumption}

This assumption, which is satisfied by standard choices such as the triangular kernel $k(x) = (1+x)1\{x \in [-1,0]\} + (1-x)1\{x \in [0,1]\}$ or uniform kernel $k(x) = 1\{x \in [-1,1]\}$, mirrors commonly assumed conditions in the scalar RDD literature (e.g., Assumption 2 in \citet{calo:14}). 

In the sharp regression discontinuity setup, we construct two LFRs, one on each side of the cutoff $c$, to estimate the conditional \f means of $Y$ given $R = c^-$ and $R = c^+$. Define $K_0(x) = k(x)1\{x < 0\}$ and $K_1(x) = k(x)1\{x \ge 0\}$ for use on the left and right-hand side of the cutoff point, respectively. By $t \in \{0,1\}$ we indicate the side of the cutoff and consider rescaled kernels  $K_{t,h_t}(x) = h_t^{-1} K_t(x/h_t)$ utilizing bandwidths $h_t > 0$ such that $h_t \to 0$ as $n \to \infty$. The LFR target on side $t$ is defined as the minimizer of the population weighted \f loss
\[
\nu_{t,\oplus} = \argmin_{\nu \in \mathcal{M}} M_{t,n}(\nu),\quad\text{where }M_{t,n}(\nu) = \E[s_t(c; R, h_t) d^2(\nu, Y)].
\]
Here the weight function $s_t(c; R, h_t)$ is given by
\[
s_t(c; R, h_t) = \frac{1}{\sigma_t^2} K_{t,h_t}(R - c)\left\{ \mu_{t,2} - \mu_{t,1}(R - c) \right\},
\]
with moment terms $\mu_{t,k} = \E[K_{t,h_t}(R - c)(R - c)^k]$ for $k = 0,1,2$ and normalization constant $\sigma_t^2 = \mu_{t,0} \mu_{t,2} - \mu_{t,1}^2$. This construction mirrors the equivalent weighted formulation of local linear regression in Euclidean settings and ensures that the weight function integrates to one. The construction and intuition behind LFR, including its connection to local linear regression in Euclidean settings, are further discussed  in Section S.1 of the Supplement.

We define the intermediate target $\tau_{\mathrm{GRDD}}^* = \gamma_{\nu_{0,\oplus},\nu_{1,\oplus}}$, which approximates the GLATE based on smoothed local \f estimates. The empirical GRDD estimator is defined as $\hat{\tau}_{\mathrm{GRDD}} = \gamma_{\hat{\nu}_{0,\oplus},\hat{\nu}_{1,\oplus}}$, where for $t \in \{0,1\}$,
\begin{align*}
\hat{\nu}_{t,\oplus} = \argmin_{\nu \in \mathcal{M}}\hat{M}_{t,n}(\nu),\quad\text{with }\hat{M}_{t,n}(\nu)= \frac{1}{n_t}\sum_{i=1}^n \hat{s}_t(c;R_i,h_t)d^2(\nu,Y_i).
\end{align*}
Here, $n_t = \sum_{i=1}^n 1\{T_i = t\}$ is the number of observations on side $t$, and the empirical weights are defined as
\[\hat{s}_t(c;R,h_t) = \frac{1}{\hat{\sigma}_t^2}K_{t,h_t}(R-c)\{\hat{\mu}_{t,2} - \hat{\mu}_{t,1}(R - c)\},\]
where $\hat{\mu}_{t,k} = n_t^{-1}\sum_{i=1}^nK_{t,h_t}(R_i-c)(R_i-c)^k$, and $\hat{\sigma}_t^2 = \hat{\mu}_{t,0}\hat{\mu}_{t,2} - \hat{\mu}_{t,1}^2$.

In the special case where $\mathcal{M} = \mathbb{R}$ and $d$ is the Euclidean distance, the LFR estimator $\hat{\nu}_{t,\oplus}$ corresponds to the intercept of the local linear regression fit, recovering the standard RDD estimator. For general metric space-valued outcomes, this construction provides a natural generalization of classical sharp RDD estimation, enabling causal inference even when subtraction is undefined and vector space structure is absent.

For $t \in \{0,1\}$, define the population target
\[
\mu_{t,\oplus}:= \E_\oplus[Y(t)|R=c] = \argmin_{\nu \in \mathcal{M}}M_{t,\oplus}(\nu,c),
\]
where $M_{t,\oplus}(\nu,c) = \E[d^2(\nu,Y(t))|R=c]$. To establish the convergence rate of the GRDD estimator, we impose the following assumptions. 
\begin{assumption}\label{ass:GSRD-rate} \quad 
\begin{itemize}
\item[(i)] For each $n$, $\nu_{0,\oplus}, \nu_{1,\oplus}, \hat{\nu}_{0,\oplus}$, and $\hat{\nu}_{1,\oplus}$ exist and are unique, the latter two almost surely. Additionally, for $t \in \{0,1\}$ and every $\varepsilon>0$, 
\begin{align*}
\inf_{d(\nu, \mu_{t,\oplus})>\varepsilon}\{M_{t,\oplus}(\nu, c) - M_{t,\oplus}(\mu_{t,\oplus},c)\} &>0,\\
\liminf_{n \to \infty} \inf_{d(\nu, \nu_{t,\oplus})>\varepsilon}\{M_{t,n}(\nu) - M_{t,n}(\nu_{t,\oplus})\} &>0.
\end{align*}
\item[(ii)] For $t \in \{0,1\}$, let $B_\delta(\mu_{t,\oplus}) \subset \mathcal{M}$ be the ball of radius $\delta$ centered at $\mu_{t,\oplus}$ and $N(\varepsilon,B_\delta(\mu_{t,\oplus}),d)$ be its covering number using balls of size $\varepsilon>0$. Then
\[
\int_0^1 \sqrt{1 + \log N(\delta \varepsilon, B_\delta(\mu_{t,\oplus}),d)}d\varepsilon = O(1)\ \text{as}\ \delta \to 0.
\]
\item[(iii)] There exists a neighborhood $\mathcal{I}$ of $c$ such that the density $f_R$ of $R$ is twice continuously differentiable on $\mathcal{I}$ with $f_R(c) > 0$. Moreover, for each $t \in \{0,1\}$, the reverse conditional densities $g_{y,t}$ of $R|Y(t)=y$ exist and are twice continuously differentiable for all $y \in \mathcal{M}$ and all $r \in \mathcal{I}$, with $\sup_{r \in \mathcal{I},\, y \in \mathcal{M}} \bigl(|g''_{y,0}(r)| + |g''_{y,1}(r)|\bigr) < \infty$. Additionally, for any open set $\mathcal{U} \subset \mathcal{M}$ and $t \in \{0,1\}$, the map $r \mapsto \int_{\mathcal{U}} dF_{Y(t)|R=r}(r,y)$ is continuous for $r \in \mathcal{I}$.
\item[(iv)] For $t \in \{0,1\}$ and $\nu\in \mathcal{M}$, there exist $\bar{\eta}>0$, $\bar{C}>0$, and $\beta_1>1$ such that
\[
M_{t,\oplus}(\nu,c) - M_{t,\oplus}(\mu_{t,\oplus},c) \geq \bar{C}d(\nu, \mu_{t,\oplus})^{\beta_1},
\]
provided $d(\nu, \mu_{t,\oplus}) <\bar{\eta}$.
\item[(v)] For $t \in \{0,1\}$ and $\nu\in \mathcal{M}$, there exist $\tilde{\eta}>0$, $\tilde{C}>0$, and $\beta_2>1$ such that
\[
\liminf_{n \to \infty}\{M_{t,n}(\nu) - M_{t,n}(\nu_{t,\oplus}) - \tilde{C}d(\nu,\nu_{t,\oplus})^{\beta_2}\}\geq 0, 
\]
provided $d(\nu,\nu_{t,\oplus})<\tilde{\eta}$. 
\end{itemize}
\end{assumption}
Assumption~\ref{ass:GSRD-rate} is standard in the LFR literature; see, for example, \citet{mull:19:6,mull:22:11}, where it is verified for Examples~\ref{exm:fun}--\ref{exm:mea}. Condition (i) is a standard separation condition that ensures consistency of the M-estimator (see, e.g., \citet[Chapter 3.2]{well:96}). Condition (ii) imposes a local entropy bound on metric balls centered at the population Fr\'echet means and is used to control the associated empirical processes. This condition is not required for functional outcomes in $L^2(\mathcal{T})$ or for univariate probability distributions equipped with the $2$-Wasserstein metric, as in Examples~\ref{exm:fun} and \ref{exm:mea}. In these settings, the outcomes admit natural Hilbert space representations through the identity map for functional data and through quantile functions for Wasserstein data. The required stochastic bounds can therefore be derived using Hilbert space methods and, when necessary, projection onto the corresponding closed convex image space; see \citet{mull:19:6}.

Condition (iii) specifies distributional smoothness requirements that ensure sufficient regularity of the population Fr\'echet objective in a neighborhood of the cutoff. The conditional distributions $F_{Y(t)|R}$ are well-defined because $(\mathcal{M},\mathcal{B}(\mathcal{M}))$ is a standard Borel space, which guarantees the existence of regular conditional distributions. Conditions (iv) and (v) characterize the bias and stochastic error rates, respectively. Together, conditions (ii), (iv), and (v) use tools from empirical process theory to control deviations of $M_{t,n}(\cdot) - M_{t,\oplus}(\cdot,c)$ and $\hat{M}_{t,n}(\cdot) - M_{t,n}(\cdot)$ around their respective minimizers $\mu_{t,\oplus}$ and $\nu_{t,\oplus}$.

To quantify the contrast among the geodesics, $\tau_{\mathrm{GRDD}}$, $\tau_{\mathrm{GRDD}}^*$ and $\hat{\tau}_{\mathrm{GRDD}}$, we define a distance between two geodesics. Let $\mathcal{G}(\mathcal{M})=\{\gamma_{\alpha, \beta}:\alpha, \beta\in\mathcal{M}\}$ be the space of geodesics on the uniquely geodesic space $(\mathcal{M},d)$. We first introduce the notion of a geodesic transport map.
\begin{assumption}\label{asp:ug}
Let $(\mathcal{M}, d)$ be a uniquely geodesic space. For any two points $\alpha, \beta \in \mathcal{M}$, there exists a \textit{geodesic transport map} $\Gamma_{\alpha,\beta}: \mathcal{M} \mapsto \mathcal{M}$ with the following property: $\Gamma_{\alpha,\beta}(\alpha)=\beta$ and for any $\omega \in \mathcal{M}$, there exists a unique point $\zeta \in \mathcal{M}$ such that $\Gamma_{\alpha, \beta}(\omega) = \zeta$.\end{assumption}
Specific constructions of geodesic transport maps for Examples \ref{exm:fun}--\ref{exm:mea} can be found in Section 3 of \citet{kuri:25:1}. We define a binary relation $\sim$ on $\mathcal{G}(\mathcal{M})$ through the geodesic transport map, where $\gamma_{\alpha_1, \beta_1}\sim\gamma_{\alpha_2, \beta_2}$ if and only if $\Gamma_{\alpha_1, \beta_1}(\omega)=\Gamma_{\alpha_2, \beta_2}(\omega)$ for all $\omega\in\mathcal{M}$. This relation is an equivalence relation on $\mathcal{G}(\mathcal{M})$ (Proposition 4.1, \citet{zhou:25}), and we denote the equivalence class of a geodesic $\gamma_{\alpha, \beta}$ by $[\gamma_{\alpha, \beta}]$. The set of all equivalence classes $\mathcal{G}(\mathcal{M})/\sim:=\{[\gamma_{\alpha, \beta}]: \gamma_{\alpha, \beta}\in\mathcal{G}(\mathcal{M})\}$ is the quotient space of $\mathcal{G}(\mathcal{M})$, for which we define a metric as follows. For any two equivalence classes $[\gamma_{\alpha_1, \beta_1}], [\gamma_{\alpha_2, \beta_2}]\in\mathcal{G}(\mathcal{M})/\sim$,
\begin{equation}\label{eq:dg}
    d_{\mathcal{G}}([\gamma_{\alpha_1, \beta_1}], [\gamma_{\alpha_2, \beta_2}])=d_{\mathcal{G},\omega_\oplus}([\gamma_{\alpha_1, \beta_1}], [\gamma_{\alpha_2, \beta_2}])=d(\Gamma_{\alpha_1, \beta_1}(\omega_\oplus), \Gamma_{\alpha_2, \beta_2}(\omega_\oplus)),
\end{equation}
where $\omega_\oplus\in\mathcal{M}$ is a fixed reference point, such as the \f mean of the outcome. Intuitively, $d_{\mathcal{G}}$ measures the distance between two geodesics by comparing where they map a common reference point $\omega_\oplus$ under their associated transport maps. In this sense, the distance reflects how differently the two geodesics act on the space, rather than comparing their endpoints directly. Note that $(\mathcal{G}(\mathcal{M})/\sim, d_{\mathcal{G}})$ is a metric space (Proposition 4.2 in \citet{zhou:25}). Additionally, we require the following regularity assumption. 
\begin{assumption}\label{asp:com}
There exists a constant $C>0$ such that for any  $\alpha_1,\alpha_2, \beta_1,\beta_2, \omega \in\mathcal{M}$, 
\[
d(\Gamma_{\alpha_1, \beta_1}(\omega), \Gamma_{\alpha_2, \beta_2}(\omega))\leq C\{d(\alpha_1, \alpha_2)+d(\beta_1, \beta_2)\}.
\]
\end{assumption}
We refer to Section S.2 of the Supplement in \citet{zhou:25} for a verification of Assumption \ref{asp:com} for Examples \ref{exm:fun}--\ref{exm:mea}. Letting $h_{\mathrm{max}}=\max\{h_0,h_1\}$ and $h_{\mathrm{min}} = \min\{h_0,h_1\}$, the convergence rate of the GRDD estimator is obtained as follows.
\begin{theorem}\label{thm:GSRD-rate}
Consider the setup of Section~\ref{subsec:GSRD-id} and suppose that Assumptions~\ref{asm:basic}, \ref{ass:ker}, \ref{asp:ug}, and \ref{asp:com} are satisfied.
\begin{itemize}
\item[(i)] Under Assumptions \ref{ass:GSRD-rate} (i)--(iv) and if $h_{\mathrm{max}} \to 0$ as $n \to \infty$, then
\begin{align*}
d_\mathcal{G}(\tau_{\mathrm{GRDD}},\tau_{\mathrm{GRDD}}^*) &= O(h_{\mathrm{max}}^{2/(\beta_1-1)}).
\end{align*} 

\item[(ii)] Under Assumptions \ref{ass:GSRD-rate}(i), (ii) and (v) and if  $h_{\mathrm{max}} \to 0$ and $nh_{\mathrm{min}} \to \infty$ as $n \to \infty$, then
\begin{align*}
d_\mathcal{G}(\tau_{\mathrm{GRDD}}^*,\hat{\tau}_{\mathrm{GRDD}}) &= O_p((nh_{\mathrm{min}})^{-1/(2(\beta_2-1))}).
\end{align*}
\end{itemize}
\end{theorem}

By an argument similar to the proof of Propositions 1--3 in \citet{mull:19:6}, one can see that Assumption \ref{ass:GSRD-rate} holds with $\beta_1 = \beta_2 = 2$ for Examples \ref{exm:fun}--\ref{exm:mea} in Section \ref{sec:pre}. Therefore, Theorem \ref{thm:GSRD-rate} yields that if $h_0 = h_1 = h$, then
\[
d_\mathcal{G}(\tau_{\mathrm{GRDD}}, \hat{\tau}_{\mathrm{GRDD}})\leq d_\mathcal{G}(\tau_{\mathrm{GRDD}}, \tau_{\mathrm{GRDD}}^*) + d_\mathcal{G}(\tau_{\mathrm{GRDD}}^*, \hat{\tau}_{\mathrm{GRDD}})= O(h^2) + O_p((nh)^{-1/2}).
\]
The right-hand side is optimized with $h=n^{-1/5}$ and the resulting convergence rate is $O_p(n^{-2/5})$, which corresponds to the optimal rate for local linear regression with scalar responses \citep{ston:80}.

\section{Inference for the magnitude of the treatment effect}\label{sec:inf}

We develop here statistical inference for the magnitude of the treatment effect, defined as $d(\mu_{1,\oplus},\mu_{0,\oplus})$, which corresponds to the length of the geodesic treatment effect $\tau_{\mathrm{GRDD}}$. While estimation of the endpoints $\mu_{t,\oplus}$ is carried out intrinsically via LFR, an intrinsic approach to inference in general metric spaces is challenging. Therefore, we adopt an extrinsic approach by embedding the metric space $(\mathcal{M}, d)$ into a Hilbert space. This makes it possible to exploit linearity to derive asymptotic distributions as a prerequisite to derive inference. We establish  asymptotic properties of the LFR estimators and develop bootstrap-based methods for testing and inference on $d(\mu_{1,\oplus},\mu_{0,\oplus})$.

\subsection{Hilbert space embedding and LFR representation}\label{subsec:LFR-H}
We impose the following assumption.
\begin{assumption}\label{ass:LFR-H} \quad 
\begin{itemize}
\item[(i)] There is a separable Hilbert space $\mathcal{H}$ equipped with inner product $\langle \cdot, \cdot \rangle_\mathcal{H}$ and a continuous injection $\Psi:\mathcal{M} \to \mathcal{H}$ such that $\Psi: \mathcal{M} \to \Psi(\mathcal{M})$ is an isometry, that is, for any $\alpha,\beta \in \mathcal{M}$, $d(\alpha,\beta)=\|\Psi(\alpha)-\Psi(\beta)\|_\mathcal{H}$ where $\|\cdot\|_\mathcal{H}$ is the norm induced by the inner product. 
\item[(ii)] The image $\Psi(\mathcal{M})$ is a nonempty closed convex subset of $\mathcal{H}$.  
\end{itemize}
\end{assumption}

Under Assumption~\ref{ass:LFR-H}, one obtains a linear representation of the metric space that is crucial for inference. In particular, it implies that conditional Fr\'echet means admit the representation $\E_\oplus[Y(t)|R=r] = \Psi^{-1}(\E[\Psi(Y(t))|R=r])$, and that the conditional expectation $\E[\Psi(Y(t))|R=r]$ remains in $\Psi(\mathcal{M})$. A useful sufficient condition for Assumption~\ref{ass:LFR-H}(i) is that $d^2$ is of negative type. To formalize this, recall that a semimetric space $(\mathcal{Z},\rho)$ is a set equipped with a symmetric, nonnegative function $\rho$ satisfying $\rho(z_1,z_2)=0$ if and only if $z_1=z_2$. The semimetric $\rho$ is of negative type if, for any $n\ge2$, points $z_1,\dots,z_n\in\mathcal{Z}$ and coefficients $\alpha_1,\dots,\alpha_n\in\mathbb{R}$ with $\sum_{i=1}^n \alpha_i=0$, one has $\sum_{i=1}^n\sum_{j=1}^n \alpha_i \alpha_j \rho(z_i,z_j)\le 0$. If $d^2$ is of negative type, then an isometric embedding into a Hilbert space exists; see \citet{scho:38} and Proposition~3 of \citet{sejd:13}.

Assumption~\ref{ass:LFR-H} is satisfied by many commonly encountered spaces, including functional data under the $L^2$ metric (Example~\ref{exm:fun}), SPD matrices under the Frobenius, power, Log-Euclidean or Log-Cholesky metrics (Example~\ref{exm:spd}), networks represented by graph Laplacians with the Frobenius metric (Example~\ref{exm:net}), and univariate probability distributions on a compact support equipped with the 2-Wasserstein metric (Example~\ref{exm:mea}). Explicit constructions of the embedding $\Psi$ are provided in Section~S.7 of the Supplement. 
For compositional data (Example~\ref{exm:com}), we embed observations on the positive orthant of the sphere into the tangent space using the logarithmic map with base point given by the estimated Fr\'echet mean $\hat{\nu}_{0,\oplus}$. Although the logarithmic map is not globally isometric, it preserves distances from the base point \citep{lee:18}, which is the relevant quantity for testing the null hypothesis of no treatment effect. 
This allows us to carry out the testing procedure described in Section~\ref{subsec:LFR-boot} in this setting. A rigorous justification of this argument is provided in Section S.7.2 of the Supplement.

Under this embedding, we consider the regression model in $\mathcal{H}$ for an i.i.d. sample $\{Y_i,T_i,R_i\}_{i=1}^n$:
\[
\Psi(Y_i(t)) = m_t^{\Psi}(R_i) + \varepsilon_{t,i}, \quad t\in\{0,1\},
\]
where $m_t^{\Psi}(r):\mathbb{R} \to \Psi(\mathcal{M})$ is the regression function and the error satisfies $\E[\langle \varepsilon_{t,i}, g \rangle_\mathcal{H}|R_i]=0$ for every $g\in\mathcal{H}$. Under Assumptions~\ref{ass:LFR-H} and \ref{ass:LFR-AN-lim} below, $m_t^\Psi(r)=\E[\Psi(Y(t))|R=r]$ and the embedded conditional mean $\E[\Psi(Y(t))|R=r]$ coincides with the image of the conditional Fr\'echet mean, so that $m_t^\Psi(r)=\Psi(\E_\oplus[Y(t)|R=r])$. In particular, at the cutoff, $m_t^\Psi(c)=\Psi(\mu_{t,\oplus})$ or $\mu_{t,\oplus}=\Psi^{-1}(m_t^\Psi(c))$. We estimate $m_t^\Psi(c)$ by applying LFR in $\mathcal H$. Specifically,
\[
\hat{\nu}_t^\Psi = \argmin_{g \in \mathcal{H}} \frac{1}{n_t}\sum_{i=1}^n \hat{s}_t(c;R_i,h_t)\|g-\Psi(Y_i)\|_\mathcal{H}^2=\frac{1}{n_t}\sum_{i=1}^n \hat{s}_t(c;R_i,h_t)\Psi(Y_i).
\]

\subsection{Asymptotic properties of the LFR estimators}
Under the Hilbert space embedding, we study the asymptotic behavior of the LFR estimators in $\mathcal{H}$. Recall that $\mu_{t,\oplus}=\E_\oplus[Y(t)|R=c]=\Psi^{-1}(m_t^\Psi(c))$. We denote its image in the Hilbert space by $\mu_{t,\oplus}^\Psi=\Psi(\mu_{t,\oplus})=m_t^\Psi(c)$. Therefore, inference on $\mu_{t,\oplus}$ can be carried out by studying $\mu_{t,\oplus}^\Psi$ in $\mathcal{H}$. Let $\{e_j\}_{j \geq 1}$ be an orthonormal basis of $\mathcal{H}$. We impose the following regularity conditions.
\begin{assumption}\label{ass:LFR-AN-lim}
For each $t\in\{0,1\}$, the following conditions hold.
\begin{itemize}
    \item[(i)] The density function $f_R$ of the running variable $R$ is continuous and positive on a neighborhood $\mathcal{I}$ of $c$. 
    \item[(ii)] For any $g \in \mathcal{H}$, the function $m_t^\Psi(r;g) = \langle m_t^\Psi(r),g \rangle_\mathcal{H}$ is twice continuously differentiable on $\mathcal{I}$. Additionally,  $\sup_{r \in \mathcal{I}}\{\|m_t^\Psi(r)\|_\mathcal{H}+\sum_{s=1}^2\sum_{j\geq 1}|\partial^sm_t^\Psi(r;e_j)|^2\}<\infty$ where $\partial^sm_t^\Psi(r;g) = \frac{d^s}{dr^s}\langle m_t^\Psi(r),g\rangle_\mathcal{H}$, $s=1,2$, $g \in \mathcal{H}$.
    \item[(iii)] For any $g \in \mathcal{H} \backslash \{0\}$, $\E[\langle\varepsilon_{t,1}, g \rangle_\mathcal{H}^2|R_1=r]$ is continuous on $\mathcal{I}$, $\inf_{r \in \mathcal{I}}\E[\langle\varepsilon_{t,1}, g \rangle_\mathcal{H}^2|R_1=r]>0$, and $\sup_{r \in \mathcal{I}}\E[\|\varepsilon_{t,1}\|_\mathcal{H}^4|R_1=r]<\infty$. Additionally, there exists a sequence of positive constants $\{b_j\}_{j \geq 1}$ such that $\sup_{r \in \mathcal{I}}\E[\langle \varepsilon_{t,1}, e_j\rangle_\mathcal{H}^2|R_1=r]\leq b_j$, $\sum_{j\geq 1}b_j<\infty$.\end{itemize}
\end{assumption}
Assumption~\ref{ass:LFR-AN-lim} parallels standard conditions for local linear regression with scalar responses \citep{fan:96}, applied here to the Hilbert space through projections onto one-dimensional directions. Condition (ii) ensures sufficient smoothness of the regression function in $\mathcal{H}$, while condition (iii) controls the stochastic variability of the error process. We use $\leadsto$ to denote weak convergence. The following result establishes the joint asymptotic normality of the LFR estimators in the Hilbert space.
\begin{theorem}\label{thm:LFR-AN-joint}
Suppose that Assumptions \ref{asm:basic}, \ref{ass:ker}, \ref{ass:LFR-H}, and \ref{ass:LFR-AN-lim} hold. Assume that $h_t=h \to 0$, $nh \to \infty$, and $nh^5 \to 0$ as $n \to \infty$. Then we have $\sqrt{nh}(\hat{\nu}_0^\Psi-\mu_{0,\oplus}^\Psi, \hat{\nu}_1^\Psi-\mu_{1,\oplus}^\Psi) \leadsto (\mathbb{G}_0,\mathbb{G}_1)$, where $\mathbb{G}_0$ and $\mathbb{G}_1$ are independent centered Gaussian elements in $\mathcal{H}$. Their covariance structure is characterized through projections:
\begin{align*}\label{eq:LFR-AVar}
\Var(\langle \mathbb{G}_t,g \rangle_\mathcal{H}) = \frac{\int(\kappa_{t,2}-\kappa_{t,1}u)^2K_t^2(u)du}{(\kappa_{t,2}\kappa_{t,0}-\kappa_{t,1}^2)^2}\frac{\E[\langle \varepsilon_{t,1},g \rangle_\mathcal{H}^2|R_1=c]}{f_R(c)},\ t\in\{0,1\},\ g \in \mathcal{H},
\end{align*}
where $\kappa_{t,p} = \int u^pK_t(u)du$. \end{theorem}
Theorem~\ref{thm:LFR-AN-joint} shows that the LFR estimators behave analogously to local linear estimators with Euclidean responses, but now for responses taking values in a possibly infinite-dimensional Hilbert space. The convergence rate $\sqrt{nh}$ corresponds to the standard boundary rate in RDD, and the limiting Gaussian elements are independent across the two sides of the cutoff due to the use of one-sided kernels. This result provides the foundation for conducting statistical inference using the extrinsic approach described in Section~\ref{subsec:LFR-H}.

\subsection{Bootstrap-based inference for the treatment effect magnitude}\label{subsec:LFR-boot}
Building on the asymptotic normality established in Theorem~\ref{thm:LFR-AN-joint}, we develop bootstrap-based procedures for statistical inference on the magnitude of the treatment effect $d(\mu_{1,\oplus},\mu_{0,\oplus})$. Since the Gaussian limit in $\mathcal{H}$ involves unknown covariance structures, we employ a multiplier bootstrap to approximate its distribution; similar resampling strategies have been considered, for example, in \citet{chia:19} and \citet{va:25}.

Let $\{\xi_i\}_{i=1}^n$ be i.i.d.\ standard normal random variables defined on a probability space $(\Omega_\xi,\mathcal{A}_\xi,\mathrm{P}_\xi)$, independent of the data $\{Y_i,T_i,R_i\}_{i=1}^n$. Define the bootstrap processes
\[\hat{\mathbb{G}}_t = \frac{\sqrt{nh}}{n_t}\sum_{i=1}^n \xi_i \hat{s}_t(c;R_i,h_t)\{\Psi(Y_i)-\hat{\nu}_t^\Psi\}, \quad t\in\{0,1\}.\]
We use $\stackrel{\mathrm{P}_\xi}{\leadsto}$ to denote conditional weak convergence given the data; see Section~1.13 of \citet{well:96}.
\begin{proposition}\label{prp:LFR-boot}
Suppose that the assumptions of Theorem~\ref{thm:LFR-AN-joint} hold. Then, conditionally on the data $\{Y_i,T_i,R_i\}_{i=1}^n$ and with probability approaching one, the process $(\hat{\mathbb{G}}_0,\hat{\mathbb{G}}_1)$ weakly converges to the limit process $(\mathbb{G}_0,\mathbb{G}_1)$, that is, $(\hat{\mathbb{G}}_0,\hat{\mathbb{G}}_1) \stackrel{\mathrm{P}_\xi}{\leadsto} (\mathbb{G}_0,\mathbb{G}_1)$.
\end{proposition}
Proposition~\ref{prp:LFR-boot} shows that the multiplier bootstrap consistently approximates the joint Gaussian limit in $\mathcal{H}$, thereby enabling inference without explicit estimation of the covariance operator.

We first consider testing for the presence of a treatment effect at the cutoff, i.e., testing $\mathbb{H}_0: d(\mu_{1,\oplus},\mu_{0,\oplus}) = 0$ versus $\mathbb{H}_1: d(\mu_{1,\oplus},\mu_{0,\oplus}) > 0$. Define the test statistic $T_n = nh\|\hat{\nu}_1^\Psi-\hat{\nu}_0^\Psi\|_\mathcal{H}^2$, and its bootstrap analogue $\hat{T}_n = \|\hat{\mathbb{G}}_1 - \hat{\mathbb{G}}_0\|_\mathcal{H}^2$. Let $q_{n,\alpha}$ denote the $(1-\alpha)$-quantile of the conditional distribution of $\hat{T}_n$ given the data, and define the test function $\hat{\phi}_n = \mathbf{1}\{T_n > q_{n,\alpha}\}$. The following result establishes the validity of this bootstrap test.

\begin{corollary}\label{cor:LFR-test}
Suppose that the assumptions of Theorem \ref{thm:LFR-AN-joint} hold. Then the test $\hat{\phi}_n$ is asymptotically level $\alpha$, i.e., $\E[\hat{\phi}_n] \to \alpha$ under $\mathbb{H}_0$ as $n \to \infty$. Moreover, $\hat{\phi}_n$ is consistent, i.e. $\E[\hat{\phi}_n] \to 1$ under $\mathbb{H}_1$ as $n \to \infty$. 
\end{corollary}
When $d(\mu_{1,\oplus},\mu_{0,\oplus})>0$, we construct confidence intervals for the magnitude of the treatment effect. Based on the linear expansion implied by Theorem~\ref{thm:LFR-AN-joint}, let $c_{n,\alpha}$ denote the $(1-\alpha)$-quantile of the conditional distribution of $2|\langle \hat{\nu}_1^\Psi - \hat{\nu}_0^\Psi,\hat{\mathbb{G}}_1-\hat{\mathbb{G}}_0\rangle_\mathcal{H}|$. A $(1-\alpha)$ confidence interval for $d(\mu_{1,\oplus},\mu_{0,\oplus})$ is given by
\[I_{n,\alpha} = \left[
\left(\max\left\{\|\hat{\nu}_1^\Psi-\hat{\nu}_0^\Psi\|_\mathcal{H}^2 - \frac{c_{n,\alpha}}{\sqrt{nh}},\,0\right\}\right)^{1/2},
\left(\|\hat{\nu}_1^\Psi-\hat{\nu}_0^\Psi\|_\mathcal{H}^2 + \frac{c_{n,\alpha}}{\sqrt{nh}}\right)^{1/2}\right].\]

\begin{corollary}\label{cor:LFR-CI}
Under the assumptions of Theorem \ref{thm:LFR-AN-joint}, if $d(\mu_{1,\oplus},\mu_{0,\oplus})>0$, Then \[\Pr(d(\mu_{1,\oplus},\mu_{0,\oplus}) \in I_{n,\alpha})\to 1-\alpha\quad\text{as }n \to \infty.\] 
\end{corollary}

\begin{remark}
Theorem \ref{thm:LFR-AN-joint} and Proposition \ref{prp:LFR-boot} also enable us to construct a confidence region of $\mu_{t,\oplus}$ for $t\in\{0,1\}$. Let $q_{n,\alpha}^{(t)}$ denote the $(1-\alpha)$-quantile of the conditional distribution of $\|\hat{\mathbb{G}}_t\|_\mathcal{H}^2$ given the data. A $(1-\alpha)$ confidence region for $\mu_{t,\oplus}$ is
\begin{align*}
R_{n,\alpha}^{(t)} = \left\{\nu \in \mathcal{M}: \|\hat{\nu}_t^\Psi-\Psi(\nu)\|_\mathcal{H} \leq \sqrt{\frac{q_{n,\alpha}^{(t)}}{nh_t}}\right\}.
\end{align*}
Indeed, under the assumptions of Theorem \ref{thm:LFR-AN-joint}, one has $\Pr(\mu_{t,\oplus} \in R_{n,\alpha}^{(t)}) \to 1-\alpha$ as $n \to \infty$. 
\end{remark}

\section{Implementation and simulations}\label{sec:sim}
\subsection{Bandwidth selection}\label{subsec:bw}Selecting an appropriate bandwidth is critical for reliable estimation in RDDs, especially when the outcome resides in a general metric space. For scalar outcomes, conventional RDD bandwidth selectors are typically based on asymptotic MSE expansions for local polynomial estimators at the cutoff \citep{imbe:12,calo:14,calo:20}. Such plug-in rules rely on Euclidean structure and derivative-based approximations, and are therefore not directly available in our general non-Euclidean setting.

We adopt a boundary-focused one-sided cross-validation approach, motivated by \citet{ludw:07}, to select the bandwidth in the GRDD framework. For a fraction $\delta\in(0,1)$, the cross-validation criterion is evaluated over a neighborhood of the cutoff, defined by the $(1-\delta)$-quantile of the running variable below $c$ and the $\delta$-quantile above $c$. At each evaluation point, LFR is fitted using only observations farther from the cutoff on the same side. The bandwidth is then selected by minimizing the resulting sum of squared metric prediction errors over a candidate grid; see Algorithm~\ref{alg:bw_select}. This combination of one-sided fitting and cutoff-focused evaluation yields a practical bandwidth selection procedure adapted to the boundary structure of GRDD in general metric spaces.

The parameter $\delta$ controls the width of the evaluation region, with larger values including more observations. In the default implementation, $\delta$ is selected automatically so that the number of evaluation observations is close to a fixed computational target. Specifically, the search is restricted to $\delta\in[0.05,0.5]$, and the target number of evaluation observations is set to $100$ throughout all simulations and empirical applications. These fixed defaults allow the evaluation region to adapt to the sample size while keeping the cross-validation procedure computationally manageable. Users may instead specify $\delta$ directly when substantive considerations favor a particular neighborhood around the cutoff.

While the cross-validation procedure provides a practical data-driven bandwidth selection rule, the asymptotic validity of the bootstrap procedure in Theorem~\ref{thm:LFR-AN-joint} is established under the undersmoothing condition $nh^5\to0$. Our simulation studies suggest that the proposed bootstrap procedure nevertheless provides reliable empirical size control with cross-validation-based bandwidth selection. Developing bandwidth selection methods specifically tailored for inference in GRDD is an interesting direction for future research.

\begin{algorithm}[tb]
\single
\caption{One-sided cross-validation bandwidth selection for GRDD}
\label{alg:bw_select}
\begin{algorithmic}[1]
\REQUIRE Sample $\{(Y_i, R_i)\}_{i=1}^n$, cutoff $c$, fraction $\delta \in (0,1)$
\ENSURE Selected bandwidth $b^\ast$
\STATE Sort the running variables $\{R_i\}_{i=1}^n$ in increasing order and let $R_{\min} = \min_i R_i$ and $R_{\max} = \max_i R_i$
\STATE Let $q_{-,\,1-\delta}$ be the $(1-\delta)$-quantile of $\{R_i:R_i<c\}$ and $q_{+,\,\delta}$ be the $\delta$-quantile of $\{R_i:R_i\ge c\}$
\STATE Define the evaluation region $\mathcal{R}=[q_{-,\,1-\delta}, q_{+,\,\delta}]$
\STATE Compute $b_{\max}$ as half the minimum of $c-R_{\min}$ and $R_{\max}-c$ and $b_{\min}$ as a data-adaptive lower bound based on the local spacing of the running variable near the cutoff
\STATE Construct a grid of bandwidths $\mathcal{B}\subset[b_{\min}, b_{\max}]$
\FOR{each $b \in \mathcal{B}$}
  \STATE Set $\mathrm{CV}(b)=0$
  \FOR{each $R_i \in \mathcal{R}$}
    \IF{$R_i<c$}
      \STATE Compute the one-sided LFR estimate $\hat{m}_b(R_i)$ at $R_i$ using only observations with $R_j<R_i$
    \ELSE
      \STATE Compute the one-sided LFR estimate $\hat{m}_b(R_i)$ at $R_i$ using only observations with $R_j>R_i$
    \ENDIF
    \STATE Update
    \[
    \mathrm{CV}(b)=\mathrm{CV}(b)+d^2\!\left(Y_i,\hat{m}_b(R_i)\right)
    \]
  \ENDFOR
\ENDFOR
\STATE $b^\ast=\argmin_{b\in\mathcal{B}} \mathrm{CV}(b)$
\end{algorithmic}
\end{algorithm}

\subsection{Simulation for network-valued outcomes}
To evaluate both estimation and inference in a non-Euclidean setting, we consider outcomes given by network structures. Specifically, each observation is a graph Laplacian taking values in the space $\mathcal{L}$ equipped with the Frobenius metric (Example~\ref{exm:net}). This setting reflects applications where the outcome is a weighted network and the objective is to detect discontinuities in its geometry at the cutoff.

Networks are generated under a weighted stochastic block model with $10$ nodes divided equally into two communities, $C_1$ and $C_2$. The binary adjacency matrix is determined by a block-structured edge probability matrix 
$\begin{pmatrix}
0.8 & 0.2 \\
0.2 & 0.8
\end{pmatrix}$,
where the entries denote the probabilities of edge formation within and between communities. The running variable $R$ is generated from a truncated normal distribution on $[-1,1]$ with mean $0$ and standard deviation $0.1$. Given $R$, each present edge is assigned weight
\[
w_{ij}(R)=
\begin{cases}
1+\cos\left(\frac{\pi}{2}R\right)+\epsilon_{ij}, & R<0,\\[4pt]
1+\cos\left(\frac{\pi}{2}R\right)+\Delta\tau_{ij}+\epsilon_{ij}, & R\ge 0,
\end{cases}
\]
where $\epsilon_{ij}\sim \mathrm{Unif}(0,0.1)$ independently. The term $\tau_{ij}$ induces heterogeneous changes across edge types: $\tau_{ij}=1$ for edges within $C_1$, $\tau_{ij}=-1$ for edges within $C_2$, and $\tau_{ij}=0$ for edges between communities. The parameter $\Delta \ge 0$ controls the magnitude of the discontinuity at the cutoff at the edge level, where $\Delta=0$ corresponds to no treatment effect. Each weighted adjacency matrix is converted into a graph Laplacian, which serves as the outcome $Y_i$.

We first assess estimation performance by fixing $\Delta=1$ and considering sample sizes $n \in \{100,200,500,1000\}$. For each $n$, we generate $500$ independent replications $\{Y_i,R_i\}_{i=1}^n$ and apply GRDD with the data-adaptive bandwidth selector described in Section~\ref{subsec:bw}. In each replication, performance is evaluated using the relative bias $d_\mathcal{G}(\hat{\tau}_{\mathrm{GRDD}},\tau_{\mathrm{GRDD}})/|\tau_{\mathrm{GRDD}}|$, where $d_\mathcal{G}$ denotes the metric on the space of geodesics defined in \eqref{eq:dg} and $|\tau_{\mathrm{GRDD}}|$ is the magnitude of the true effect. As this metric aggregates discrepancies across all edge weights in the network, it reflects a global measure of estimation error.

To evaluate inference, we use the same data-generating mechanism but vary $\Delta$ over the grid $\{0,0.1,\ldots,1\}$. For each pair $(n,\Delta)$, we again generate $500$ independent replications and apply the bootstrap-based inference procedure developed in Section~\ref{subsec:LFR-boot}. Specifically, we test $\mathbb{H}_0:d(\mu_{1,\oplus},\mu_{0,\oplus})=0$ versus $\mathbb{H}_1:d(\mu_{1,\oplus},\mu_{0,\oplus})>0$, using the multiplier bootstrap with $B=1000$ replications and nominal level $\alpha=0.05$. The empirical rejection frequency over replications estimates the size of the test when $\Delta=0$ and its power when $\Delta>0$.

\begin{figure}[tb]
    \single
    \centering
    \begin{subfigure}{0.45\linewidth}
        \centering
        \includegraphics[width=\linewidth]{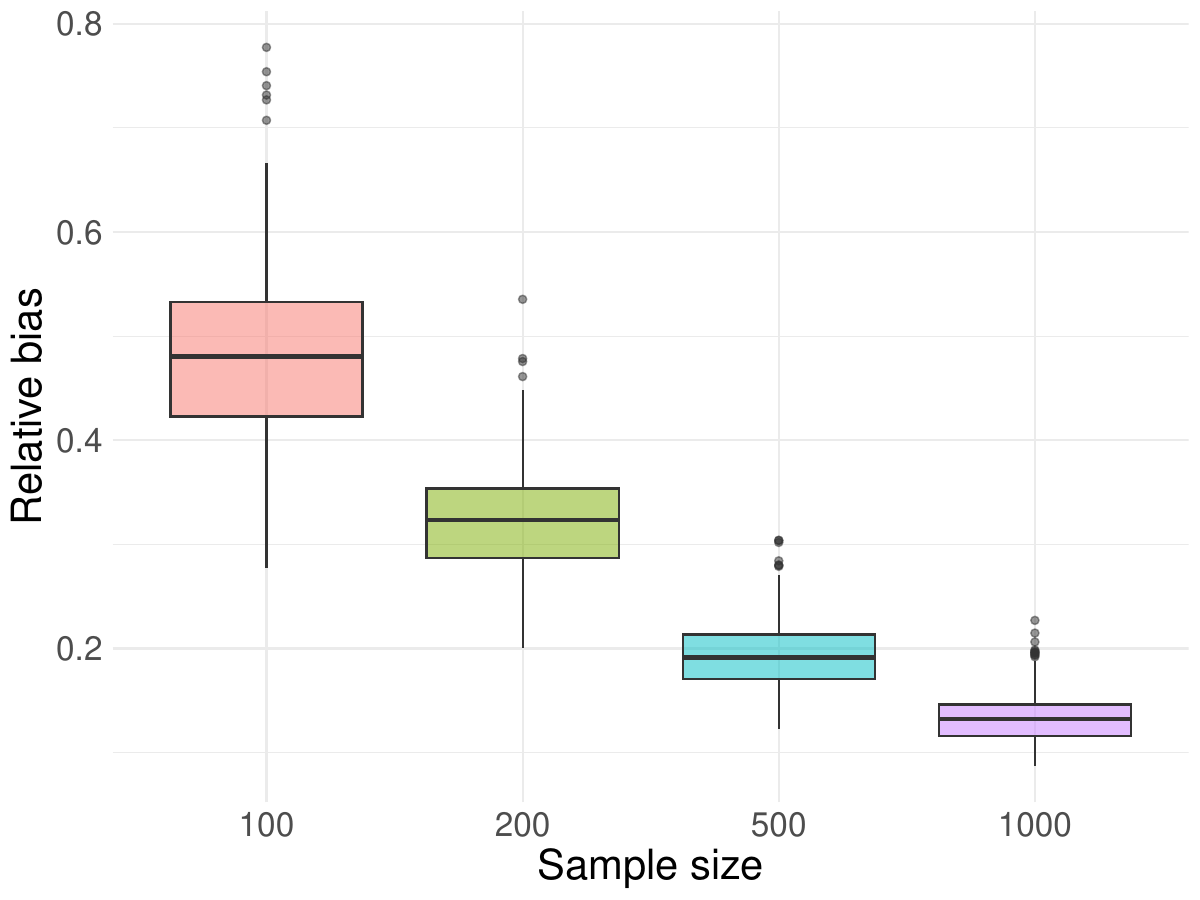}
        \caption{}
    \end{subfigure}
    \hfill
    \begin{subfigure}{0.45\linewidth}
        \centering
        \includegraphics[width=\linewidth]{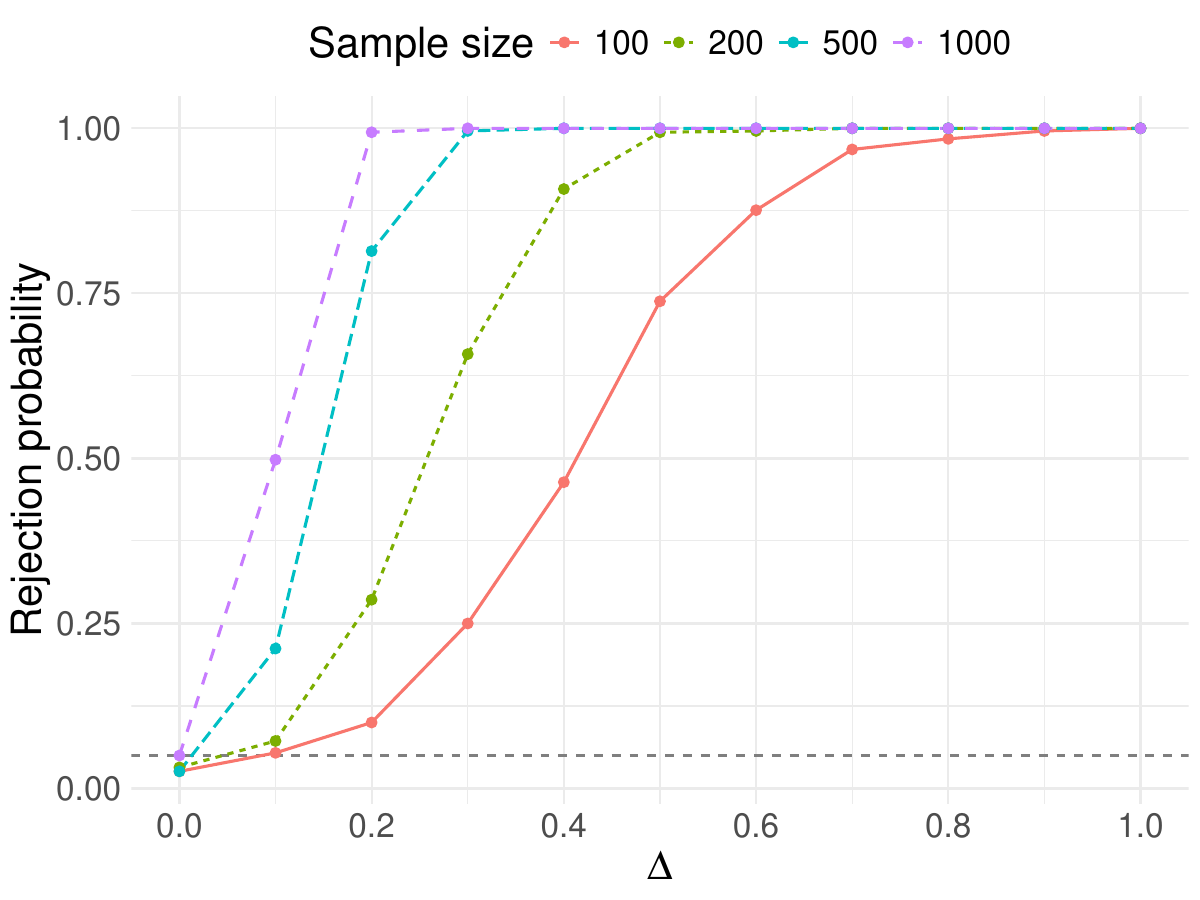}
        \caption{}
    \end{subfigure}
    \caption{Simulation results for network-valued outcomes. (a) Boxplots of the relative bias across $500$ replications for each sample size $n \in \{100,200,500,1000\}$. (b) Empirical rejection probabilities of the bootstrap test for $\mathbb{H}_0: d(\mu_{1,\oplus},\mu_{0,\oplus})=0$ at level $\alpha=0.05$.}
    \par\smallskip\noindent\textit{Alt text:} The left panel shows decreasing estimation error as the sample size grows. The right panel shows rejection probability increasing with the discontinuity magnitude, with higher power for larger samples.
    \label{fig:sim}
\end{figure}

Figure~\ref{fig:sim} summarizes the results. Panel~(a) displays boxplots of the relative bias for $\Delta=1$. The bias decreases steadily with increasing sample size, indicating consistency of the GRDD estimator for network-valued outcomes. Panel~(b) displays the empirical rejection probabilities as a function of $\Delta$ for each sample size, with the dashed horizontal line indicating the nominal level $\alpha=0.05$. At $\Delta=0$, the rejection rates across all sample sizes are close to the nominal level, suggesting that the bootstrap procedure provides reliable empirical size control in the simulation settings considered. As $\Delta$ increases, the rejection probability rises rapidly, with larger sample sizes exhibiting substantially higher power. In particular, for moderate values of $\Delta$ (e.g., $\Delta=0.2$ or $0.3$), the power improves markedly as $n$ increases, and for $\Delta \ge 0.5$, the test attains near-perfect power even at moderate sample sizes. These results suggest that the proposed inference procedure yields empirical rejection probabilities close to the nominal level under the null while remaining sensitive to increasingly strong treatment effects.

An additional null simulation with unequal curvature of the regression functions on the two sides of the cutoff is provided in Section~S.13 of the Supplement. This setting further examines empirical size under cross-validation-based bandwidth selection in a more general configuration.

\section{Incumbency effects in multiparty elections}\label{sec:app}
We use the proposed GRDD method to investigate the impact of incumbency in a multiparty electoral system, focusing on data from UK parliamentary elections compiled by \citet{egge:17}. The dataset covers all general elections in the United Kingdom from 1950 to 2010, including constituency-level results for each of the three major parties: Conservative, Labour, and Liberal (or its historical equivalents, such as the Liberal Democrats or SDP-Liberal Alliance). The full sample contains 8,170 constituency-level contests, with information on vote totals, margins of victory, and party affiliations for every candidate in each election. The UK employs a single-member district plurality system, in which candidates compete in geographically defined constituencies and the winner is determined by a simple plurality rule. This institutional setting provides a compelling environment for studying incumbency effects, as each contest produces one winner and allows clear identification of party continuity or change.

For each constituency, we define the running variable as the Conservative margin in the previous election, i.e., the difference in vote share between the Conservative candidate and the top rival. The outcome is the compositional vector of vote shares in the subsequent election, restricted to the three major parties. Following Example~\ref{exm:com}, these compositions are mapped to the positive orthant of the unit sphere via the square-root transformation and analyzed using the arc-length distance. In this representation, the geodesic between two compositions corresponds to great-circle interpolation on the unit sphere. As a benchmark, we also estimate separate scalar RDDs for each vote-share component using the \texttt{rdrobust} package \citep{calo:17}.

The proposed data-adaptive procedure selects $\delta=0.05$ for the boundary evaluation region and yields a bandwidth of $0.245$. Using this bandwidth, the estimated vote-share composition at the cutoff on the left side (where the Conservative candidate narrowly lost) is $(43.5\%,\,37.7\%,\,18.8\%)$ for the Conservative, Labour, and Liberal parties, respectively. On the right side (where the Conservative candidate narrowly won), the estimated composition shifts to $(44.6\%,\,37.6\%,\,17.8\%)$. Thus, the estimated GRDD discontinuity corresponds to a gain of about $1.1$ percentage points for the Conservatives, essentially no change for Labour, and a decline of about $1.0$ percentage point for the Liberals. This pattern is consistent with a modest incumbency advantage for the Conservatives, operating primarily through a reallocation of support away from the Liberal party.

The componentwise scalar RDD analysis yields a broadly similar pattern. The bias-corrected \texttt{rdrobust} estimates \citep{calo:20} indicate a discontinuity of about $1.6$ percentage points for the Conservatives, essentially no change for Labour, and a decline of about $1.5$ percentage points for the Liberals. These results are consistent with the GRDD estimates, suggesting that the primary shifts occur in the Conservative and Liberal shares, with little change for Labour. However, the scalar approach treats the components independently, does not provide a unified assessment of the compositional effect, and fails to preserve the sum-to-zero constraint, which arises because the treatment effect is defined as the difference between two compositions.

For statistical inference, GRDD analyzes the outcome jointly as a composition and provides a global inferential statement. The estimated effect magnitude is $0.0147$. Using the bootstrap-based inference procedure described in Section~\ref{subsec:LFR-boot}, we obtain a one-sided $p$-value of $0.190$. Thus, while the estimated shift is directionally consistent with an incumbency effect, we do not reject the null of no overall compositional treatment effect at the $5\%$ level. In contrast, the scalar approach conducts inference separately for each component. The componentwise \texttt{rdrobust} analysis yields $p$-values of $0.001$ for the Conservatives, $0.928$ for Labour, and $0.081$ for the Liberals, indicating statistical significance only for the Conservative component at the $5\%$ level. This illustrates that, while componentwise analyses may suggest localized changes, the joint analysis provides a different perspective on whether there is a statistically meaningful shift in the composition as a whole.

\begin{figure}[tb]
    \single
    \centering
    \includegraphics[width=0.9\linewidth]{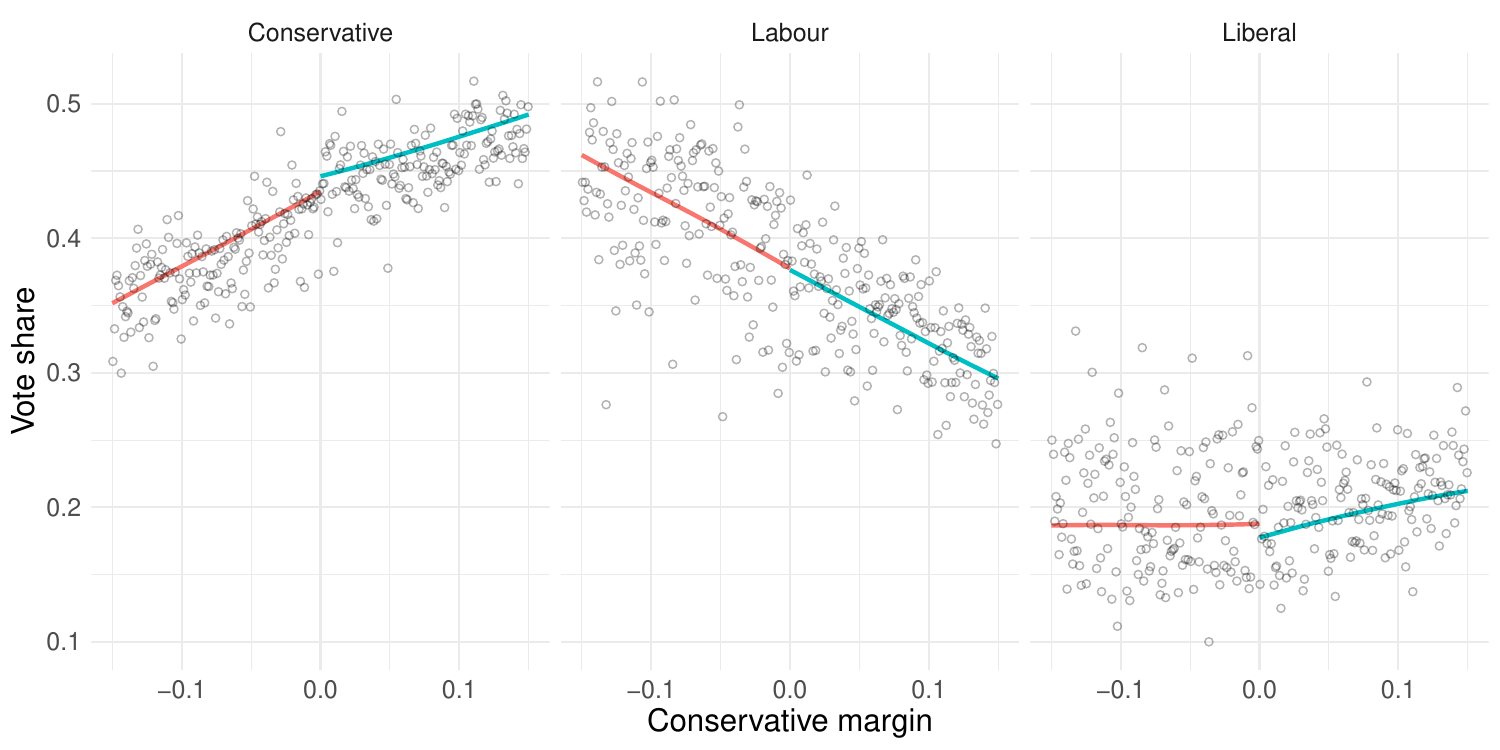}
    \caption{Estimated GRDD fits and bin-averaged outcomes by Conservative margin using UK parliamentary election data from 1950 to 2010.}
    \par\smallskip\noindent\textit{Alt text:} Three panels display Conservative, Labour, and Liberal vote shares against the previous Conservative margin. The fitted curves show a small upward discontinuity for Conservatives, almost no discontinuity for Labour, and a small downward discontinuity for Liberals.
    \label{fig:uk}
\end{figure}

Figure~\ref{fig:uk} illustrates the estimated results. Each panel corresponds to one of the three major parties. The horizontal axis displays the Conservative margin in the previous election, and the vertical axis shows the party's current vote share. Points represent bin-averaged outcomes (in 0.1 percentage point bins), while the smooth curves show the GRDD estimates on either side of the cutoff. The estimated discontinuity at the cutoff is modest and is most apparent in the Conservative and Liberal panels.

An additional application to functional air pollution data is provided in Section S.2 of the Supplement.

\section{Fuzzy regression discontinuity design for random objects}\label{sec:FRD}
In the sharp RDD considered so far,  the treatment is assigned based on whether a threshold is met, assuming full compliance where everyone adheres to the assigned treatment. However, in reality, some individuals might not follow the assigned treatment. The fuzzy RDD addresses such situations of imperfect compliance, which are practically relevant. This motivates extending the GRDD framework to the fuzzy design, where the average treatment effect at the cutoff is obtained by amplifying the discontinuity in the outcome by the reciprocal discontinuity in the treatment probability. 

We first develop this extension for outcome spaces that admit an isometric embedding into a Hilbert space, leading to an extrinsic definition of the causal effect with corresponding identification and estimation procedures. An intrinsic formulation, where the causal effect is defined as the geodesic between the relevant Fr\'echet means in the outcome space, is presented in Section S.3 of the Supplement. The special case of Riemannian manifold-valued outcomes is discussed in Section S.4, where logarithmic and exponential maps are used to represent the causal effect in the tangent space. An intrinsic, geodesic-based formulation for manifold-valued outcomes is provided in Section S.5 of the Supplement.

\subsection{Setup} \label{subsec:FRD-setup}
Since the fuzzy RDD case is quite distinct from the sharp RDD case studied in Section \ref{sec:GSRD}, we first reintroduce notation. Suppose we observe an independent and identically distributed sample $\{Y_i, T_i, R_i\}_{i=1}^n$ of $(Y,T,R) \in \mathcal{M} \times \{0,1\} \times \mathbb{R}$, where $Y\in \mathcal{M}$ denotes the outcome, $T$ is the binary treatment indicator, and $R$ is a running variable that yields the assignment variable $Z=1\{R \ge c\}$ with a known constant $c$. Then the observed treatment is determined by $T=ZT(1)+(1-Z)T(0)$, where $T(z)$ is the indicator of the potential treatment status for $z \in \{0,1\}$. The outcome variable satisfies $Y=Y(0)$ if $T=0$ and $Y=Y(1)$ if $T=1$, where $Y(t) \in \mathcal{M}$ is a potential outcome for treatment $T=t$. Here we assume that there is no direct effect of $Z$ on the potential outcomes (exclusion restriction). 

In fuzzy RDDs with scalar outcomes, a causal effect of interest is the local average treatment effect (LATE) of the compliers, given by $\E[Y(T(1))|T(1)>T(0),R=c]-\E[Y(T(0))|T(1)>T(0),R=c]$. In the case of scalar outcomes, compliers' LATE is identified by the difference of the expected outcomes at the cutoff divided by the difference of the treatment probabilities at the cutoff (\citealp{hahn:01}). However, when outcomes take values in a general metric space, not only differencing but also division operators may not be applicable. Thus, the fuzzy RDD analysis for random objects is fundamentally different from the sharp RDD analysis considered in the previous sections.

Denote by $\mu_{z,\oplus}=\E_\oplus[Y(T(z))|T(1)>T(0),R=c]$ for $z \in \{0,1\}$ the unique Fr\'echet mean of the potential outcome $Y(T(z))$ of compliers at the cutoff. To extend the notion of compliers' causal effect to outcomes in geodesic metric spaces, we introduce a mapping $\Psi$ of $\mu_{z,\oplus}$ to a Hilbert space, where the linear structure makes it possible to evaluate a contrast for $\Psi(\mu_{z,\oplus})$.

\subsection{Identification}\label{subsec:FRD-id}
To define the causal estimand in the fuzzy RDD setting, we build on the Hilbert space embedding framework introduced in Section~\ref{subsec:LFR-H}. Specifically, we assume that the metric space $(\mathcal{M},d)$ admits an isometric embedding into a separable Hilbert space $\mathcal{H}$ via a map $\Psi:\mathcal{M} \to \mathcal{H}$ satisfying Assumption~\ref{ass:LFR-H}. 

In particular, $\Psi$ is injective and distance-preserving, and its image $\Psi(\mathcal{M})$ is a closed convex subset of $\mathcal{H}$. This structure allows us to represent Fr\'echet means and causal effects in terms of linear operations in $\mathcal{H}$. Based on this representation, we define the causal estimand as follows.

\begin{definition}[Compliers' LATE at the cutoff]\label{def:LATE-Object}
Suppose that $(\mathcal{M},d)$ satisfies Assumption~\ref{ass:LFR-H}. Then the compliers' local average treatment effect at the cutoff is defined as
\[
\tau_{\mathrm{FRDD}}:=\Psi(\mu_{1,\oplus}) - \Psi(\mu_{0,\oplus}).
\]
\end{definition}
When $\mathcal{M} = \mathbb{R}$, one can take $\Psi$ as the identity map, and the above definition reduces to the classical complier LATE:
\[
\tau_{\mathrm{FRDD}}=\mu_{1,\oplus} - \mu_{0,\oplus} = \E[Y(T(1))|T(1)>T(0),R=c] - \E[Y(T(0))|T(1)>T(0),R=c].
\]
The above construction requires the existence of an isometric embedding $\Psi$, which excludes certain spaces such as compositional data equipped with the arc-length metric (Example~\ref{exm:com}), due to their positive curvature. For such spaces, alternative intrinsic formulations based on tangent space representations are provided in Section~S.5 of the Supplement.

For the identification of $\tau_{\mathrm{FRDD}}$, we assume the following conditions.
\begin{assumption}\label{ass:FRD-id} \quad
\begin{itemize}
\item[(i)] $\Pr(T(1)\ge T(0)|R=r)=1$ for all $r$ in a neighborhood of $c$.
\item[(ii)] $\Pr(T(1) > T(0)|R=c) \in (\eta_0, 1-\eta_0)$ for some $\eta_0 \in (0,1/2)$.
\item[(iii)] $\E[T(1)|R=r]$ and $\E[T(0)|R=r]$ are continuous over a neighborhood of $c$.
\item[(iv)] $\E_\oplus[Y(T(0))|R=r]$ and $\E_\oplus[Y(T(1))|R=r]$ exist uniquely for each $r$ over a neighborhood of $c$, and they are continuous on that neighborhood. 
\end{itemize}
\end{assumption}

Condition (i) is a monotonicity condition on the potential treatment variable, which rules out defiers having $(T(0),T(1))=(1,0)$. Condition (ii) is a non-overlapping condition for compliers and other types (i.e., always/never-takers). Conditions (iii) and (iv) are continuity requirements for the conditional Fr\'echet means. Define $\mu^\dag_{z,\oplus} = \E_\oplus[Y(T(z))|R=c]$ for $z \in \{0,1\}$ and assume existence and  uniqueness. The next result provides identification of $\tau_{\mathrm{FRDD}}$. 

\begin{theorem}\label{thm:LATE-id}
Consider the setup of Section \ref{subsec:FRD-setup}. Under Assumption \ref{ass:FRD-id}, 
\begin{align*}
\tau_{\mathrm{FRDD}} 
= \frac{\Psi(\lim_{r \downarrow c}\E_\oplus[Y|R=r]) - \Psi(\lim_{r \uparrow c}\E_\oplus[Y|R=r])}{\lim_{r \downarrow c}\E[T|R=r] - \lim_{r \uparrow c}\E[T|R=r]}.
\end{align*}
\end{theorem}

Note that for the space of univariate distributions with the 2-Wasserstein metric (Example \ref{exm:mea}), we can take $\Psi(\alpha)=F_\alpha^{-1}(\cdot)$ where $F_\alpha^{-1}(\cdot)$ is the quantile function of $\alpha \in \mathcal{W}$. In this case, $\tau_{\mathrm{FRDD}}$ corresponds to the fuzzy local average quantile treatment effect $F_{\mu_{1,\oplus}}^{-1}(\cdot) - F_{\mu_{0,\oplus}}^{-1}(\cdot)$
that was considered in \citet{va:25}. Theorem \ref{thm:LATE-id} gives
\[
\tau_{\mathrm{FRDD}} = \frac{F_{\mu^\dag_{1,\oplus}}^{-1}(\cdot) - F_{\mu^\dag_{0,\oplus}}^{-1}(\cdot)}{\E[T(1)|R=c] - \E[T(0)|R=c]},
\]
which corresponds to Lemma 2 in \citet{va:25}. Therefore, our framework for fuzzy RDD with non-Euclidean outcomes can be seen as a generalization of the fuzzy design in \citet{va:25} to general geodesic metric spaces.

\subsection{Estimation}\label{subsec:FRD-est}
We now construct an estimator of $\tau_{\mathrm{FRDD}}$ and provide its asymptotic properties. A natural estimator is \begin{align}\label{eq:LL-est}
\hat{\tau}_{\mathrm{FRDD}} = \frac{\Psi(\hat{\nu}_{1,\oplus}) - \Psi(\hat{\nu}_{0,\oplus})}{\hat{m}_1 - \hat{m}_0},
\end{align}
where $\hat{\nu}_{0,\oplus}$ and $\hat{\nu}_{1,\oplus}$ are the LFR estimators defined in Section \ref{subsec:GSRD-est}, and $\hat{m}_0$ and $\hat{m}_1$ are (nonparametric) estimators for $m_0=\E[T(0)|R=c]$ and $m_1=\E[T(1)|R=c]$, respectively. For example, one can use local linear regression for the construction of $\hat{m}_z = \hat{m}_{z,0}$, leading to 
\begin{equation}\label{eq:first-stage-LL}
(\hat{m}_{z,0}, \hat{m}_{z,1})^\top = \argmin_{\beta_0,\beta_1 \in \mathbb{R}}\frac{1}{n_z}\sum_{i=1}^n(T_i - \beta_0 - \beta_1(R_i - c))^2K_{z,h_z}(R_i-c)\quad \mbox{for} \quad z \in \{0,1\}.
\end{equation}

Define $M_{z,\oplus}^\dag(\nu,c) = \E[d^2(\nu,Y(T(z)))|R=c]$ for $z \in \{0,1\}$. 
Letting $h_{\mathrm{max}}=\max\{h_0,h_1\}$ and $h_{\mathrm{min}} = \min\{h_0,h_1\}$, the convergence rate of $\hat{\tau}_{\mathrm{FRDD}}$ is obtained as follows.

\begin{theorem}\label{thm:LATE-rate}
Consider the setup of Sections \ref{subsec:FRD-setup} and \ref{subsec:FRD-id}. Assume that $\hat{m}_z - m_z = O_p(r_n)$ for $z \in \{0,1\}$,
where $r_n \to 0$ as $n \to \infty$. Under Assumptions \ref{ass:ker} and \ref{ass:FRD-id} and under Assumption \ref{ass:GSRD-rate}, replacing $\mu_{t,\oplus}$, $M_{t,\oplus}(\nu,c)$, $Y(t)$ for $t \in \{0,1\}$ with $\mu_{z,\oplus}^\dag$, $ M_{z,\oplus}^\dag(\nu,c)$, $Y(T(z))$ for $z \in \{0,1\}$, if $h_{\mathrm{max}} \to 0$ and $nh_{\mathrm{min}}\to\infty$ as $n \to \infty$, then
\begin{align*}
d_\mathcal{H}(\tau_{\mathrm{FRDD}},\hat{\tau}_{\mathrm{FRDD}}) = O(h_{\mathrm{max}}^{2/(\beta_1-1)}) + O_p(r_n + (nh_{\mathrm{min}})^{-\frac{1}{2(\beta_2-1)}}).
\end{align*}
\end{theorem}

Many common metric spaces satisfy Assumption \ref{ass:GSRD-rate} with $\beta_1 = \beta_2=2$. If one uses the local linear estimator \eqref{eq:first-stage-LL} for estimating $m_z$, its typical convergence rate is $\tilde{r}_n = O(h_{\mathrm{max}}^2) + O_p((nh_{\mathrm{min}})^{-1/2})$ (cf. Theorems 19.1 and 19.2 in \citet{ha22}). Then letting $\beta_1=\beta_2=2$, $h_0=h_1=h$ and $r_n  = \tilde{r}_n$, we have
\begin{align*}
d_\mathcal{H}(\tau_{\mathrm{FRDD}},\hat{\tau}_{\mathrm{FRDD}}) &= O(h^2) + O_p((nh)^{-1/2}).
\end{align*}
The right-hand side is optimized with $h=n^{-1/5}$ and the resulting convergence rate is 
the nonparametric rate $d_\mathcal{H}(\tau_{\mathrm{FRDD}}, \hat{\tau}_{\mathrm{FRDD}}) = O_p(n^{-2/5})$. In the fuzzy RDD setting, we can also perform a test for no treatment effect by implementing the test under the sharp RDD discussed in Section \ref{subsec:LFR-boot}. Indeed, under the null hypothesis $\mathbb{H}_0: d_\mathcal{H}(\Psi(\mu_{1,\oplus}),\Psi(\mu_{0,\oplus}))=d(\mu_{1,\oplus},\mu_{0,\oplus})=0$, we have $d(\mu_{1,\oplus},\mu_{0,\oplus})=0$ if and only if $d(\E_\oplus[Y(T(1))|R=c],\E_\oplus[Y(T(0))|R=c])=0$.

\section{Concluding remarks}\label{sec:concl}
We introduced a general framework for RDD for complex non-Euclidean outcomes, including functional data, compositional data and networks, where causal effects are quantified via geodesics between local Fr\'echet means. A key feature of the proposed framework is the distinction between intrinsic estimation and extrinsic inference: estimation is carried out within the original metric space using LFR, ensuring interpretability, while inference is conducted via Hilbert space embedding, which enables asymptotic analysis and hypothesis testing. For implementation, we develop a data-adaptive bandwidth selection method based on one-sided cross-validation near the cutoff. The resulting methodology is supported by theoretical guarantees and is demonstrated through simulations and an application.

Several promising directions remain for future research. First, extending GRDD to accommodate multiple cutoff points would enhance its utility in settings involving staggered interventions or tiered eligibility rules; see \citet{bert:20} and \citet{catt:21} for the scalar case. Second, generalizing GRDD to handle multivariate running variables would make the method applicable to more complex treatment assignment mechanisms  \citep{black:99,keel:15:2,sawa:24,catt:25}. This extension introduces challenges related to neighborhood construction and bandwidth selection in higher-dimensional spaces. 

\section*{Acknowledgements}
We wish to thank the editor, the associate editor, and referees for their valuable comments, which have significantly improved the quality of the paper.

\noindent{\it Conflicts of interest:}  None declared.

\section*{Funding}
D.K. was supported in part by JSPS KAKENHI Grant Numbers 23K12456,  25K00624, and 26K00323. H.G.M. was partially supported by NSF grant DMS-2310450. We thank Takuya Ishihara and Yasumasa Matsuda for their helpful comments and discussion.

\section*{Data availability}
All source code required to implement the proposed methods and reproduce the simulation and empirical results in this paper is publicly available at
\url{https://github.com/yidongzhou/GRDD}. The UK parliamentary election data analyzed in Section~\ref{sec:app} were compiled by \citet{egge:17} and are available through the JOP Dataverse at \url{https://doi.org/10.7910/DVN/7QLCV1}.

\single
\bibliographystyle{rss}
\bibliography{collection.bib}
\double

\clearpage
\begin{center}
{\large\bf SUPPLEMENTARY MATERIAL}
\end{center}

\setcounter{section}{0}
\setcounter{subsection}{0}
\setcounter{equation}{0}
\setcounter{figure}{0}
\setcounter{table}{0}
\setcounter{proposition}{0}
\setcounter{definition}{0}
\setcounter{assumption}{0}
\setcounter{theorem}{0}
\setcounter{corollary}{0}
\setcounter{remark}{0}
\setcounter{example}{0}

\makeatletter
\@removefromreset{equation}{section}
\@removefromreset{proposition}{section}
\@removefromreset{definition}{section}
\@removefromreset{assumption}{section}
\@removefromreset{theorem}{section}
\@removefromreset{corollary}{section}
\@removefromreset{remark}{section}
\@removefromreset{example}{section}
\makeatother

\renewcommand{\thesection}{S.\arabic{section}}
\renewcommand{\thesubsection}{S.\arabic{section}.\arabic{subsection}}
\renewcommand{\theequation}{S\arabic{equation}}
\renewcommand{\thefigure}{S\arabic{figure}}
\renewcommand{\thetable}{S\arabic{table}}
\renewcommand{\theproposition}{S\arabic{proposition}}
\renewcommand{\thedefinition}{S\arabic{definition}}
\renewcommand{\theassumption}{S\arabic{assumption}}
\renewcommand{\thetheorem}{S\arabic{theorem}}
\renewcommand{\thecorollary}{S\arabic{corollary}}
\renewcommand{\theremark}{S\arabic{remark}}

\def\theHsection{supp.\arabic{section}}
\def\theHsubsection{supp.\arabic{section}.\arabic{subsection}}
\def\theHequation{supp.\arabic{equation}}
\def\theHfigure{supp.\arabic{figure}}
\def\theHtable{supp.\arabic{table}}
\def\theHproposition{supp.\arabic{proposition}}
\def\theHdefinition{supp.\arabic{definition}}
\def\theHassumption{supp.\arabic{assumption}}
\def\theHtheorem{supp.\arabic{theorem}}
\def\theHcorollary{supp.\arabic{corollary}}
\def\theHremark{supp.\arabic{remark}}

\section{Local Fr\'echet regression}\label{app:lfr}

Local Fr\'echet regression (LFR) \citep{mull:19:6,mull:22:8} generalizes local linear regression \citep{fan:96} to settings where the outcome lies in a metric space. Here we outline the intuition and construction of LFR by first revisiting local linear regression, then drawing parallels that motivate the metric-space extension.

\subsection{Local linear regression as a weighted Fr\'echet mean}

Let $Y \in \mathbb{R}$ be a scalar outcome and $R \in \mathbb{R}$ a scalar predictor. The local linear regression function $m(r)$ estimates the conditional mean $\E[Y|R = r]$ by solving a localized least squares problem centered at $r$:
\[
(\beta_0, \beta_1) = \argmin_{\beta_0, \beta_1} \E\left[ K_h(R - r)\left\{Y - \beta_0 - \beta_1(R - r)\right\}^2 \right],
\]
where $K_h(u) = h^{-1}K(u/h)$ is a kernel function with bandwidth $h > 0$.

The solution for $\beta_0$ admits a closed-form expression. Let $\mu_k = \E[K_h(R - r)(R - r)^k]$ for $k = 0,1,2$ and define the weighted cross-moments:
\[
\xi_k = \E[K_h(R - r)(R - r)^k Y], \quad k = 0,1,
\]
and set $\sigma^2 = \mu_0 \mu_2 - \mu_1^2$. Then:
\[
\beta_0 = \frac{\mu_2 \xi_0 - \mu_1 \xi_1}{\sigma^2}, \quad \beta_1 = \frac{\mu_0 \xi_1 - \mu_1 \xi_0}{\sigma^2}.
\]
This yields the regression function:
\[
m(r) = \beta_0 = \E\left[ s(r; R, h)\, Y \right],
\]
where the weight function is defined by:
\[
s(r; R, h) = \frac{1}{\sigma^2} K_h(R - r)\left( \mu_2 - \mu_1(R - r) \right).
\]

Moreover, $m(r)$ can also be characterized as a weighted mean:
\[
m(r) = \argmin_{y \in \mathbb{R}} \E\left[ s(r; R, h) (Y - y)^2 \right].
\]
That is, local linear regression solves a local weighted Fr\'echet mean problem in the Euclidean space, where squared Euclidean distance is used and the weights adapt to the proximity of $R$ to $r$ via a kernel function.

\subsection{Local Fr\'echet regression}

The insight from the above formulation extends naturally to non-Euclidean settings. Suppose now that the outcome $Y$ resides in a metric space $(\mathcal{M}, d)$, and $R \in \mathbb{R}$ remains the scalar running variable. The local Fr\'echet regression function $m(r)$ is defined as:
\[
m(r) = \argmin_{\nu \in \mathcal{M}} \E\left[ s(r; R, h)\, d^2(Y, \nu) \right],
\]
where the weight function $s(r; R, h)$ is the same as in the Euclidean case:
\[
s(r; R, h) = \frac{1}{\sigma^2} K_h(R - r)\left( \mu_2 - \mu_1(R - r) \right),
\]
with $\mu_k = \E[K_h(R - r)(R - r)^k]$ and $\sigma^2 = \mu_0 \mu_2 - \mu_1^2$ as before.

Thus, LFR generalizes local linear regression by replacing the Euclidean outcome $Y \in \mathbb{R}$ and the squared error loss $(Y - y)^2$ with a metric space-valued outcome $Y \in \mathcal{M}$ and the squared metric loss $d^2(Y, \nu)$.

\subsection{Sample estimator}

Given a sample $\{(Y_i, R_i)\}_{i=1}^n$, we define the empirical moments:
\[
\hat{\mu}_k = \frac{1}{n} \sum_{i=1}^n K_h(R_i - r)(R_i - r)^k, \quad k = 0,1,2,
\]
and $\hat{\sigma}^2 = \hat{\mu}_0 \hat{\mu}_2 - \hat{\mu}_1^2$. The sample weight function becomes:
\[
\hat{s}(r; R_i, h) = \frac{1}{\hat{\sigma}^2} K_h(R_i - r)\left( \hat{\mu}_2 - \hat{\mu}_1(R_i - r) \right).
\]
The local Fr\'echet regression estimator is then:
\[
\hat{m}(r) = \argmin_{\nu \in \mathcal{M}} \frac{1}{n} \sum_{i=1}^n \hat{s}(r; R_i, h)\, d^2(Y_i, \nu).
\]

\section{Air pollution and urban rail in Taipei}
As an illustration of the proposed GRDD methodology for functional outcomes, we revisit the Taipei air pollution setting studied by \citet{chen:12} and consider a functional extension of their analysis. The data consist of hourly carbon monoxide (CO) measurements from ten monitoring stations in the Taipei metropolitan area, obtained from the Taiwanese Environmental Protection Administration for the period 1994--2007. Among these stations, seven are located near major highways and three are located farther away, allowing a comparison of areas with different exposure to traffic-related pollution.

The Taipei Metro underwent a major expansion in March 1996. Following \citet{chen:12}, we take March 28, 1996, the official opening date as the cutoff and use calendar time, measured in days relative to this date, as the running variable. The outcome for each observational unit is a daily CO concentration curve, which we view as an element of a Hilbert space of functions (Example~\ref{exm:fun}). In this setting, the geodesic between two curves corresponds to pointwise linear interpolation. Since the measurements are recorded hourly and exhibit substantial missingness and noise, the functional observations are not directly observed in continuous form but must be reconstructed from discrete data. We therefore first smooth the hourly measurements to obtain station-day curves. This preprocessing step falls within the broader problem of empirical function estimation from discrete and potentially incomplete data, which is common in functional data analysis \citep{rams:05, mull:05:4, mull:16:3}.

These reconstructed station-level curves are then aggregated within the near-highway and far-from-highway groups to obtain group-level daily curves, yielding a sequence of functional observations suitable for GRDD analysis. Aggregation is motivated by the presence of missingness and measurement noise in the station-level data, which can vary across stations and over time and may affect the stability of the reconstructed curves. Constructing group-level curves provides a more stable representation of temporal pollution patterns near the cutoff, although it may smooth over some station-level heterogeneity. The results should therefore be interpreted with this trade-off in mind \citep{gend:22}.

This application serves primarily to demonstrate how GRDD can be implemented in a functional data setting. The empirical design follows a regression discontinuity in time framework, with daily pollution profiles treated as curve-valued outcomes. As in related time-based RDD settings \citep{haus:18}, interpretation of the estimated discontinuity may be affected by features such as gradual policy implementation and temporal dependence in the data. Accordingly, the results are best viewed as summarizing local changes in pollution profiles around the cutoff, rather than as definitive causal evidence. Extending GRDD to dependent data settings (e.g., under mixing conditions) is an important direction for future work.

\begin{figure}[tb]
    \single
    \centering
    \includegraphics[width=0.9\linewidth]{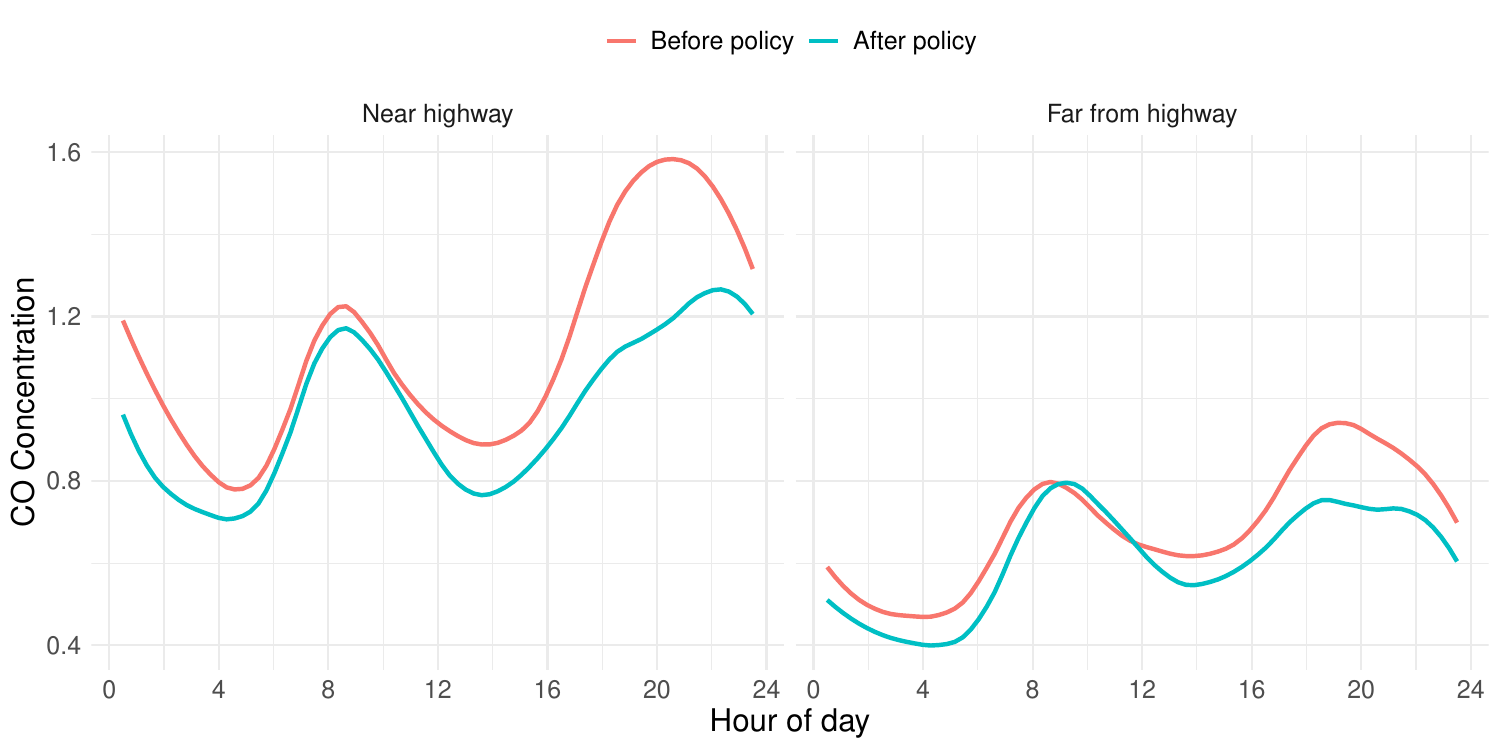}
    \caption{Estimated daily CO concentration curves before and after the Taipei Metro expansion. Each panel shows the average 24-hour CO concentration curves estimated for stations near highways (left) and far from highways (right) using geodesic sharp RDD. The blue and red curves represent the estimated CO concentration trajectories just before and just after the March 28, 1996 policy cutoff.}
    \label{fig:metro}
\end{figure}

While one could instead estimate station-level curves and apply pointwise scalar RDD methods at each time point, such approaches treat each hour independently and do not account for the inherent smoothness and functional structure of the data. In contrast, GRDD models the discontinuity at the level of entire curves, capturing cross-time dependence in daily pollution patterns and facilitating interpretation of changes in overall curve shape. At the same time, GRDD should be viewed as complementary to, rather than a replacement for, pointwise approaches in this setting.

We apply GRDD separately to the near-highway and far-from-highway groups using local Fr\'echet regression with the data-adaptive bandwidth selector described in Section~5.1. The selected bandwidths are 60 days for the near-highway and far-from-highway groups. Figure~\ref{fig:metro} displays the estimated average 24-hour CO concentration curves immediately before and after the cutoff. For stations near highways, the estimated post-cutoff curve lies below the pre-cutoff curve over much of the day, with the largest differences occurring during peak traffic hours. For stations farther from highways, the differences are smaller and more diffuse.

\begin{figure}[tb]
    \single
  \centering
  \begin{subfigure}{0.45\textwidth}
    \centering
    \includegraphics[width=1\linewidth]{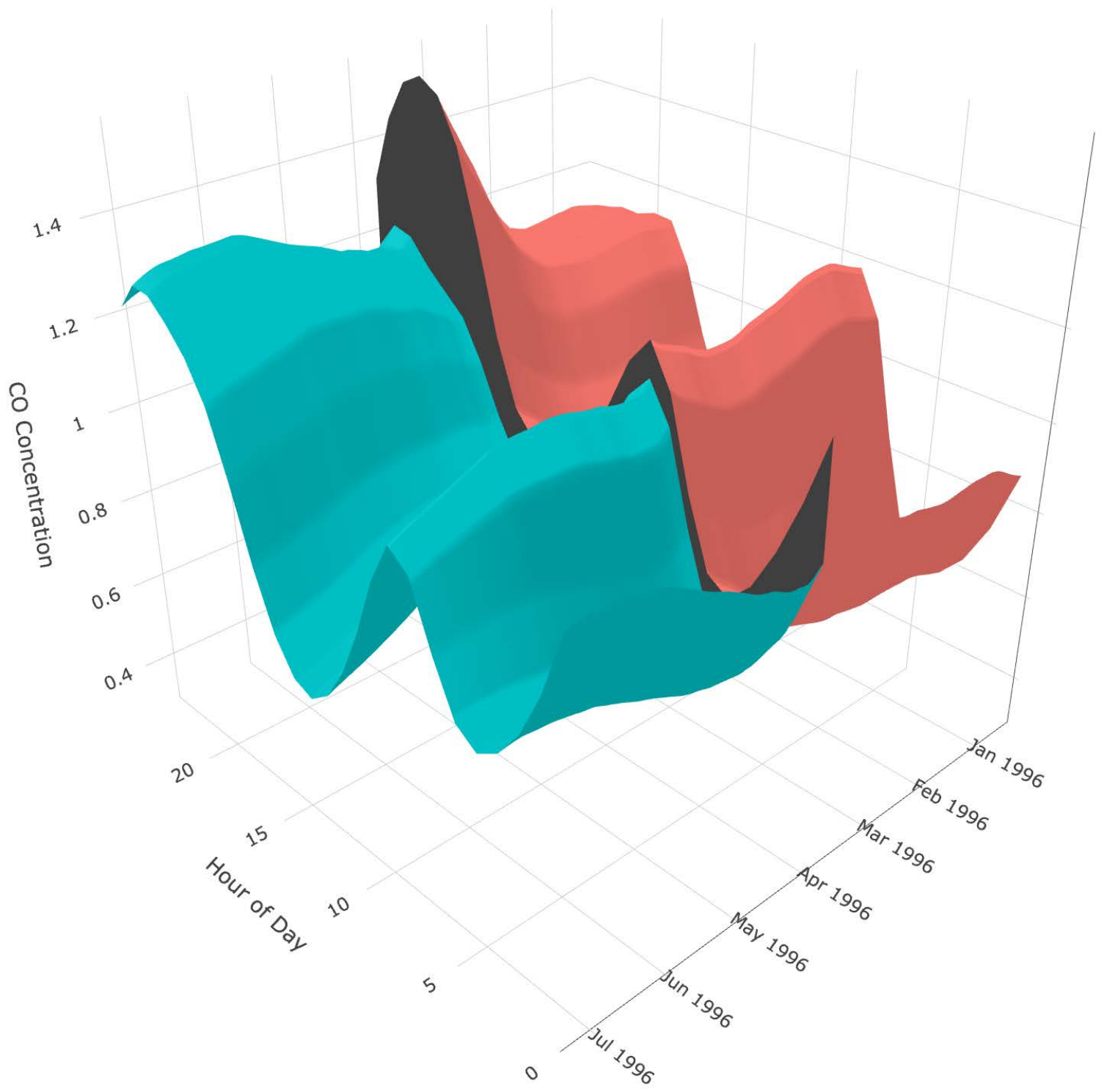}
    \caption{}
  \end{subfigure}\hfill
  \begin{subfigure}{0.45\textwidth}
    \centering
    \includegraphics[width=1\linewidth]{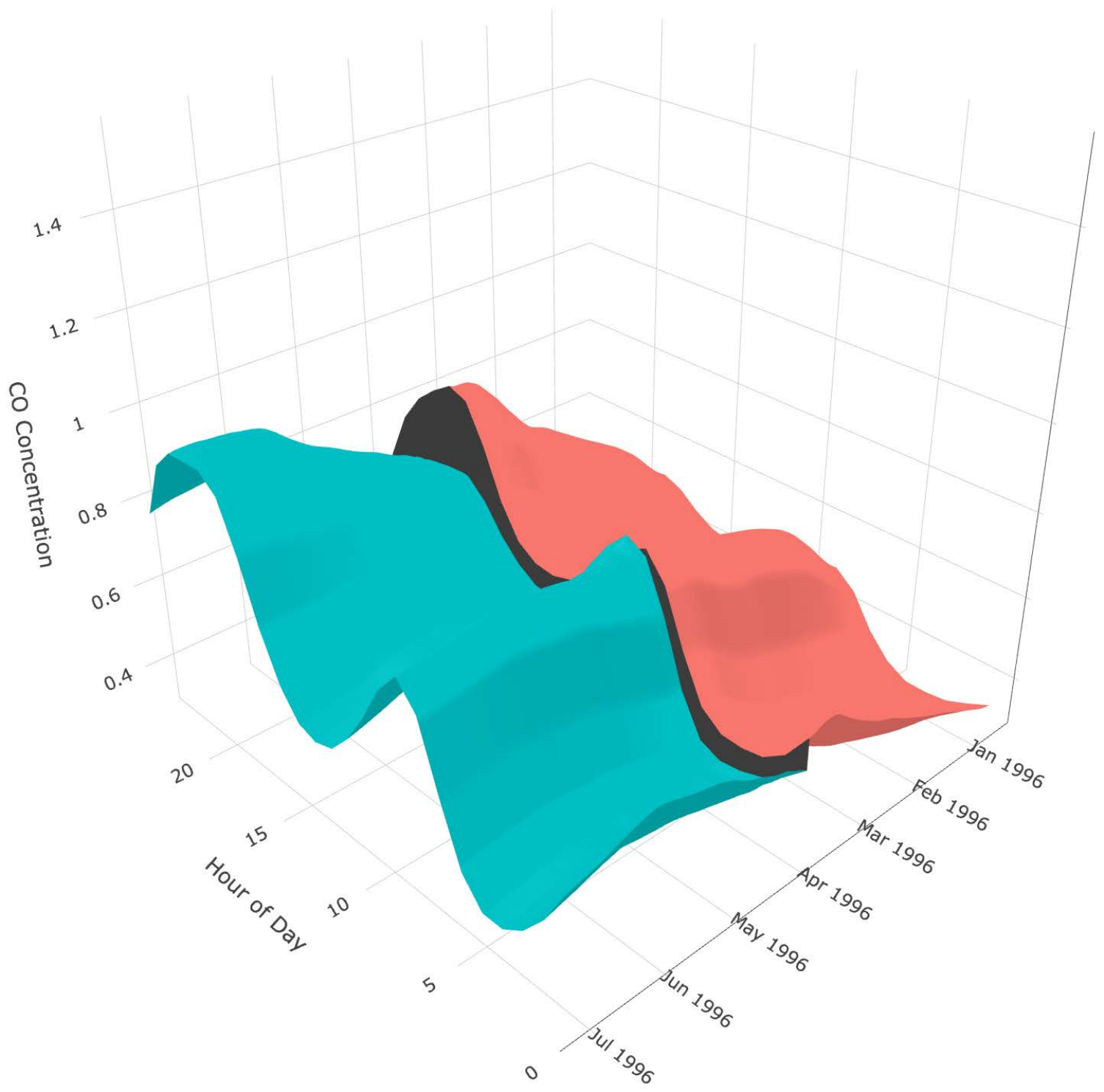}
    \caption{}
  \end{subfigure}
  \caption{Daily CO concentration surfaces around the metro policy cutoff for monitoring stations located (a) near and (b) far from highways. Dates before the policy cutoff are shown in red, and those after the cutoff are shown in blue. The policy cutoff date (March 28, 1996) is marked by a gray vertical wall. Both panels use a shared vertical axis scale to facilitate comparison.}
  \label{fig:metronf}
\end{figure}

To further visualize the evolution of the fitted curves around the policy date, Figure~\ref{fig:metronf} displays the local Fr\'echet regression surfaces over a 200-day window centered at the cutoff. The results suggest a more pronounced post-cutoff decline in the near-highway group relative to the far-from-highway group \citep{chen:12, jaya:17}. Overall, this application illustrates how GRDD can be used to summarize local changes in functional outcomes around a cutoff.

\section{Geodesic fuzzy RDD}\label{subsec:GFRD}
One might prefer to treat compliers' LATE at the cutoff as a quantity defined within the metric space $\mathcal{M}$, rather than involving an extrinsic Hilbert space. For such an intrinsic approach,  the geodesic from $\mu_{0,\oplus}$ to $\mu_{1,\oplus}$ can be considered as the causal effect of interest instead of $\tau_{\mathrm{FRDD}}$.
\begin{definition}[Compliers' GLATE at the cutoff]
The compliers' GLATE is defined as the geodesic from $\mu_{0,\oplus}$ to $\mu_{1,\oplus}$, that is, \[\tau_{\mathrm{GFRD}} = \gamma_{\mu_{0,\oplus}, \mu_{1,\oplus}}.\]
\end{definition}

\begin{figure}[tb]
    \single
    \centering
    \includegraphics[width=0.9\linewidth]{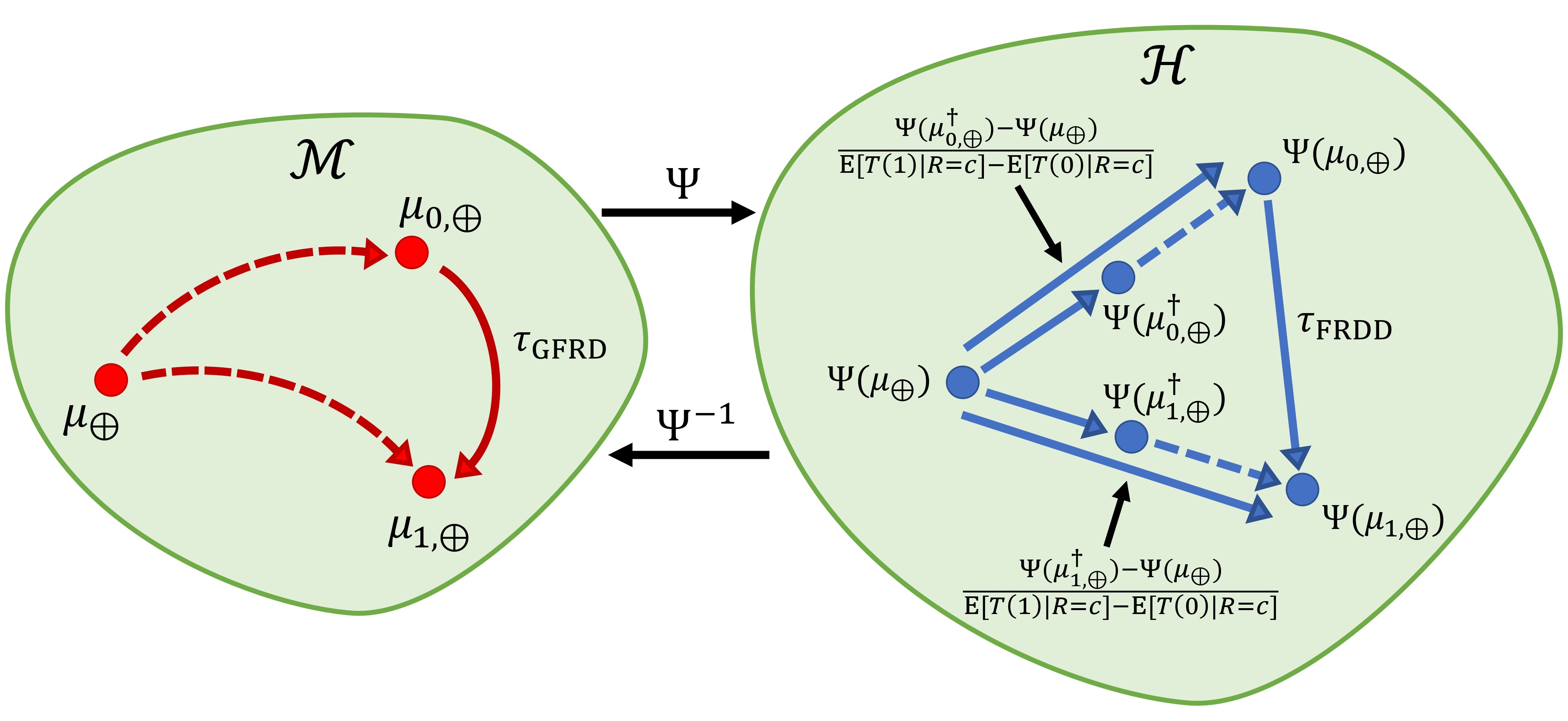}
    \caption{Illustration of the identification of $\tau_{\mathrm{FRDD}}$ and $\tau_{\mathrm{GFRD}}$. The circles symbolize objects in the geodesic space $\mathcal{M}$, and the arrows emanating from these circles represent the geodesic paths connecting them. The solid blue arrow from $\Psi(\mu_{0,\oplus})$ to $\Psi(\mu_{1,\oplus})$ represents the compliers' LATE. The solid red arrow represents the compliers' GLATE.}
    \label{fig:GFRD}
\end{figure}

In contrast to compliers' LATE in Definition 7.1, compliers' GLATE $\tau_{\mathrm{GFRD}}$ requires identification of the starting and ending points of the geodesic, i.e., $\mu_{z,\oplus}$. To this end, define $\mu_\oplus$ as the Fr\'echet mean of the outcome for always/never-takers at the cutoff, that is, $\mu_\oplus = \E_\oplus[Y(T)|T(1)=T(0),R=c]$. When the metric space $(\mathcal{M},d)$ admits a map $\Psi$ satisfying Assumption 4.1, one can see that
\begin{align}\label{eq:GFRD-id-z-main}
\Psi(\mu_{z,\oplus}) &= \Psi(\mu_\oplus) + \frac{\Psi(\mu_{z,\oplus}^\dag) - \Psi(\mu_\oplus)}{\E[T(1)|R=c]- \E[T(0)|R=c]}\quad \mbox{for} \quad z \in \{0,1\},
\end{align}
where $\mu_{z,\oplus}^\dag = \E_\oplus[Y(T(z))|R=c]$ (see (\ref{eq:GFRD-id-z}) for the derivation of (\ref{eq:GFRD-id-z-main})). Since $\Psi: \mathcal{M} \to \Psi(\mathcal{M})$ is bijective, we have
\begin{align*}
\mu_{z,\oplus} &= \Psi^{-1}\left(\Psi(\mu_\oplus) + \frac{\Psi(\mu_{z,\oplus}^\dag) - \Psi(\mu_\oplus)}{\E[T(1)|R=c]- \E[T(0)|R=c]}\right)\quad \mbox{for} \quad z \in \{0,1\},
\end{align*}
where $\Psi^{-1}:\Psi(\mathcal{M}) \to \mathcal{M}$ is the inverse map of $\Psi$. In the following, we will assume that the argument of $\Psi^{-1}$ is in the image space $\Psi(\mathcal{M})$ and the results are conditional on the event that the corresponding estimates introduced below also satisfy this condition. Figure \ref{fig:GFRD} illustrates the identification of $\tau_{\mathrm{FRDD}}$ and $\tau_{\mathrm{GFRD}}$. Note that we can estimate $\mu_{z,\oplus}^\dag$ and $m_z$ by $\hat{\nu}_{z,\oplus}$ and $\hat{m}_z$, respectively. Therefore, once $\mu_\oplus$ is identified, we can also identify $\mu_{z,\oplus}$. 

However, in the present setup, $\mu_{\oplus}$ cannot be identified in general. As an important special case where $\mu_{\oplus}$ can be identified, we hereafter focus on the case of one-sided noncompliance:
\begin{assumption}\label{ass:AT-NT}
For any unit, either one of the following holds.
\begin{itemize}
\item[(i)] $(Z,T) \in \{(1,1), (0,1),(0,0)\}$.
\item[(ii)] $(Z,T) \in \{(1,1), (1,0),(0,0)\}$.
\end{itemize}
\end{assumption}
Condition (i) implies that only compliers and always-takers exist. In this case, $\mu_\oplus$ becomes the Fr\'echet mean of the outcome for always-takers at the cutoff. On the other hand, Condition (ii) implies that only compliers and never-takers exist. In this case, $\mu_\oplus$ becomes the Fr\'echet mean of the outcome for never-takers at the cutoff.

Assume that $\E_\oplus[Y(T)|T(1)=T(0), R=r]$ exists uniquely for any $R=r$ over a neighborhood of $c$, and that it is continuous over that neighborhood. We refer to Assumption 7.1 with this condition as Assumption 7.1'. 
Under one-sided noncompliance, we can identify $\mu_\oplus$ as follows. 
\begin{proposition}\label{prp:GFRD-id}
Consider the setup of Sections 7.1 and 7.2 and suppose that Assumption 7.1' holds true. Then
\begin{align*}
\mu_\oplus &= 
\begin{cases}
\lim_{r \uparrow c}\E_\oplus[Y|T=1,Z=0, R=r] & \text{under Assumption \ref{ass:AT-NT}(i)}\\[5pt]
\lim_{r \downarrow c}\E_\oplus[Y|T=0,Z=1, R=r] & \text{under Assumption \ref{ass:AT-NT}(ii)}
\end{cases}.
\end{align*}
\end{proposition}

From this proposition, we can estimate $\tau_{\mathrm{GFRD}}$ as follows. 
\begin{align*}
\hat{\tau}_{\mathrm{GFRD}} &= \gamma_{\hat{\mu}_{0,\oplus}, \hat{\mu}_{1,\oplus}},\quad \hat{\mu}_{z,\oplus}= \Psi^{-1}\left(\Psi(\hat{\mu}_\oplus) + \frac{\Psi(\hat{\nu}_{z,\oplus}) - \Psi(\hat{\mu}_\oplus)}{\hat{m}_1- \hat{m}_0}\right)\quad \mbox{for} \quad z \in \{0,1\},
\end{align*}
where $\hat{\nu}_{z,\oplus}$ and $\hat{m}_z$ are the estimators considered in Section 3.2, and $\hat{\mu}_\oplus$ is the LFR estimator defined as 
\begin{align*}
\hat{\mu}_\oplus &= 
\begin{cases}
 \hat{\mu}_{\mathrm{A},\oplus} := \argmin_{\nu \in \mathcal{M}}\frac{1}{n_\mathrm{A}}\sum_{i\in I_\mathrm{A}} \hat{s}_0(c;R_i,h_0)d^2(\nu,Y_i) & \text{under Assumption \ref{ass:AT-NT}(i)}\\
\hat{\mu}_{\mathrm{N},\oplus} := \argmin_{\nu \in \mathcal{M}}\frac{1}{n_\mathrm{N}}\sum_{i\in I_\mathrm{N}} \hat{s}_1(c;R_i,h_1)d^2(\nu,Y_i) & \text{under Assumption \ref{ass:AT-NT}(ii)}
\end{cases},
\end{align*} 
where $I_\mathrm{A} = \{1\leq i \leq n: T_i=1, Z_i = 0\}$, $I_\mathrm{N} = \{1\leq i \leq n: T_i=0, Z_i = 1\}$, $n_\mathrm{S}$ is the cardinality of $I_\mathrm{S}$ for $\mathrm{S} \in \{\mathrm{A}, \mathrm{N}\}$, and $\hat{s}_z(c;r,h_z)$ is the estimated weight function of the LFR defined in Section 3.2. 

For $z \in \{0,1\}$, define $M_\oplus(\nu,c) = \E[d^2(\nu,Y(T))|T(1)=T(0),R=c]$ and $\mu_{\mathrm{S},\oplus} =\argmin_{\nu \in \mathcal{M}}M_{\mathrm{S}, n}(\nu)$, $\mathrm{S} \in \{\mathrm{A},\mathrm{N}\}$ where   
\begin{align*}
&M_{\mathrm{A}, n}(\nu) = \E[s_0(c;R,h_0)d^2(\nu,Y(T))|T=1,Z=0],\\
&M_{\mathrm{N}, n}(\nu) = \E[s_1(c;R,h_1)d^2(\nu,Y(T))|T=0,Z=1].
\end{align*}

Letting $h_{\mathrm{max}}=\max\{h_0,h_1\}$ and $h_{\mathrm{min}} = \min\{h_0,h_1\}$, the convergence rate of $\hat{\tau}_\mathrm{GFRD}$ is obtained as follows.
\begin{proposition}\label{prp:GLATE-rate}
Consider the setup of Sections 7.1 and 7.2. Under Assumptions 3.4, 3.5, 7.1', \ref{ass:AT-NT} and the assumptions in Theorem 7.2, except for Assumption 7.1, as well as Assumption 3.3 when replacing $\mu_{t,\oplus}$, $\nu_{t,\oplus}$, $\hat{\nu}_{t,\oplus}$, $M_{t,n}(\nu)$, $M_{t,\oplus}(\nu,c)$, $Y(t)$ for $t \in \{0,1\}$ with $\mu_\oplus$, $\mu_{\mathrm{S},\oplus}$, $\hat{\mu}_{\mathrm{S},\oplus}$, $M_{\mathrm{S},n}(\nu)$, $M_\oplus(\nu,c)$, $Y(T)$ for $\mathrm{S} \in \{\mathrm{A},\mathrm{N}\}$, it holds that  
\begin{align*}
d_\mathcal{G}(\tau_{\mathrm{GFRD}},\hat{\tau}_{\mathrm{GFRD}}) = O(h_{\mathrm{max}}^{2/(\beta_1-1)}) + O_p(r_n + (nh_{\mathrm{min}})^{-\frac{1}{2(\beta_2-1)}}).
\end{align*}
\end{proposition}

\section{Fuzzy RDD for outcomes in Riemannian manifolds}\label{subsec:FRD-manifold}
In this subsection, we consider fuzzy RDD for outcomes in (a subset of) a Riemannian manifold $\mathcal{M}\subset \mathbb{R}^p$, for which the Riemannian logarithmic and exponential maps are well-defined. Let $T_\mu \mathcal{M}$ be the $(p-1)$-dimensional tangent space at $\mu \in \mathcal{M}$. We assume that for any $\mu \in \mathcal{M}$, the logarithmic map $\mathrm{Log}_\mu(\cdot):\mathcal{M} \to T_\mu\mathcal{M}$ is bijective and $\mathrm{Log}_\mu(\mathcal{M})$, the image space of $\mathcal{M}$ under $\mathrm{Log}_\omega$, is a convex subset of $\mathbb{R}^{p-1}$.  A typical example for $\mathcal{M}$ is the space of compositional data with the arc-length distance (Example~\ref{exm:com}), which is the positive orthant of a finite-dimensional sphere.

For $z \in \{0,1\}$, define $\mu_{z,\omega }^\sharp = \E[\mathrm{Log}_{\omega}\left(Y(T(z))\right)|T(1)>T(0),R=c]$ where $\omega \in \mathcal{M}$ is a fixed reference point that can be chosen arbitrarily; a simple choice is the sample Fr\'echet mean of the outcomes $\{Y_i\}_{i=1}^n$.
Due to the linear structure of the tangent space, one may consider a contrast of the means $\mu_{z,\omega }^\sharp$ of potential outcomes (transformed by the logarithmic map). Specifically, our causal estimand of interest is defined as follows. 

\begin{definition}[Compliers' LATE at the cutoff for outcomes in a Riemannian manifold]
For outcomes taking values in a Riemannian manifold considered in this subsection, we define the compliers' LATE at the cutoff with a reference point $\omega$ as 
\begin{align*}
\tau_{\mathrm{FRDD}}^\sharp &:= \mu_{1,\omega}^\sharp - \mu_{0,\omega}^\sharp.
\end{align*}
\end{definition}

When $\mathcal{M}$ is (a subset of) the Euclidean space $\mathbb{R}$, $\tau_{\mathrm{FRDD}}^\sharp$ is independent of $\omega$ and is reduced to 
\[
\tau_{\mathrm{FRDD}}^\sharp = \E[Y(T(1))|T(1)>T(0),R=c] - \E[Y(T(0))|T(1)>T(0),R=c],
\] 
which is the conventional compliers' LATE at the cutoff for Euclidean outcomes. 

For the identification of $\tau_{\mathrm{FRDD}}^\sharp$, we add the following conditions.
\begin{assumption}\label{ass:FRD-id-manifold} \quad 
\begin{itemize}
\item[(i)] $T(0) \leq T(1)$.
\item[(ii)] $\Pr(T(1) > T(0)|R=c) \in (\eta_0, 1-\eta_0)$ for some $\eta_0 \in (0,1/2)$.
\item[(iii)] $\E[T(1)|R=r]$, $\E[T(0)|R=r]$, $\E[\mathrm{Log}_{\omega}(Y(T(0)))|R=r]$, and $\E[\mathrm{Log}_{\omega}(Y(T(1)))|R=r]$ are continuous over a neighborhood of $c$. 
\end{itemize}
\end{assumption}

By a similar argument to Theorem 7.1, the causal effect $\tau_{\mathrm{FRDD}}^\sharp$ can be identified as follows.
\begin{theorem}\label{thm:FRD-comp-id}
Consider the setup of Sections 7.1 and \ref{subsec:FRD-manifold} and suppose that Assumption \ref{ass:FRD-id-manifold} holds true. Then 
\begin{align*}
\tau_{\mathrm{FRDD}}^\sharp 
&= 
\frac{\lim_{r \downarrow c}\E[\mathrm{Log}_{\omega}(Y)|R=r] - \lim_{r \uparrow c}\E[\mathrm{Log}_{\omega}(Y)|R=r]}{\lim_{r \downarrow c}\E[T|R=r] - \lim_{r \uparrow c}\E[T|R=r]}.
\end{align*}
\end{theorem}

From Theorem \ref{thm:FRD-comp-id}, we can construct an estimator of $\tau_{\mathrm{FRDD}}^\sharp$ as follows. 
\begin{align*}
\hat{\tau}_{\mathrm{FRDD}}^\sharp &=
\frac{\hat{\nu}_{1,\omega}^\sharp - \hat{\nu}_{0,\omega}^{\sharp}}{\hat{m}_1 - \hat{m}_0},
\end{align*} 
where $\hat{m}_z$ is the estimator considered in Section 3.2 and $\hat{\nu}_{z,\omega}^\sharp$ is the LFR estimator defined as 
\begin{align*}
\hat{\nu}_{z,\omega}^\sharp = \argmin_{\nu \in T_\omega \mathcal{M}}\hat{M}_{z,n}(\nu),\ \hat{M}_{z,n}(\nu)=\frac{1}{n_z}\sum_{i=1}^n \hat{s}_z(c;R_i,h_z)d_E^2(\nu,\mathrm{Log}_\omega(Y_i))\quad \mbox{for} \quad z \in \{0,1\},
\end{align*}
where $d_E$ is the standard Euclidean metric and $\hat{s}_z(c;r,h_z)$ is the estimated weight function of the LFR defined in Section 3.2.

For $z \in \{0,1\}$, define $\nu_{z,\omega}^\sharp = \argmin_{\nu \in T_\omega \mathcal{M}}M_{z,\oplus}^\sharp(\nu,c)$ and $\tilde{\nu}_{z,\omega}^\sharp= \argmin_{\nu \in T_\omega \mathcal{M}}M_{z,n}^\sharp(\nu)$ where
\begin{align*}
&M_{z,\oplus}^\sharp(\nu,c) = \E[d_E^2(\nu,\mathrm{Log}_\omega(Y(T(z))))|R=c],\\
&M_{z,n}^\sharp(\nu) = \E[s_z(c;R,h_z)d_E^2(\nu,\mathrm{Log}_\omega(Y(T(z))))].
\end{align*}
Letting $h_{\mathrm{max}}=\max\{h_0,h_1\}$ and $h_{\mathrm{min}} = \min\{h_0,h_1\}$, the convergence rate of $\hat{\tau}_{\mathrm{FRDD}}^\sharp$ is obtained as follows.
\begin{theorem}\label{thm:LATE-rate-manifold}
Consider the setup of Sections 7.1 and \ref{subsec:FRD-manifold}. Assume that $\hat{m}_z - m_z = O_p(r_n)$, $z \in \{0,1\}$,
where $r_n$ is a sequence of constants such that $r_n \to 0$ as $n \to \infty$. Under Assumptions 3.2 and \ref{ass:FRD-id-manifold} and  Assumption 3.3 when replacing $(\mathcal{M},d)$, $\mu_{t,\oplus}$, $\nu_{t,\oplus}$, $\hat{\nu}_{t,\oplus}$, $M_{t,n}(\nu)$, $M_{t,\oplus}(\nu,c)$, $Y(t)$ for $t \in \{0,1\}$ with $(T_\omega\mathcal{M},d_E)$, $\nu_{z,\omega}^\sharp$, $\tilde{\nu}_{z,\omega}^\sharp$, $\hat{\nu}_{z,\omega}^\sharp$, $M_{z,n}^\sharp(\nu)$, $M_{z,\oplus}^\sharp(\nu,c)$, $\mathrm{Log}_\omega(Y(T(z)))$ for $z \in \{0,1\}$, if  $h_{\mathrm{max}} \to 0$ and $nh_{\mathrm{min}}\to\infty$ as $n \to \infty$, then 
\begin{align*}
d_E(\tau_{\mathrm{FRDD}}^\sharp,\hat{\tau}_{\mathrm{FRDD}}^\sharp) = O(h_{\mathrm{max}}^{2/(\beta_1-1)}) + O_p(r_n + (nh_{\mathrm{min}})^{-\frac{1}{2(\beta_2-1)}}).
\end{align*}
\end{theorem}

If one uses the sample Fr\'echet mean $\hat{m}_\oplus$ of the outcomes $\{Y_i\}_{i=1}^n$ as the reference point $\omega$, then the techniques used to derive Theorem \ref{thm:LATE-rate-manifold} cannot be directly applied since the resulting estimator involves the logarithmic map at the estimated point $\hat{m}_\oplus$. \citet{chen:23:1} investigate the theoretical properties of estimators in a similar setting. By leveraging the techniques developed in that paper, it is expected that the asymptotic properties of the estimator can be established. However,  additional substantial work is required and hence we leave this as an open problem for future research.

\section{Geodesic fuzzy RDD for outcomes in Riemannian manifolds}\label{subsec:GFRD-manifold}
As discussed in Section \ref{subsec:GFRD}, one may prefer to consider the compliers' LATE at the cutoff as a quantity defined on the manifold $\mathcal{M}$, i.e., the compliers' GLATE. 

\begin{figure}[tb]
    \single
    \centering
    \includegraphics[width=0.75\linewidth]{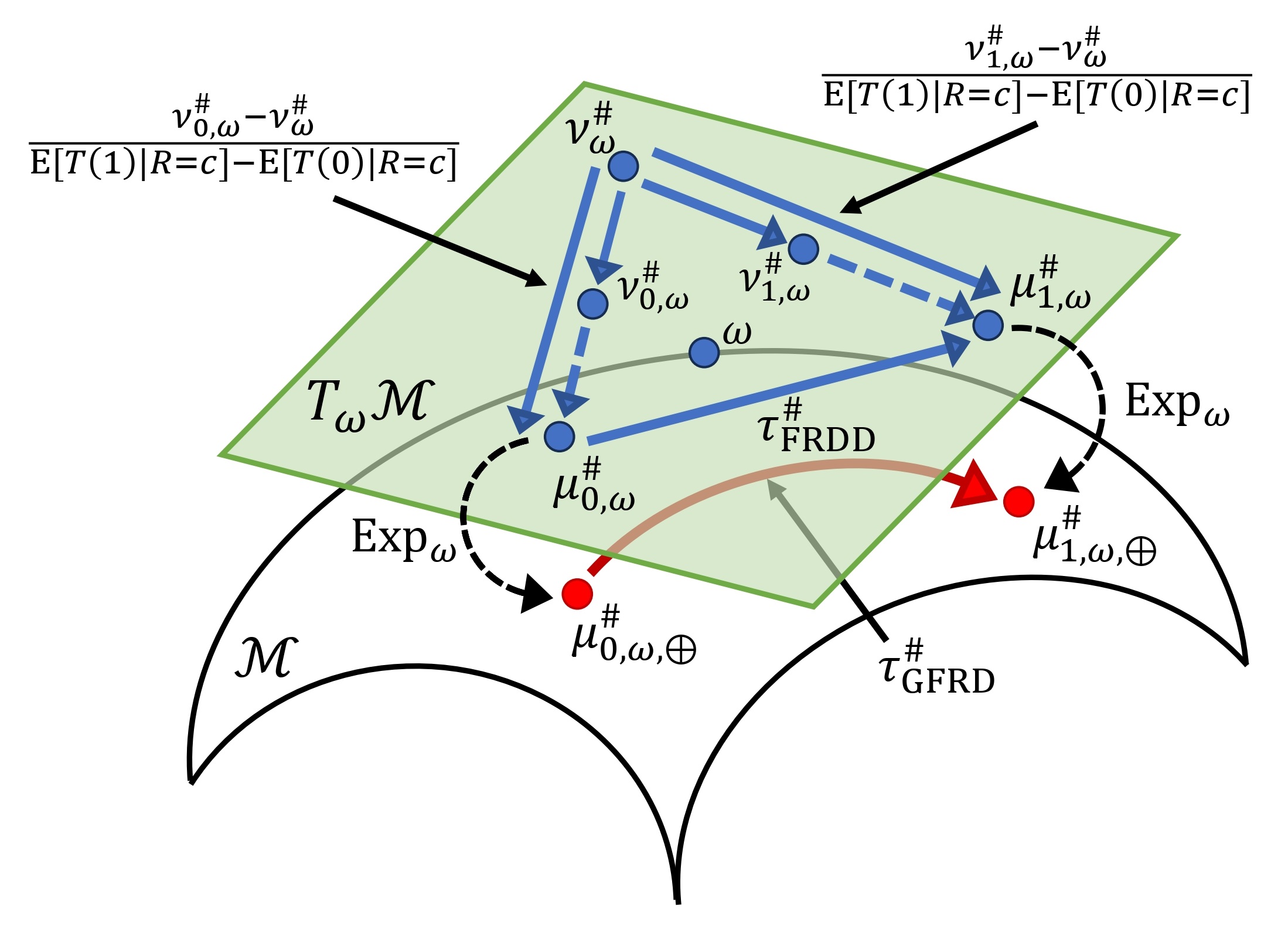}
    \caption{Illustration of the identification of $\tau_{\mathrm{FRDD}}^\sharp$ and $\tau_{\mathrm{GFRD}}^\sharp$. The circles symbolize objects in the geodesic space $\mathcal{M}$, and the arrows emanating from these circles represent the geodesic paths connecting them. The solid blue arrow from $\mu_{0,\omega}$ to $\mu_{1,\omega}$ represents the compliers' LATE. The solid red arrow represents the compliers' GLATE.}
    \label{fig:GFRD-M}
\end{figure}

\begin{definition}[Compliers' GLATE at the cutoff for outcomes in a Riemannian manifold]
For the outcomes in a Riemannian manifold considered in Section \ref{subsec:FRD-manifold}, compliers' GLATE is defined as the geodesic from $\mu_{0,\omega,\oplus}^\sharp$ to $\mu_{1,\omega,\oplus}^\sharp$, that is,
\[\tau_{\mathrm{GFRD}}^\sharp := \gamma_{\mu_{0,\omega, \oplus}^\sharp,\mu_{1,\omega, \oplus}^\sharp},\]
where $\mu_{z,\omega, \oplus}^\sharp = \mathrm{Exp}_\omega(\mu_{z,\omega}^\sharp)$, $\mu_{z,\omega}^\sharp=\E[\mathrm{Log}_{\omega}\left(Y(T(z))\right)|T(1)>T(0),R=c]$, and $\mathrm{Exp}_\omega(\cdot): T_\omega \mathcal{M} \to \mathcal{M}$ is the exponential map at a fixed reference point $\omega \in \mathcal{M}$. 
\end{definition}
Let $\nu_{z,\omega}^\sharp=\E[\mathrm{Log}_\omega(Y(T(z)))|R=c]$ and let $\nu_\omega^\sharp$ be the expectation of $\mathrm{Log}_\omega(Y)$ for always/never-takers at the cutoff, that is, $\nu_\omega^\sharp=\E[\mathrm{Log}_\omega(Y)|T(1)=T(0),R=c]$. For the Riemannian manifold considered in Section \ref{subsec:FRD-manifold}, 
\begin{align}\label{eq:GFRD-id-manifold}
\mu_{z,\omega, \oplus}^\sharp &= \mathrm{Exp}_\omega\left(\nu_\omega^\sharp + \frac{\nu_{z,\omega}^\sharp - \nu_\omega^\sharp}{m_1-m_0}\right)\quad \mbox{for} \quad  z\in \{0,1\}, 
\end{align}
where $m_z = \E[T(z)|R=c]$, assuming this is well-defined, i.e., the argument of Exp$_{\omega}$ is in the image space Log$_{\omega}$; see (\ref{eq:FRD-id-manifold-z}) for the derivation of \eqref{eq:GFRD-id-manifold}. An illustration of the identification of $\tau_{\mathrm{FRDD}}^\sharp$ and $\tau_{\mathrm{GFRD}}^\sharp$ is shown in Figure \ref{fig:GFRD-M}. Note that $\nu_{z,\omega}^\sharp$ and $m_z$ can be estimated by $\hat{\nu}_{z,\omega}^\sharp$ and $\hat{m}_z$, respectively. Therefore, if $\nu_\omega^\sharp$ can be estimated, we can construct an estimator of $\mu_{z,\omega,\oplus}^\sharp$. 

Assume that $\E[\mathrm{Log}_\omega(Y)|T(1)=T(0), R=r]$ is continuous over a neighborhood of $c$ and we refer to Assumption \ref{ass:FRD-id-manifold} with this condition as Assumption \ref{ass:FRD-id-manifold}'. Under one-sided noncompliance, we can identify $\mu_\oplus$ as follows. 

\begin{proposition}\label{prp:GFRD-id-manifold}
Consider the setup of Sections 7.1 and \ref{subsec:FRD-manifold}. Under Assumption \ref{ass:FRD-id-manifold}' 
\begin{align*}
\nu_\omega^\sharp &= 
\begin{cases}
\lim_{r \uparrow c}\E[\mathrm{Log}_\omega(Y)|T=1,Z=0, R=r] & \text{under Assumption \ref{ass:AT-NT}(i)}\\[5pt]
\lim_{r \downarrow c}\E[\mathrm{Log}_\omega(Y)|T=0,Z=1, R=r] & \text{under Assumption \ref{ass:AT-NT}(ii)}
\end{cases}.
\end{align*}
\end{proposition}

Proposition \ref{prp:GFRD-id-manifold} motivates estimators \begin{align*}
\hat{\tau}_{\mathrm{GFRD}}^\sharp &= \gamma_{\hat{\mu}_{0,\omega,\oplus}^\sharp,\hat{\mu}_{1,\omega,\oplus}^\sharp},\quad 
\hat{\mu}_{z,\omega,\oplus}^\sharp = \mathrm{Exp}_\omega\left(\hat{\nu}_\omega^\sharp + \frac{\hat{\nu}_{z,\omega}^\sharp - \hat{\nu}_\omega^\sharp}{\hat{m}_1-\hat{m}_0}\right)\quad \mbox{for} \quad  z \in \{0,1\},
\end{align*}
where $\hat{\nu}_{z,\omega}^\sharp$ and $\hat{m}_z$ are estimators considered in Sections \ref{subsec:FRD-manifold} and 3.2, respectively, and $\hat{\nu}_\omega^\sharp$ is the LFR estimator defined as 
\begin{align*}
\hat{\nu}_\omega^\sharp &= \left\{
\begin{aligned}
 \hat{\nu}_{\omega,\mathrm{A}}^\sharp
 &:= \argmin_{\nu \in T_\omega\mathcal{M}}\frac{1}{n_\mathrm{A}}\sum_{i\in I_\mathrm{A}} \hat{s}_0(c;R_i,h_0)d_E^2(\nu,\mathrm{Log}_\omega(Y_i)) \\
 &\hspace{2em}\text{under Assumption \ref{ass:AT-NT}(i)},\\[0.4em]
 \hat{\nu}_{\omega,\mathrm{N}}^\sharp
 &:= \argmin_{\nu \in T_\omega\mathcal{M}}\frac{1}{n_\mathrm{N}}\sum_{i\in I_\mathrm{N}} \hat{s}_1(c;R_i,h_1)d_E^2(\nu,\mathrm{Log}_\omega(Y_i)) \\
 &\hspace{2em}\text{under Assumption \ref{ass:AT-NT}(ii)}
\end{aligned}
\right. .
\end{align*}

For $z \in \{0,1\}$, define $M^\sharp(\nu,c) = \E[d_E^2(\nu,\mathrm{Log}_\omega(Y(T)))|T(1)=T(0),R=c]$, 
\begin{align*}
&\nu_{\omega, \mathrm{S}}^\sharp=\argmin_{\nu \in T_\omega \mathcal{M}}M_{\mathrm{S}, n}^\sharp(\nu),\ \mathrm{S} \in \{\mathrm{A},\mathrm{N}\},\\
&M_{\mathrm{A}, n}^\sharp(\nu) = \E[s_0(c;R,h_0)d_E^2(\nu,\mathrm{Log}_\omega(Y(T)))|T=1,Z=0],\\
&M_{\mathrm{N}, n}^\sharp(\nu) = \E[s_1(c;R,h_1)d_E^2(\nu,\mathrm{Log}_\omega(Y(T)))|T=0,Z=1].
\end{align*}

Letting $h_{\mathrm{max}}=\max\{h_0,h_1\}$ and $h_{\mathrm{min}} = \min\{h_0,h_1\}$, the convergence rate of $\hat{\tau}_{\mathrm{GFRD}}^\sharp$ is obtained as follows.
\begin{proposition}\label{prp:GLATE-rate-manifold}
Consider the setup of Sections 7.1 and \ref{subsec:FRD-manifold}. Suppose that Assumptions 3.4, 3.5, \ref{ass:AT-NT}, \ref{ass:FRD-id-manifold}', and assumptions in Theorem \ref{thm:LATE-rate-manifold} except for Assumption \ref{ass:FRD-id-manifold} are satisfied  and that there exists a constant $C_\omega>0$ such that for any $\mu_1,\mu_2 \in T_\omega \mathcal{M}$, $d(\mathrm{Exp}_\omega(\mu_1),\mathrm{Exp}_\omega(\mu_2)) \leq C_\omega d_E(\mu_1,\mu_2)$. Additionally, suppose that Assumption 3.3 is satisfied when  replacing $(\mathcal{M},d)$, $\mu_{t,\oplus}$, $\nu_{t,\oplus}$, $\hat{\nu}_{t,\oplus}$, $M_{t,n}(\nu)$, $M_{t,\oplus}(\nu,c)$, $Y(t)$ for $t \in \{0,1\}$ with $(T_\omega\mathcal{M},d_E)$, $\nu_\omega^\sharp$, $\nu_{\omega,\mathrm{S}}^\sharp$, $\hat{\nu}_{\omega,\mathrm{S}}^\sharp$, $M_{\mathrm{S},n}^\sharp(\nu)$, $M^\sharp(\nu,c)$, $\mathrm{Log}_\omega(Y(T))$ for $\mathrm{S} \in \{\mathrm{A},\mathrm{N}\}$. Then 
\begin{align*}
d_\mathcal{G}(\tau_{\mathrm{GFRD}}^\sharp,\hat{\tau}_{\mathrm{GFRD}}^\sharp) = O(h_{\mathrm{max}}^{2/(\beta_1-1)}) + O_p(r_n + (nh_{\mathrm{min}})^{-\frac{1}{2(\beta_2-1)}}).
\end{align*}
\end{proposition}

\section{Proofs for Section 3}
\subsection{Proof of Theorem 3.1}

For the proof of (1), it suffices to show
\begin{align}
\E_\oplus[Y(1)|R=c] &= \lim_{r \downarrow c}\E_\oplus[Y|R=r]\ \label{eq:GSRD-id1},\\
\E_\oplus[Y(0)|R=c] &= \lim_{r \uparrow c}\E_\oplus[Y|R=r]\ \label{eq:GSRD-id2}.
\end{align}
Here we only prove (\ref{eq:GSRD-id1}) since the proof of (\ref{eq:GSRD-id2}) is almost the same. Observe that
\begin{align*}
\E_\oplus[Y(1)|R=c]&= \lim_{r \downarrow c}\E_\oplus[Y(1)|R=r]\\
&= \lim_{r \downarrow c}\E_\oplus[Y(1)|T=1,R=r]\\
&= \lim_{r \downarrow c}\E_\oplus[Y|T=1,R=r]\\
&= \lim_{r \downarrow c}\E_\oplus[Y|R=r],
\end{align*}
where the first equality follows from Assumption~3.1(ii) and the third equality follows from Assumption~3.1(i). Therefore, we obtain the conclusion.

\subsection{Proof of Theorem 3.2}
For each side $t \in \{0,1\}$ of the cutoff, the population target $\nu_{t,\oplus}$
and the estimator $\hat{\nu}_{t,\oplus}$ are defined through one-sided local Fr\'echet regression. Since these are exactly local Fr\'echet regression objects with Euclidean predictor $R \in \mathbb{R}$, the rates established in Theorems 3 and 4 of \citet{mull:19:6} apply directly under our Assumptions 3.2 and 3.3. For completeness, we restate these results in our notation.

\begin{theorem}[Theorem 3 of \citet{mull:19:6}]\label{thm:pete3}
If Assumptions 3.2 and 3.3(i)--(iv) hold, then for each $t \in \{0,1\}$,
\[
d(\mu_{t,\oplus},\nu_{t,\oplus}) = O\bigl(h_t^{2/(\beta_1-1)}\bigr)
\quad \text{as } h_t \to 0.
\]
\end{theorem}

\begin{theorem}[Theorem 4 of \citet{mull:19:6}]\label{thm:pete4}
If Assumptions 3.2 and 3.3(i), (ii), and (v) hold, then for each $t \in \{0,1\}$,
\[
d(\nu_{t,\oplus},\hat{\nu}_{t,\oplus})
= O_p\bigl((n h_t)^{-1/(2(\beta_2-1))}\bigr)
\quad \text{as } h_t \to 0 \text{ and } n h_t \to \infty.
\]
\end{theorem}

\noindent\textbf{Proof of (i).}
By definition of $d_\mathcal{G}$ and Assumption 3.5,
\begin{align}
d_\mathcal{G}(\tau_{\mathrm{GRDD}}, \tau_{\mathrm{GRDD}}^*)
&= d_\mathcal{G}\bigl(\gamma_{\mu_{0,\oplus},\mu_{1,\oplus}},
\gamma_{\nu_{0,\oplus},\nu_{1,\oplus}}\bigr) \nonumber\\
&= d\bigl(
\Gamma_{\mu_{0,\oplus},\mu_{1,\oplus}}(\omega_\oplus),
\Gamma_{\nu_{0,\oplus},\nu_{1,\oplus}}(\omega_\oplus)
\bigr) \nonumber\\
&\le C\bigl\{
d(\mu_{0,\oplus},\nu_{0,\oplus})
+
d(\mu_{1,\oplus},\nu_{1,\oplus})
\bigr\}. \label{ineq:GSRD-rate-bias}
\end{align}
Under Assumptions 3.2 and 3.3(i)--(iv), Theorem~\ref{thm:pete3} yields
\[
d(\mu_{t,\oplus},\nu_{t,\oplus})
= O\bigl(h_t^{2/(\beta_1-1)}\bigr),
\qquad t \in \{0,1\}.
\]
Substituting these bounds into \eqref{ineq:GSRD-rate-bias}, we obtain
\[
d_\mathcal{G}(\tau_{\mathrm{GRDD}}, \tau_{\mathrm{GRDD}}^*)
= O\bigl(h_{\max}^{2/(\beta_1-1)}\bigr).
\]

\medskip
\noindent\textbf{Proof of (ii).}
Again by Assumption 3.5,
\begin{align}
d_\mathcal{G}(\tau_{\mathrm{GRDD}}^*, \hat{\tau}_{\mathrm{GRDD}})
&=
d_\mathcal{G}\bigl(\gamma_{\nu_{0,\oplus},\nu_{1,\oplus}},
\gamma_{\hat{\nu}_{0,\oplus},\hat{\nu}_{1,\oplus}}\bigr) \nonumber\\
&\le
C\bigl\{
d(\nu_{0,\oplus},\hat{\nu}_{0,\oplus})
+
d(\nu_{1,\oplus},\hat{\nu}_{1,\oplus})
\bigr\}. \label{ineq:GSRD-stoch}
\end{align}
Under Assumptions 3.2 and 3.3(i), (ii), and (v), Theorem~\ref{thm:pete4} gives
\[
d(\nu_{t,\oplus},\hat{\nu}_{t,\oplus})
=
O_p\bigl((n h_t)^{-1/(2(\beta_2-1))}\bigr),
\qquad t \in \{0,1\}.
\]
Combining these bounds with \eqref{ineq:GSRD-stoch}, we obtain
\[
d_\mathcal{G}(\tau_{\mathrm{GRDD}}^*, \hat{\tau}_{\mathrm{GRDD}})
=
O_p\bigl((n h_{\min})^{-1/(2(\beta_2-1))}\bigr).
\]
This completes the proof.

\section{Hilbert space embedding}\label{sec:ex-iso-map}
\subsection{Functional data}

Let $L^2(\mathcal{T})$ be the Hilbert space of square integrable functions where $\mathcal{T}$ is a compact interval. In this example, we can take $\Psi=\mathrm{id}$ since $\mathcal{H}:=L^2(\mathcal{T})$ is the Hilbert space with the inner product $\langle f,g \rangle_\mathcal{H}=\int_\mathcal{T} f(t)g(t)dt$ which induces the $L^2$ metric $d_{L^2}(f,g)=(\int_\mathcal{T}\{f(t)-g(t)\}^2dt)^{1/2}$. 

\subsection{Compositional data}
For compositional data, we work with the square-root transformation so that the observations take values in the positive orthant of the unit sphere $\mathcal S^{d-1}_+$. To simplify notation, we write $Y_i$ and $Y_i(t)$ for the square-root transformed compositions. Let $\mathcal{S}^{d-1} = \{\mathbf{z} \in \mathbb{R}^d: \|\mathbf{z}\|=1\}$ be the unit sphere and let $T_{\mathbf{z}_0} \mathcal{S}^{d-1} = \{\mathbf{z} \in \mathbb{R}^d: \mathbf{z}_0^\top\mathbf{z} = 0\}$ denote the tangent space of $\mathcal{S}^{d-1}$ at the fixed point $\mathbf{z}_0 \in \mathcal{S}^{d-1}$. Using the logarithmic map $\mathrm{Log}_{\mathbf{z}_0}: \mathcal{S}^{d-1}\backslash\{-\mathbf{z}_0\} \to T_{\mathbf{z}_0} \mathcal{S}^{d-1}$ defined as
\[
\mathrm{Log}_{\mathbf{z}_0}(\mathbf{z}) = \frac{\theta}{\sin (\theta)}(\mathbf{z}-\cos(\theta)\mathbf{z}_0),\quad \theta=d_g(\mathbf{z}_0,\mathbf{z}),
\]
compositional data can be embedded into $\mathrm{Log}_{\mathbf{z}_0}(\mathcal{S}_+^{d-1})$, a subset of the linear space $T_{\mathbf{z}_0} \mathcal{S}^{d-1}$.

Under the arc-length metric $d_g$, the null hypothesis of no treatment effect is
\[
\mathbb H_0: d_g(\mu_{0,\oplus},\mu_{1,\oplus})=0.
\]
Since $d_g(\mu_{0,\oplus},\mu_{1,\oplus})=0$ is equivalent to $\mu_{1,\oplus}=\mu_{0,\oplus}$, and the Fr\'echet mean on the positive orthant of the unit sphere is characterized by the first-order optimality condition $\E\bigl[\mathrm{Log}_{\mu_{1,\oplus}}(Y(1))|R=c\bigr]=0$ \citep{lee:18}, it follows that $\mu_{1,\oplus}=\mu_{0,\oplus}$ if and only if $\E\bigl[\mathrm{Log}_{\mu_{0,\oplus}}(Y(1))|R=c\bigr]=0$. Therefore, for compositional outcomes, testing for no treatment effect is equivalent to testing
\[
\mathbb H_0': \E\bigl[\mathrm{Log}_{\mu_{0,\oplus}}(Y(1))|R=c\bigr]=0.
\]

In practice, the base point $\mu_{0,\oplus}$ is unknown and is replaced by its local Fr\'echet regression estimator $\hat{\nu}_{0,\oplus}$. Consider the estimator
\[
\hat{\nu}_t^\Psi=
\frac{1}{n_t}\sum_{i=1}^n \hat{s}_t(c;R_i,h_t)\,\mathrm{Log}_{\hat{\nu}_{0,\oplus}}(Y_i),
\qquad t=0,1.
\]

To analyze $\hat{\nu}_1^\Psi$, write
\begin{align*}
\hat{\nu}_1^\Psi
&=
\frac{1}{n_1}\sum_{i=1}^n \hat{s}_1(c;R_i,h_1)\,\mathrm{Log}_{\hat{\nu}_{0,\oplus}}(Y_i(1))\\
&=
\frac{1}{n_1}\sum_{i=1}^n \hat{s}_1(c;R_i,h_1)\,\mathrm{Log}_{\mu_{0,\oplus}}(Y_i(1))
\\&\hspace{1em}+
\frac{1}{n_1}\sum_{i=1}^n \hat{s}_1(c;R_i,h_1)
\bigl\{\mathrm{Log}_{\hat{\nu}_{0,\oplus}}(Y_i(1))-\mathrm{Log}_{\mu_{0,\oplus}}(Y_i(1))\bigr\}\\
&=: D_1 + D_2.
\end{align*}

The first term $D_1$ is the ideal statistic constructed with the true base point. Under the null hypothesis $\mathbb H_0'$ and $n_1h_1^5 \to 0$, $\sqrt{n_1h_1}D_1$ is asymptotically normal.

For the second term, we use the Lipschitz continuity of the logarithmic map on the positive orthant of the unit sphere to obtain
\begin{align*}
\|D_2\|
&\le
\frac{1}{n_1}\sum_{i=1}^n |\hat{s}_1(c;R_i,h_1)|
\,\bigl\|\mathrm{Log}_{\hat{\nu}_{0,\oplus}}(Y_i(1))-\mathrm{Log}_{\mu_{0,\oplus}}(Y_i(1))\bigr\|\\
&\lesssim
d_g(\hat{\nu}_{0,\oplus},\mu_{0,\oplus})\frac{1}{n_1}\sum_{i=1}^n |\hat{s}_1(c;R_i,h_1)|.
\end{align*}
Since the local Fr\'echet regression estimator satisfies
\[
d_g(\hat{\nu}_{0,\oplus},\mu_{0,\oplus}) = O_p\!\left(h_0^2 + (n_0h_0)^{-1/2}\right),
\]
see \citet{mull:19:6}, and
\[
\frac{1}{n_1}\sum_{i=1}^n |\hat{s}_1(c;R_i,h_1)| = O_p(1),
\]
it follows that
\[
\|D_2\| = O_p\!\left(h_0^2 + (n_0h_0)^{-1/2}\right).
\]

Now assume that $n_1/n_0\to c \in (0,\infty), h_0\asymp n^{-1/5}$ and $h_1 \ll n^{-1/5}$. Then
\begin{align*}
\sqrt{n_1h_1}\,\|D_2\|
&= O_p\!\left(\sqrt{n_1h_1}\,h_0^2\right) + O_p\!\left(\sqrt{\frac{n_1h_1}{n_0h_0}}\right)\\
&= O_p\!\left(\sqrt{n_1h_1}\,n^{-2/5}\right) + O_p\!\left(\sqrt{\frac{h_1}{h_0}}\right).
\end{align*}
Both terms are $o_p(1)$, and hence
\[
\sqrt{n_1h_1}\,\|D_2\| = o_p(1).
\]
Consequently,
\[
\sqrt{n_1h_1}\,\hat{\nu}_1^\Psi = \frac{\sqrt{n_1h_1}}{n_1}\sum_{i=1}^n \hat{s}_1(c;R_i,h_1)\,\mathrm{Log}_{\mu_{0,\oplus}}(Y_i(1)) + o_p(1),
\]

Therefore, the effect of estimating the base point is asymptotically negligible, so that the test statistic behaves as if the true embedding $\mathrm{Log}_{\mu_{0,\oplus}}$ were used. It follows that the bootstrap-based testing procedure described in Section~4.3, when implemented with
\[
\Psi = \mathrm{Log}_{\hat{\nu}_{0,\oplus}},
\]
is asymptotically equivalent to the procedure based on the oracle embedding $\mathrm{Log}_{\mu_{0,\oplus}}$. Consequently, it yields a valid test of
\[
\mathbb H_0': \E\bigl[\mathrm{Log}_{\mu_{0,\oplus}}(Y(1))|R=c\bigr]=0.
\]
As established above, this condition is equivalent to
\[
\mathbb H_0: d_g(\mu_{0,\oplus},\mu_{1,\oplus})=0,
\]
which corresponds to the null hypothesis of no treatment effect for compositional outcomes under the arc-length metric.

\subsection{SPD matrices}
First, we consider the space of SPD matrices $\mathrm{Sym}_m^+$ with the Frobenius metric $d_F$ defined as $d_F(A,B)=\|A-B\|_F$ where $\|A\|_F$ is the Frobenius norm of a real matrix $A$. In this example, we can take $\Psi=\mathrm{id}$ where $\mathrm{id}$ is the identity map since $A \in \mathrm{Sym}_m^+$ is also an element of the Hilbert space $\mathcal{H}$ with the inner product $\langle A,B \rangle_\mathcal{H}=\mathrm{tr}(A^\top B)$ where $\mathrm{tr}(A)$ is the trace of a real matrix $A$ and this inner product induces the Frobenius metric. Let $\lambda_{\min}(A)$ be the minimum eigenvalue of $A \in \mathrm{Sym}_m^+$. For $\varepsilon>0$, define the space $\mathrm{Sym}_{m,\varepsilon}^+=\{A \in \mathrm{Sym}_m^+: \lambda_{\min}(A) \geq \varepsilon \}$. We can see that $\Psi(\mathrm{Sym}_{m,\varepsilon}^+)=\mathrm{Sym}_{m,\varepsilon}^+$ is a closed convex subset of $\mathrm{Sym}_m^+$.

Second, we consider the space of SPD matrices with the power metric family defined as 
\[
d_{F,p}(A,B) = d_F(F_p(A),F_p(B)),
\]
where $p>0$ is a constant and $F_p$ is a matrix power map defined as 
\[
F_p(A) = A^p = U\Lambda^p U^\top: \mathrm{Sym}^+_m \to \mathrm{Sym}^+_m,
\]
where $U\Lambda U^\top$ is the usual spectral decomposition of $A \in \mathrm{Sym}^+_m$. Applying almost the same arguments for the space of SPD matrices with the Frobenius metric, one can verify Assumption 4.1 with $\Psi(A)=F_p(A)$, $\mathcal{M}=\mathrm{Sym}_{m,\varepsilon}^+$. 

Third, we consider the space of SPD matrices with the Log-Euclidean metric defined as 
\[
d_{\mathrm{LE}}(A,B) = d_F(\log (A),  \log (B)),
\]
where $\log(A)$ is the matrix logarithmic map for a real matrix $A$. Again applying similar arguments as for the space of SPD matrices with the Frobenius metric, one can verify Assumption 4.1 with $\Psi(A)=\log(A)$, $\mathcal{M}=\mathrm{Sym}_m^+$. 

Fourth, we consider the space of SPD matrices with the Log-Cholesky metric. Let $\mathcal{L}^+_m$ be the set of lower triangular matrices with positive diagonal entries. For any $S \in \mathrm{Sym}^+_m$, there exists a unique $L \in \mathcal{L}^+_m$ such that $S = LL^\top$ and a diffeomorphism $\mathscr{L}$ between $\mathrm{Sym}^+_m$ and $\mathcal{L}^+_m$. For any $L \in \mathcal{L}^+_m$, let $\lfloor L \rfloor$ be the strictly lower triangular part and $\mathbb{D}(L)$ the diagonal part of $L$. The Log-Cholesky metric is defined as
\begin{align*}
d_{\mathrm{LC}}^2(L_1,L_2) &= d_F^2(\lfloor L_1 \rfloor, \lfloor L_2 \rfloor) + d_F^2(\log (\mathbb{D}(L_1)),\log (\mathbb{D}(L_2)))\\
&= d_F^2(\lfloor L_1 \rfloor + \log (\mathbb{D}(L_1)), \lfloor L_2 \rfloor + \log (\mathbb{D}(L_2))). 
\end{align*}
Note that the space $(\mathcal{L}^+_m, d_{\mathrm{LC}})$ is a geodesic metric space \citep{lin:19:1},  and consider it as the space of SPD matrices under the Log-Cholesky metric. One can then   verify Assumption 4.1 with $\Psi(L)= \lfloor L \rfloor + \log (\mathbb{D}(L))$, $\mathcal{M}=\mathcal{L}_m^+$.

\subsection{Networks}
Let $\mathcal{L}$ be the space of networks represented by graph Laplacians with the Frobenius metric $d_F$. Applying almost the same arguments for the space of SPD matrices with the Frobenius metric, one can verify Assumption 4.1 with $\Psi = \mathrm{id}$. Likewise, one can verify Assumption 4.1 with $\Psi(A)=A^p$ for the space of networks represented by graph Laplacians with the power metric.

\subsection{Univariate probability distributions}
Let $\mathcal{W}$ be the space of univariate probability distributions on a closed interval $\mathcal{I}\subset\mathbb{R}$ with finite second moments equipped with the Wasserstein metric $d_\mathcal{W}$ defined as $d_\mathcal{W}(\mu,\nu)=d_{L^2}(F_\mu^{-1}(\cdot),F_\nu^{-1}(\cdot))$ where $F_\mu^{-1}(\cdot)$ is the quantile function of $\mu \in \mathcal{W}$. It is easy to verify Assumption 4.1 with $\Psi(\mu)=F_\mu^{-1}(\cdot)$.

\section{Proofs for Section 4}
\subsection{Proof of Theorem 4.1}
Recall that $\{e_j\}_{j \geq 1}$ is an orthonormal basis of $\mathcal{H}$. We will show 
\begin{align*}\sqrt{nh}(\hat{\nu}_0^\Psi-\mu_{0,\oplus}^\Psi, \hat{\nu}_1^\Psi-\mu_{1,\oplus}^\Psi) \leadsto (\mathbb{G}_0, \mathbb{G}_1).
\end{align*}
From  Theorem 1.8.4 in \citet{well:96}, it suffices to verify the following conditions. 
\begin{itemize}
\item[(i)] Finite-dimensional weak convergence: For all $g_0,g_1 \in \mathcal{H} \backslash \{0\}$, 
\begin{align*}
\sqrt{nh}(\langle \hat{\nu}_0^\Psi-\mu_{0,\oplus}^\Psi,g_0\rangle_\mathcal{H}, \langle \hat{\nu}_1^\Psi-\mu_{1,\oplus}^\Psi,g_1\rangle_\mathcal{H}) \leadsto (\langle \mathbb{G}_0,g_0\rangle_\mathcal{H},\langle \mathbb{G}_1,g_1\rangle_\mathcal{H}).
\end{align*}
\item[(ii)] $\sqrt{nh}(\hat{\nu}_0^\Psi-\mu_{0,\oplus}^\Psi)$ and $\sqrt{nh} (\hat{\nu}_1^\Psi-\mu_{1,\oplus}^\Psi)$ are asymptotically finite dimensional. 
\end{itemize}
Note that Condition (i) characterizes the asymptotic distribution of $\sqrt{nh}(\hat{\nu}_0^\Psi-\mu_{0,\oplus}^\Psi, \hat{\nu}_1^\Psi-\mu_{1,\oplus}^\Psi)$, and Conditions (i) and (ii) imply asymptotic tightness of $\sqrt{nh}(\hat{\nu}_0^\Psi-\mu_{0,\oplus}^\Psi, \hat{\nu}_1^\Psi-\mu_{1,\oplus}^\Psi)$ \citep[Lemmas 1.8.1 and 1.8.3,][]{well:96}.

\noindent
{\bf Verification of Condition (i)}: 
Observe that 
\begin{align}\label{eq:AB-decomp}
&\sqrt{nh}\langle \hat{\nu}_t^\Psi-\mu_{t,\oplus}^\Psi,g \rangle_\mathcal{H} \nonumber \\
&= \frac{\sqrt{nh}}{n_t}\sum_{i=1}^n\hat{s}_t(c;R_i,h)\langle \Psi(Y_i),g\rangle_\mathcal{H} - \sqrt{nh}\langle m_t^\Psi(c),g \rangle_\mathcal{H} \nonumber\\
&= \frac{\sqrt{nh}}{n_t}\sum_{i=1}^n\hat{s}_t(c;R_i,h)\langle \varepsilon_{t,i},g \rangle_\mathcal{H} + \frac{\sqrt{nh}}{n_t}\sum_{i=1}^n\hat{s}_t(c;R_i,h)\langle m_t^\Psi(R_i)-m_t^\Psi(c),g\rangle_\mathcal{H} \nonumber\\
&\quad + \sqrt{nh}\langle m_t^\Psi(c),g \rangle_\mathcal{H}\underbrace{\left(\frac{1}{n_t}\sum_{i=1}^n \hat{s}_t(c;R_i,h)-1\right)}_{=0} \nonumber\\
&= \frac{\sqrt{nh}}{n_t}\sum_{i=1}^n\hat{s}_t(c;R_i,h)\langle \varepsilon_{t,i},g \rangle_\mathcal{H}  + \frac{\sqrt{nh}}{n_t}\sum_{i=1}^n\hat{s}_t(c;R_i,h)\langle m_t^\Psi(R_i)-m_t^\Psi(c),g\rangle_\mathcal{H} \nonumber\\
&=: A_{t,n}(g) + B_{t,n}(g). 
\end{align}

For $B_{t,n}(g)$, we have
\begin{align*}
B_{t,n}(g) 
&= \frac{\sqrt{nh}}{n_t}\sum_{i=1}^n\hat{s}_t(c;R_i,h)\left\{\partial m_t^\Psi(c;g)(R_i-c) + \frac{1}{2}\partial^2m_t^\Psi(\tilde{c}_i;g)(R_i - c)^2\right\}\\
&= \partial m_t^\Psi(c;g)\underbrace{\frac{\sqrt{nh}}{n_t}\sum_{i=1}^n\hat{s}_t(c;R_i,h)(R_i-c)}_{=0} +\frac{\sqrt{nh}}{2n_t}\sum_{i=1}^n\hat{s}_t(c;R_i,h)\partial^2m_t^\Psi(\tilde{c}_i;g)(R_i - c)^2\\
&= \frac{\sqrt{nh}}{2n_t}\sum_{i=1}^n\hat{s}_t(c;R_i,h)\partial^2m_t^\Psi(\tilde{c}_i;g)(R_i - c)^2,
\end{align*}
where $\tilde{c}_i$ lies between $c$ and $R_i$. Then we have
\begin{align}\label{ineq:Bg}
|B_{t,n}(g)| 
&\leq \sup_{r \in \mathcal{I}}|\partial^2 m_t^\Psi(r;g)|\frac{\sqrt{nh}}{n_t}\sum_{i=1}^n|\hat{s}_t(c;R_i,h)||R_i - c|^2 \nonumber \\
&\leq  \|g\|_\mathcal{H}\left(\sup_{r \in \mathcal{I}}\sum_{j \geq 1}|\partial^2 m_t^\Psi(r;e_j)|^2\right)^{1/2}\frac{\sqrt{nh}}{n_t}\sum_{i=1}^n|\hat{s}_t(c;R_i,h)||R_i - c|^2 \nonumber \\
&\lesssim  O(h^2\sqrt{nh})\underbrace{\frac{1}{n_t}\sum_{i=1}^n |\hat{s}_t(c;R_i,h)|}_{=O_p(1)}= O_p(\sqrt{nh^5})=o_p(1). 
\end{align}

For $A_{t,n}(g)$, observe that
\begin{align}\label{eq:A12-decomp}
A_{t,n}(g) 
&= \frac{\sqrt{nh}}{n_t}\sum_{i=1}^n\hat{s}_t(c;R_i,h)\langle \varepsilon_{t,i},g \rangle_\mathcal{H} \nonumber \\
&=  \frac{\sqrt{nh}}{n_t}\sum_{i=1}^ns_t(c;R_i,h)\langle \varepsilon_{t,i},g \rangle_\mathcal{H} + \frac{\sqrt{nh}}{n_t}\sum_{i=1}^n\{\hat{s}_t(c;R_i,h)-s_t(c;R_i,h)\}\langle \varepsilon_{t,i},g \rangle_\mathcal{H} \nonumber \\
&=: A_{t,n,1}(g) + A_{t,n,2}(g). 
\end{align}

For $A_{t,n,2}(g)$, we have
\begin{align}\label{ineq:A2g}
\E[|A_{t,n,2}(g)|^2|R_1,\dots,R_n] 
&\lesssim \|g\|_\mathcal{H}^2\frac{nh}{n_t^2}\sum_{i=1}^n\{\hat{s}_t(c;R_i,h)-s_t(c;R_i,h)\}^2 \nonumber \\
&\quad \times \E[\|\varepsilon_{t,i}\|_\mathcal{H}^2|R_i] \nonumber \\
&\lesssim \sup_{r \in \mathcal{I}}\E[\|\varepsilon_{t,i}\|_\mathcal{H}^2|R_i=r] \nonumber \\
&\quad \times \underbrace{\frac{nh}{n_t^2}\sum_{i=1}^n\{\hat{s}_t(c;R_i,h)-s_t(c;R_i,h)\}^2}_{=o_p(1)} = o_p(1).
\end{align}
This yields 
\begin{align}\label{eq:A2-asy-neg}
A_{t,n,2}(g) = o_p(1). 
\end{align}

Combining (\ref{eq:AB-decomp})--(\ref{eq:A12-decomp}), (\ref{eq:A2-asy-neg}), one can see
\begin{align*}
\sqrt{nh}\langle \hat{\nu}_t^\Psi-\mu_{t,\oplus}^\Psi,g \rangle_\mathcal{H} &= \frac{\sqrt{nh}}{n_t}\sum_{i=1}^ns_t(c;R_i,h)\langle \varepsilon_{t,i},g \rangle_\mathcal{H}  + o_p(1)\\ 
&= \frac{1}{\sqrt{nh}}\sum_{i=1}^n \left\{\frac{\kappa_{t,2}-\kappa_{t,1}(R_i-c)/h}{(\kappa_{t,2}\kappa_{t,0}-\kappa_{t,1}^2)f_R(c)}\right\} \\
&\quad \times K_t\left(\frac{R_i - c}{h}\right)\langle \varepsilon_{t,i},g\rangle_\mathcal{H} + o_p(1).
\end{align*}
Applying the standard asymptotic theory for local linear estimators with real-valued outcomes, as developed in Theorems~3.1 and~6.1 of \citet{fan:96}, to the preceding linear representation verifies Condition~(i).

\noindent
{\bf Verification of Condition (ii)}: Applying the same argument to obtain (\ref{eq:A2-asy-neg}), we have 
\begin{align*}
\sqrt{nh}( \hat{\nu}_t^\Psi-\mu_{t,\oplus}^\Psi)
&= \frac{\sqrt{nh}}{n_t}\sum_{i=1}^n\hat{s}_t(c;R_i,h) \varepsilon_{t,i}  + \frac{\sqrt{nh}}{n_t}\sum_{i=1}^n\hat{s}_t(c;R_i,h)\{m_t^\Psi(R_i)-m_t^\Psi(c)\} \nonumber\\
&=: A_{t,n} + B_{t,n}. 
\end{align*}
By an argument analogous to obtaining (\ref{ineq:Bg}) and (\ref{ineq:A2g}), we have $\|B_{t,n}\|_\mathcal{H} = O_p(\sqrt{nh^5}) = o_p(1)$ and $\|A_{t,n} - \tilde{A}_{t,n}\|_\mathcal{H} = o_p(1)$ where
\[
\tilde{A}_{t,n}= \frac{1}{\sqrt{nh}}\sum_{i=1}^n \left\{\frac{\kappa_{t,2}-\kappa_{t,1}(R_i-c)/h}{(\kappa_{t,2}\kappa_{t,0}-\kappa_{t,1}^2)f_R(c)}\right\}K_t\left(\frac{R_i - c}{h}\right)\varepsilon_{t,i}=:\frac{1}{\sqrt{nh}}\sum_{i=1}^n W_{n,i}\varepsilon_{t,i}.
\]
Then it suffices to verify that $\tilde{A}_{t,n}$ is asymptotically finite dimensional, that is, 
\begin{align}\label{lim:AFD}
\lim_{J \to \infty}\limsup_n\Pr\left(\sum_{j > J}\langle \tilde{A}_{t,n}, e_j\rangle_\mathcal{H}^2>\delta\right)=0
\end{align}
for any $\delta>0$. Observe that  
\begin{align*}
\E\left[\sum_{j > J}\langle \tilde{A}_{t,n}, e_j\rangle_\mathcal{H}^2\right]
&= \sum_{j > J}\E\left[\langle \tilde{A}_{t,n}, e_j\rangle_\mathcal{H}^2\right] 
= \sum_{j > J}\frac{1}{nh}\sum_{i=1}^n\E[W_{n,i}^2 \langle \varepsilon_{t,i}, e_j \rangle_\mathcal{H}^2]\\
&=\sum_{j > J}\frac{1}{h}\E[W_{n,i}^2 \E[\langle \varepsilon_{t,i}, e_j \rangle_\mathcal{H}^2|R_i]] \leq \frac{1}{h}\E[W_{n,i}^2]\sum_{j > J}\sup_{r \in \mathcal{I}}\E[\langle \varepsilon_{t,i}, e_j \rangle_\mathcal{H}^2|R_i=r]\\ &\lesssim \sum_{j >J}\sup_{r \in \mathcal{I}}\E[\langle \varepsilon_{t,i}, e_j \rangle_\mathcal{H}^2|R_i=r]. 
\end{align*}
Since $\lim_{J \to \infty}\sum_{j >J}\sup_{r \in \mathcal{I}}\E[\langle \varepsilon_{t,i}, e_j \rangle_\mathcal{H}^2|R_i=r]=0$, we have 
\[
\lim_{J \to \infty}\limsup_n \E\left[\sum_{j > J}\langle \tilde{A}_{t,n}, e_j\rangle_\mathcal{H}^2\right]=0.
\]
Therefore, using Markov's inequality, we obtain (\ref{lim:AFD}).

\subsection{Proof of Proposition 4.1}
Following the characterization of weak convergence in Theorem 1.8.4 of \citet{well:96}, it suffices to verify its conditional analogues for the conditional weak convergence of the multiplier bootstrap process $(\hat{\mathbb{G}}_0,\hat{\mathbb{G}}_1)$.

From the argument in the proof of Theorem 4.1, one can see
\begin{align*}
\sqrt{nh}(\hat{\nu}_t^\Psi-\mu_{t,\oplus}^\Psi) 
&= \frac{1}{\sqrt{nh}}\sum_{i=1}^n \left\{\frac{\kappa_{t,2}-\kappa_{t,1}(R_i-c)/h}{(\kappa_{t,2}\kappa_{t,0}-\kappa_{t,1}^2)f_R(c)}\right\}K_t\left(\frac{R_i - c}{h}\right)\varepsilon_{t,i} + o_p(1).
\end{align*}
Further, applying a similar argument, one can also see 
\begin{align}
\hat{\mathbb{G}}_t 
&= \frac{\sqrt{nh}}{n_t}\sum_{i=1}^n \xi_i\hat{s}_t(c;R_i,h)\varepsilon_{t,i} \nonumber \\ 
&\quad + \frac{\sqrt{nh}}{n_t}\sum_{i=1}^n\xi_i\hat{s}_t(c;R_i,h)\{m_t^\Psi(R_i)-\mu_{t,\oplus}^\Psi\} - \left(\frac{1}{n_t}\sum_{i=1}^n \xi_i\hat{s}_t(c;R_i,h)\right)\sqrt{nh}(\hat{\nu}_t^\Psi-\mu_{t,\oplus}^\Psi)\nonumber\\
&= \frac{\sqrt{nh}}{n_t}\sum_{i=1}^n \xi_i\hat{s}_t(c;R_i,h)\varepsilon_{t,i} + o_{\mathrm{P}_\xi}(1)\nonumber\\
&=\frac{1}{\sqrt{nh}}\sum_{i=1}^n \xi_i\left\{\frac{\kappa_{t,2}-\kappa_{t,1}(R_i-c)/h}{(\kappa_{t,2}\kappa_{t,0}-\kappa_{t,1}^2)f_R(c)}\right\}K_t\left(\frac{R_i - c}{h}\right)\varepsilon_{t,i} + o_{\mathrm{P}_\xi}(1).\label{eq:LFR-linear}
\end{align}
Combining (\ref{eq:LFR-linear}) and the conditional multiplier central limit theorem for triangular arrays \citep[Theorem 2.1,][]{pauly:11}, we have the conditional finite-dimensional weak convergence: For all $g_0,g_1 \in \mathcal{H} \backslash \{0\}$, $(\langle \hat{\mathbb{G}}_0,g_0\rangle_\mathcal{H}, \langle\hat{\mathbb{G}}_1,g_1 \rangle_\mathcal{H}) \stackrel{\mathrm{P}_\xi}{\leadsto} (\langle \mathbb{G}_0,g_0\rangle_\mathcal{H}, \langle \mathbb{G}_1,g_1 \rangle_\mathcal{H})$. To obtain the conclusion, it remains to show the conditional asymptotic tightness of $(\hat{\mathbb{G}}_0,\hat{\mathbb{G}}_1)$, that is, to prove that $\hat{\mathbb{G}}_0$ and $\hat{\mathbb{G}}_1$ are conditionally asymptotically finite-dimensional: For any $\delta,\eta>0$,
\begin{align}\label{lim:condi-AFD}
\lim_{J \to \infty}\limsup_{n}\Pr\left({\Pr}_\xi\left(\sum_{j>J}\langle \hat{\mathbb{G}}_t, e_j \rangle_\mathcal{H}^2>\delta\right)>\eta\right)=0, 
\end{align}
where ${\Pr}_\xi$ denotes the probability with respect to the bootstrap multipliers $\{\xi_i\}_{i=1}^n$. Let $\E_\xi$ denote the expectation with respect to $\{\xi_i\}_{i=1}^n$. Observe that 
\begin{align*}
\E_\xi\left[\sum_{j>J}\langle \hat{\mathbb{G}}_t,e_j \rangle_\mathcal{H}^2\right]
&= \frac{nh}{n_t^2}\sum_{i=1}^n \hat{s}_t^2(c;R_i,h)\sum_{j>J}\langle \Psi(Y_i) - \hat{\nu}_t^\Psi,e_j\rangle_\mathcal{H}^2\\
&\leq 2\frac{nh}{n_t^2}\sum_{i=1}^n \hat{s}_t^2(c;R_i,h)\sum_{j>J}\langle \Psi(Y_i) - m_t^\Psi(c),e_j\rangle_\mathcal{H}^2\\
&\quad + 2\frac{nh}{n_t^2}\sum_{i=1}^n \hat{s}_t^2(c;R_i,h)\sum_{j>J}\langle m_t^\Psi(c) - \hat{\nu}_t^\Psi,e_j\rangle_\mathcal{H}^2 =: 2A_{n,J}^* + 2B_{n,J}^*.
\end{align*}
For $B_{n,J}^*$, we have
\begin{align}\label{ineq:BJ*}
B_{n,J}^* \leq \| m_t^\Psi(c) - \hat{\nu}_t^\Psi\|_\mathcal{H}^2\frac{nh}{n_t^2}\sum_{i=1}^n \hat{s}_t^2(c;R_i,h) = o_p(1)\times O_p(1)=o_p(1). 
\end{align}
For $A_{n,J}^*$, we have
\begin{align}\label{ineq:AJ*}
A_{n,J}^* 
&= \frac{nh}{n_t^2}\sum_{i=1}^n s_t^2(c;R_i,h)\sum_{j>J}\langle \Psi(Y_i) - m_t^\Psi(c),e_j\rangle_\mathcal{H}^2 + o_p(1) \nonumber \\
&\lesssim \sum_{j>J}\sup_{r \in \mathcal{I}}\E\left[\left.\langle \Psi(Y_i) - m_t^\Psi(c),e_j\rangle_\mathcal{H}^2\right|R_i = r\right] + o_p(1) \nonumber \\
&= \sum_{j>J}\sup_{r \in \mathcal{I}}\E[\langle \varepsilon_{t,1}, e_j\rangle_\mathcal{H}^2|R_1=r] + o_p(1).
\end{align}
Since $\lim_{J \to \infty}\sum_{j >J}\sup_{r \in \mathcal{I}}\E[\langle \varepsilon_{t,i}, e_j \rangle_\mathcal{H}^2|R_i=r] = 0$, from Markov's inequality together with (\ref{ineq:BJ*}) and (\ref{ineq:AJ*}), we obtain (\ref{lim:condi-AFD}).

\subsection{Proof of Corollary 4.1}
By applying the (conditional) continuous mapping theorem \citep[Theorems 7.7 and  10.8,][]{koso:08}, we have
$nh\|\hat{\nu}_1^\Psi - \hat{\nu}_0^\Psi\|_\mathcal{H}^2 
\leadsto \|\mathbb{G}_1-\mathbb{G}_0\|_\mathcal{H}^2$ and $\hat{T}_n \stackrel{\mathrm{P}_\xi}{\leadsto} \|\mathbb{G}_1-\mathbb{G}_0\|_\mathcal{H}^2$. Then, the conclusion follows by applying almost the same argument in the proof of Theorem 3.1 in \citet{pauly:11}. 

\subsection{Proof of Corollary 4.2}
By applying the (conditional) continuous mapping theorem, we have
\begin{align*}
\sqrt{nh}\{\|\hat{\nu}_1^\Psi - \hat{\nu}_0^\Psi\|_\mathcal{H}^2 -d^2(\mu_{1,\oplus},\mu_{0,\oplus})\} 
&\leadsto 2\langle \mu_{1,\oplus}^\Psi - \mu_{0,\oplus}^\Psi,\mathbb{G}_1-\mathbb{G}_0 \rangle_\mathcal{H},\\
2\langle \hat{\nu}_1^\Psi - \hat{\nu}_0^\Psi,\hat{\mathbb{G}}_1-\hat{\mathbb{G}}_0\rangle_\mathcal{H}
&\stackrel{\mathrm{P}_\xi}{\leadsto} 2\langle \mu_{1,\oplus}^\Psi - \mu_{0,\oplus}^\Psi,\mathbb{G}_1-\mathbb{G}_0 \rangle_\mathcal{H}.
\end{align*}
The conclusion follows by applying almost the same argument in the proof of Theorem 3.1 in \citet{pauly:11}.

\section{Proofs for Section 7}
\subsection{Proof of Theorem 7.1}\label{subsec:LATE-id-proof}
{\bf (Step 1)} In this step, we show 
\begin{align}
&\mu_{1,\oplus}^\dag = \E_\oplus[Y(T(1))|R=c] = \lim_{r \downarrow c}\E_\oplus[Y|R=r], \label{eq:FRD-id1}\\
&\mu_{0,\oplus}^\dag = \E_\oplus[Y(T(0))|R=c] = \lim_{r \uparrow c}\E_\oplus[Y|R=r], \label{eq:FRD-id2}\\
&\E[T(1)|R=c]-\E[T(0)|R=c] = \lim_{r \downarrow c}\E[T|R=r] - \lim_{r \uparrow c}\E[T|R=r].\label{eq:FRD-id3}
\end{align}

\noindent
{\bf Proof of (\ref{eq:FRD-id1})}: Observe that for $r \geq c$,
\begin{align*}
\E_\oplus[Y|R=r] 
&= \E_\oplus[Y|Z=1,R=r]\\
&= \E_\oplus[Y(T(1))|Z=1,R=r]\\
&= \E_\oplus[Y(T(1))|R=r].
\end{align*}
Then we have $\lim_{r \downarrow c}\E_\oplus[Y|R=r] 
=  \lim_{r \downarrow c}\E_\oplus[Y(T(1))|R=r] = \E_\oplus[Y(T(1))|R=c]$. Likewise, one can show (\ref{eq:FRD-id2}). 

\noindent
{\bf Proof of (\ref{eq:FRD-id3})} Observe that
\begin{align*}
\lim_{r \downarrow c}\E[T|R=r] - \lim_{r \uparrow c}\E[T|R=r]  
&= \lim_{r \downarrow c}\E[T|Z=1,R=r] - \lim_{r \uparrow c}\E[T|Z=0,R=r] \\
&= \lim_{r \downarrow c}\E[T(1)|Z=1,R=r] - \lim_{r \uparrow c}\E[T(0)|Z=0,R=r] \\
&= \lim_{r \downarrow c}\E[T(1)|R=r] - \lim_{r \uparrow c}\E[T(0)|R=r] \\
&= \E[T(1)|R=c] - \E[T(0)|R=c]. 
\end{align*}

\noindent
{\bf (Step 2)}  Observe that for $z \in \{0,1\}$,
\begin{align*}
\Psi(\E_\oplus[Y(T(z))|R=c])
&= \Psi(\argmin_{\nu \in \mathcal{M}}\E[d^2(\nu,Y(T(z)))|R=c])\\
&= \Psi(\argmin_{\nu \in \mathcal{M}}\E[d_\mathcal{H}^2(\Psi(\nu),\Psi(Y(T(z))))|R=c])\\
&= \argmin_{\omega \in \Psi(\mathcal{M})}\E[d_\mathcal{H}^2(\omega,\Psi(Y(T(z))))|R=c]\\
&=\E[\Psi(Y(T(z)))|R=c],
\end{align*}
where we used the closedness and convexity of the image space $\Psi(\mathcal{M})$ for the third and fourth equalities. Likewise, one can see 
\begin{align*}
\Psi(\E_\oplus[Y(T(z))|T(1)>T(0),R=c]) &= \E[\Psi(Y(T(z)))|T(1)>T(0),R=c],\\
\Psi(\E_\oplus[Y(T(z))|T(1)=T(0),R=c]) &= \E[\Psi(Y(T(z)))|T(1)=T(0),R=c].
\end{align*}
Note that
\begin{align*}
\E[\Psi(Y(T(z)))|R=c]
&= \E[\Psi(Y(T(z)))|T(1)>T(0),R=c] \times \Pr(T(1)>T(0)|R=c) \\
&\quad + \E[\Psi(Y(T))|T(1)=T(0),R=c] \times \Pr(T(1)=T(0)|R=c).
\end{align*}
Combining these identities with the fact that $\Pr(T(1)=T(0)|R=c)=1-\Pr(T(1)>T(0)|R=c)$ (by Assumption~7.1(i)), we have
\begin{align}\label{eq:GFRD-id-z}
\Psi(\mu_{z,\oplus})&=\Psi(\E_\oplus[Y(T(z))|T(1)>T(0),R=c]) \nonumber \\
&= \Psi(\E_\oplus[Y(T)|T(1)=T(0),R=c]) \nonumber \\ 
&\quad + \frac{\Psi(\E_\oplus[Y(T(z))|R=c]) - \Psi(\E_\oplus[Y(T)|T(1)=T(0),R=c])}{\Pr(T(1)>T(0)|R=c)} \nonumber \\
&= \Psi(\E_\oplus[Y(T)|T(1)=T(0),R=c]) \nonumber \\ 
&\quad + \frac{\Psi(\mu_{z,\oplus}^\dag) - \Psi(\E_\oplus[Y(T)|T(1)=T(0),R=c])}{\Pr(T(1)>T(0)|R=c)}.
\end{align}
Therefore, 
\begin{align*}
\tau_{\mathrm{FRDD}} 
&= \Psi(\mu_{1,\oplus}) - \Psi(\mu_{0,\oplus}) = \frac{\Psi(\mu_{1,\oplus}^\dag) - \Psi(\mu_{0,\oplus}^\dag)}{\Pr(T(1)>T(0)|R=c)}= \frac{\Psi(\mu_{1,\oplus}^\dag) - \Psi(\mu_{0,\oplus}^\dag)}{\E[T(1)|R=c]-\E[T(0)|R=c]}.
\end{align*}
Combining this with (\ref{eq:FRD-id1})--(\ref{eq:FRD-id3}) completes the proof.

\subsection{Proof of Theorem 7.2}
Define 
\[\tau_{\mathrm{FRDD}}^* = \frac{\Psi(\nu_{1,\oplus}) - \Psi(\nu_{0,\oplus})}{m_1 - m_0}.\] 
Observe that 
\begin{align}\label{ineq:FRD-dist-bound}
d_\mathcal{H}(\tau_{\mathrm{FRDD}},\hat{\tau}_{\mathrm{FRDD}})
&\leq d_\mathcal{H}(\tau_{\mathrm{FRDD}},\tau_{\mathrm{FRDD}}^*)  + d_\mathcal{H}(\tau_{\mathrm{FRDD}}^*,\hat{\tau}_{\mathrm{FRDD}}) \nonumber \\ 
&\leq \frac{\left\{d_\mathcal{H}(\Psi(\mu_{1,\oplus}^\dag),\Psi(\nu_{1,\oplus})) + d_\mathcal{H}(\Psi(\mu_{0,\oplus}^\dag),\Psi(\nu_{0,\oplus}))\right\}}{|m_1 - m_0|}  \nonumber \\
&\quad + \frac{\left\{d_\mathcal{H}(\Psi(\nu_{1,\oplus}),\Psi(\hat{\nu}_{1,\oplus})) + d_\mathcal{H}(\Psi(\nu_{0,\oplus}),\Psi(\hat{\nu}_{0,\oplus}))\right\} }{|\hat{m}_1 - \hat{m}_0|} \nonumber \\
&\quad + \left|\frac{1}{\hat{m}_1 - \hat{m}_0} - \frac{1}{m_1 - m_0}\right|d_\mathcal{H}(\Psi(\nu_{0,\oplus}),\Psi(\nu_{1,\oplus})) \nonumber \\
&=\frac{\left\{d(\mu_{1,\oplus}^\dag,\nu_{1,\oplus}) + d(\mu_{0,\oplus}^\dag,\nu_{0,\oplus})\right\}}{|m_1 - m_0|} \nonumber \\ 
&\quad + \frac{\left\{d(\nu_{1,\oplus},\hat{\nu}_{1,\oplus}) + d(\nu_{0,\oplus},\hat{\nu}_{0,\oplus})\right\}}{|\hat{m}_1 - \hat{m}_0|} + \left|\frac{1}{\hat{m}_1 - \hat{m}_0} - \frac{1}{m_1 - m_0}\right|d(\nu_{0,\oplus},\nu_{1,\oplus}).
\end{align}
Furthermore,
\begin{align}\label{eq:FRD-rate-LL}
\frac{1}{\hat{m}_1 - \hat{m}_0} - \frac{1}{m_1 - m_0} 
&= \frac{(m_1 -\hat{m}_1) - (m_0-\hat{m}_0)}{(\hat{m}_1 - \hat{m}_0)(m_1 - m_0)} \nonumber \\ 
&= \frac{O_p(r_n)}{(m_1- m_0 + o_p(1))(m_1 - m_0)} = O_p(r_n).
\end{align}
Applying a similar argument as in the proof of  Theorem 3.2,  
\begin{align}\label{eq:FRD-rate-bv}
d(\mu_{z,\oplus}^\dag,\nu_{z,\oplus}) &= O(h_z^{2/(\beta_1-1)}),\ d(\hat{\nu}_{z,\oplus},\nu_{z,\oplus}) = O_p((nh_z)^{-\frac{1}{2(\beta_2-1)}}),\ z \in \{0,1\}. 
\end{align}
Combining (\ref{ineq:FRD-dist-bound}), (\ref{eq:FRD-rate-LL}), and (\ref{eq:FRD-rate-bv}), we obtain
\[
d_\mathcal{H}(\tau_{\mathrm{FRDD}},\hat{\tau}_{\mathrm{FRDD}}) = O(h_{\mathrm{max}}^{2/(\beta_1-1)}) + O_p(r_n + (nh_{\mathrm{min}})^{-\frac{1}{2(\beta_2-1)}}).
\]

\section{Proofs for Section \ref{subsec:GFRD}}
\subsection{Proof of Proposition \ref{prp:GFRD-id}}
Under Assumption \ref{ass:AT-NT}(i), 
\[
\mu_\oplus = \E_\oplus[Y(T)|T(1)=T(0)=1,R=c] = \E_\oplus[Y(1)|T(0)=1,R=c].
\]
For $r < c$, 
\begin{align*}
\E_\oplus[Y(1)|T(0)=1,R=r]
&= \E_\oplus[Y(T(0))|T(0)=1, R=r]\\
&= \E_\oplus[Y(T(0))|T(0)=1,Z=0, R=r]\\
&=\E_\oplus[Y|T=1,Z=0,R=r]. 
\end{align*}
Hence, 
\begin{align*}
\lim_{r \uparrow c}\E_\oplus[Y|T=1,Z=0,R=r] 
&= \lim_{r \uparrow c}\E_\oplus[Y(1)|T(0)=1,R=r]\\
&= \E_\oplus[Y(1)|T(0)=1,R=c]\\
&= \mu_\oplus.
\end{align*}
Likewise, under Assumption \ref{ass:AT-NT}(ii), we can show 
\begin{align*}
\lim_{r \downarrow c}\E_\oplus[Y|T=0,Z=1,R=r] 
&= \E_\oplus[Y(0)|T(1)=0,R=c] = \mu_\oplus.
\end{align*}

\subsection{Proof of Proposition \ref{prp:GLATE-rate}}
Observe that $d_\mathcal{G}(\tau_{\mathrm{GFRD}},\hat{\tau}_{\mathrm{GFRD}}) \leq C\{d(\mu_{0,\oplus},\hat{\mu}_{0,\oplus}) + d(\mu_{1,\oplus},\hat{\mu}_{1,\oplus})\}$ and 
\begin{align*}
d(\mu_{z,\oplus},\hat{\mu}_{z,\oplus}) &= d_\mathcal{H}\left(\Psi(\mu_\oplus) + \frac{\Psi(\nu_{z,\oplus}) - \Psi(\mu_\oplus)}{m_1- m_0},\Psi(\hat{\mu}_\oplus) + \frac{\Psi(\hat{\nu}_{z,\oplus}) - \Psi(\hat{\mu}_\oplus)}{\hat{m}_1- \hat{m}_0}\right)\\
&\leq \left(1 + \frac{1}{|\hat{m}_1 - \hat{m}_0|}\right)d_\mathcal{H}(\Psi(\mu_\oplus),\Psi(\hat{\mu}_\oplus)) + \frac{1}{|\hat{m}_1 - \hat{m}_0|}d_\mathcal{H}(\Psi(\nu_{z,\oplus}),\Psi(\hat{\nu}_{z,\oplus}))\\
&\quad + \left|\frac{1}{\hat{m}_1 - \hat{m}_0} - \frac{1}{m_1 - m_0}\right|d_\mathcal{H}(\Psi(\nu_{z,\oplus}),\Psi(\mu_\oplus))\\
&\leq \left(1 + \frac{1}{|\hat{m}_1 - \hat{m}_0|}\right)d(\mu_\oplus,\hat{\mu}_\oplus) + \frac{1}{|\hat{m}_1 - \hat{m}_0|}d(\nu_{z,\oplus},\hat{\nu}_{z,\oplus})\\
&\quad + \left|\frac{1}{\hat{m}_1 - \hat{m}_0} - \frac{1}{m_1 - m_0}\right|d(\nu_{z,\oplus},\mu_\oplus). 
\end{align*}
Therefore, 
\begin{align*}
d_\mathcal{G}(\tau_{\mathrm{GFRD}},\hat{\tau}_{\mathrm{GFRD}})&\leq C\left(1 + \frac{1}{|\hat{m}_1- \hat{m}_0|}\right)\{2d(\mu_\oplus,\hat{\mu}_\oplus) + d(\nu_{0,\oplus},\hat{\nu}_{0,\oplus}) + d(\nu_{1,\oplus},\hat{\nu}_{1,\oplus})\}\\
&\quad + C \left|\frac{1}{\hat{m}_1 - \hat{m}_0} - \frac{1}{m_1 - m_0}\right|\{d(\nu_{1,\oplus},\mu_\oplus) + d(\nu_{0,\oplus},\mu_\oplus)\}.
\end{align*}
Applying a similar argument as in the proof of Theorem 7.2 yields  the desired result. 

\section{Proofs for Section \ref{subsec:FRD-manifold}}
\subsection{Proof of Theorem \ref{thm:FRD-comp-id}}\label{subsec:App-FRD-id-manifold}
Applying the same argument as used to obtain (\ref{eq:FRD-id1}) and (\ref{eq:FRD-id2}),  
\begin{align}
&\nu_{1,\omega}^\sharp = \E[\mathrm{Log}_\omega(Y(T(1)))|R=c] = \lim_{r \downarrow c}\E[\mathrm{Log}_\omega(Y)|R=r], \label{eq:FRD-id1-2}\\
&\nu_{0,\omega}^\sharp = \E[\mathrm{Log}_\omega(Y(T(0)))|R=c] = \lim_{r \uparrow c}\E[\mathrm{Log}_\omega(Y)|R=r]. \label{eq:FRD-id2-2}
\end{align}
Observe 
\begin{align*}
\nu_{z,\omega}^\sharp 
&= \E[\mathrm{Log}_\omega(Y(T(z)))|T(1)>T(0), R=c]\Pr(T(1)>T(0)|R=c)\\
&\quad + \E[\mathrm{Log}_\omega(Y(T(z)))|T(1)=T(0), R=c]\Pr(T(1)=T(0)|R=c)
\end{align*}
and 
\begin{align}\label{eq:FRD-id-manifold-z}
\mu_{z,\omega}^\sharp 
&= \E[\mathrm{Log}_\omega(Y(T(z)))|T(1)>T(0), R=c] \nonumber \\
&= \E[\mathrm{Log}_\omega(Y(T(z)))|T(1)=T(0), R=c] \nonumber \\
&\quad + \frac{\nu_{z,\omega}^\sharp - \E[\mathrm{Log}_\omega(Y(T(z)))|T(1)=T(0), R=c]}{\Pr(T(1)>T(0)|R=c)}. 
\end{align}
This yields
\begin{align*}
\tau_{\mathrm{FRDD}}^\sharp
&= \mu_{1,\omega}^\sharp - \mu_{0,\omega}^\sharp \\
&= \frac{\nu_{1,\omega}^\sharp - \nu_{0,\omega}^\sharp}{\Pr(T(1)>T(0)|R=c)} \\
&= \frac{\nu_{1,\omega}^\sharp - \nu_{0,\omega}^\sharp}{\E[T(1)|R=c] - \E[T(0)|R=c]}.
\end{align*}
Combining this with (\ref{eq:FRD-id3}), (\ref{eq:FRD-id1-2}), and (\ref{eq:FRD-id2-2}) leads to the desired result. 

\subsection{Proof of Theorem \ref{thm:LATE-rate-manifold}}
The proof is omitted as it is similar to that of Theorem 7.2.

\section{Proofs for Section \ref{subsec:GFRD-manifold}}
\subsection{Proof of Proposition \ref{prp:GFRD-id-manifold}}
The proof is omitted as it is similar to that of Proposition \ref{prp:GFRD-id}. 

\subsection{Proof of Proposition \ref{prp:GLATE-rate-manifold}}
Observe that $d_\mathcal{G}(\tau_{\mathrm{GFRD}}^\sharp,\hat{\tau}_{\mathrm{GFRD}}^\sharp) 
\leq C\{d(\mu_{0,\omega,\oplus}^\sharp,\hat{\mu}_{0,\omega,\oplus}^\sharp) + d(\mu_{1,\omega,\oplus}^\sharp,\hat{\mu}_{1,\omega,\oplus}^\sharp)\}$ and 
\begin{align*}
&d(\mu_{z,\omega,\oplus}^\sharp,\hat{\mu}_{z,\omega,\oplus}^\sharp)\\
&= d\left(\mathrm{Exp}_\omega\left(\nu_\omega^\sharp + \frac{\nu_{z,\omega}^\sharp - \nu_\omega^\sharp}{m_1-m_0}\right),\mathrm{Exp}_\omega\left(\hat{\nu}_\omega^\sharp + \frac{\hat{\nu}_{z,\omega}^\sharp - \hat{\nu}_\omega^\sharp}{\hat{m}_1-\hat{m}_0}\right)\right)\\
&\leq C_\omega d_E\left(\nu_\omega^\sharp + \frac{\nu_{z,\omega}^\sharp - \nu_\omega^\sharp}{m_1-m_0}, \hat{\nu}_\omega^\sharp + \frac{\hat{\nu}_{z,\omega}^\sharp - \hat{\nu}_\omega^\sharp}{\hat{m}_1-\hat{m}_0}\right)\\
&\leq C_\omega\left\{\left(1 + \frac{1}{|\hat{m}_1 - \hat{m}_0|}\right)d_E(\nu_\omega^\sharp,\hat{\nu}_\omega^\sharp) + \frac{1}{|\hat{m}_1 - \hat{m}_0|}d_E(\nu_{z,\omega}^\sharp, \hat{\nu}_{z,\omega}^\sharp)\right\} \\ 
&\ \quad \quad \quad + C_\omega \left|\frac{1}{m_1 - m_0} - \frac{1}{\hat{m}_1 - \hat{m}_0}\right|d_E(\nu_{z,\omega}^\sharp, \nu_\omega^\sharp),\ z \in \{0,1\}. 
\end{align*}
Then 
\begin{align*}
&d_\mathcal{G}(\tau_{\mathrm{GFRD}}^\sharp,\hat{\tau}_{\mathrm{GFRD}}^\sharp) \\
&\leq CC_\omega\left(1 + \frac{1}{|\hat{m}_1 - \hat{m}_0|}\right)\{2d_E(\nu_\omega^\sharp,\hat{\nu}_\omega^\sharp) + d_E(\nu_{0,\omega}^\sharp, \hat{\nu}_{0,\omega}^\sharp) + d_E(\nu_{1,\omega}^\sharp, \hat{\nu}_{1,\omega}^\sharp)\}\\
&\quad + CC_\omega \left|\frac{1}{m_1 - m_0} - \frac{1}{\hat{m}_1 - \hat{m}_0}\right|\{d_E(\nu_{1,\omega}^\sharp, \nu_\omega^\sharp) + d_E(\nu_{0,\omega}^\sharp, \nu_\omega^\sharp)\}.
\end{align*}
Applying a similar argument in the proof of Theorem 7.2 then implies the desired result.

\section{Additional simulation with unequal curvature}
\label{sec:supp_unequal_curvature}
In the simulation study presented in the main text, the regression functions on the two sides of the cutoff have identical curvature. Under the null hypothesis, this implies that the leading-order local linear bias terms cancel. To assess the finite-sample behavior of the proposed inference procedure when this property does not hold, we consider an additional simulation in which the regression functions have different curvature on the two sides of the cutoff.

The simulation setup is identical to that in Section~5.2 of the main text except for the conditional mean edge weights. Specifically, for each edge $(i,j)$, we generate
\[
w_{ij}(R)=
\begin{cases}
1+0.5R^2+\epsilon_{ij}, & R<0,\\[2mm]
1+R^2+\Delta\tau_{ij}+\epsilon_{ij}, & R\ge0,
\end{cases}
\]
where $\epsilon_{ij}\sim\mathrm{Unif}(0,0.1)$ independently across edges and $\tau_{ij}$ is defined as in Section~5.2. The remainder of the simulation setup, including the cross-validation bandwidth selection procedure, is unchanged. Throughout this experiment, we set $\Delta=0$, so that the null hypothesis holds. Under the null, the conditional mean function is continuous at the cutoff with equal first derivatives from the left and right, but different second derivatives. Consequently, unlike the simulation considered in the main text, the leading-order local linear bias terms do not cancel.

Table~\ref{tab:unequal_curvature_size} reports the empirical rejection probabilities of the proposed bootstrap test based on 500 Monte Carlo replications.

\begin{table}[ht]
\single
\centering
\caption{Empirical rejection probabilities under the null hypothesis for the additional simulation with unequal curvature across the cutoff. The nominal significance level is $0.05$.}
\label{tab:unequal_curvature_size}
\begin{tabular}{ccccc}
\hline
Sample size & 100 & 200 & 500 & 1000 \\
\hline
Empirical rejection probability & 0.026 & 0.032 & 0.052 & 0.058 \\
\hline
\end{tabular}
\end{table}

The empirical rejection probabilities are close to the nominal significance level for moderate and large sample sizes ($n=500$ and $1000$), while the procedure is somewhat conservative for smaller sample sizes. Overall, this additional simulation provides further empirical evidence that the proposed bootstrap inference procedure maintains accurate empirical size under a setting where the leading-order bias terms no longer cancel.
\end{document}